\documentclass[fleqn,usenatbib]{mnras}

\usepackage{newtxtext,newtxmath}
\usepackage[T1]{fontenc}

\DeclareRobustCommand{\VAN}[3]{#2}
\let\VANthebibliography\thebibliography
\def\thebibliography{\DeclareRobustCommand{\VAN}[3]{##3}\VANthebibliography}

\usepackage{graphicx}	% Including figure files
\usepackage{amsmath}	% Advanced maths commands
\usepackage{hyperref}
\usepackage{color}
\definecolor{Blue}{rgb}{0,0.08,0.65}
\definecolor{Red}{rgb}{0.65,0.08,0.05}
\definecolor{Green}{rgb}{0.35,0.45,0.25}
\definecolor{Orange}{rgb}{1.0,0.5,0.15}
\definecolor{Purple}{rgb}{0.5,0.0,0.5}

\usepackage{ulem} 
\newcommand\redout{\bgroup\markoverwith
{\textcolor{red}{\rule[0.5ex]{2pt}{0.8pt}}}\ULon}

\title[CNN photometric redshifts in the SDSS at $r\leq 20$]{CNN photometric redshifts in the SDSS at $r\leq 20$}

\author[M. Treyer et al. ]{
M. Treyer,$^{1}$ R. Ait-Ouahmed,$^{1}$ J. Pasquet,$^{2,3}$  S. Arnouts,$^{1}$ E. Bertin,$^{4,5}$ and D. Fouchez$^{6}$
\\
$^{1}$Aix Marseille Universit\'e, CNRS, CNES, LAM, Marseille, France\\ $^{2}$AMIS - Université Paul-Valéry - Montpellier 3\\
$^{3}$UMR TETIS - Inrae, AgroParisTech, Cirad, CNRS, Univ. Montpellier, France\\
$^{4}$Sorbonne Universit\'e, CNRS, IAP, Paris, France\\
$^{5}$CFHT, Hawaii\\ 
$^{6}$Aix Marseille Univ, CNRS/IN2P3, CPPM, Marseille, France
}

 \date{}

\begin{document} 
\label{firstpage}
\pagerange{\pageref{firstpage}--\pageref{lastpage}}
\maketitle
 
\begin{abstract}
We release photometric redshifts, reaching $\sim$0.7, for $\sim$14M galaxies at $r\leq 20$ in the 11,500 deg$^2$ of the SDSS north and south galactic caps. These estimates were inferred from a convolution neural network (CNN) trained on $ugriz$ stamp images of galaxies labelled with a spectroscopic redshift from the SDSS, GAMA and BOSS surveys. Representative training sets of $\sim$370k galaxies were constructed from the much larger combined spectroscopic data to limit biases, particularly those arising from the over-representation of Luminous Red Galaxies. The CNN outputs a redshift classification that offers all the benefits of a well-behaved PDF, with a width efficiently signaling unreliable estimates due to poor photometry or stellar sources. The dispersion, mean bias and rate of catastrophic failures of the median point estimate are of order $\sigma_{\rm MAD}=0.014$, <$\Delta z_{\rm norm}$>$=0.0015$, $\eta(|\Delta z_{\rm norm}|>0.05)=4\%$ on a representative test sample at $r<19.8$, out-performing currently published estimates. The distributions in narrow intervals of magnitudes of the redshifts inferred for the photometric sample are in good agreement with the results of tomographic analyses. The inferred redshifts also match the photometric redshifts of the redMaPPer galaxy clusters for the probable cluster members. The CNN input and output are available at: \url{https://deepdip.iap.fr/treyer+2023}. 
\end{abstract}


\section{Introduction}\label{sec:intro}
As photometric redshifts have become inescapable in most cosmological endeavors, so have machine learning techniques to predict them. Spectroscopy alone can no longer fulfil the task of measuring the distances to the millions of sources detected in current photometric sky surveys, e.g. DES \citep[]{DES2016} and KIDS \citep[]{deJong2013}, let alone future ones such as Euclid \citep[]{Laureijs2011} and Vera Rubin/LSST \citep[]{Ivezic2019}. 
Spectral energy distribution (SED) template fitting techniques have been widely used for several decades to estimate the redshifts of galaxies from multi-band photometry, the so-called photometric redshifts. This technique relies on a set of observed or modeled SEDs assumed to represent the diversity of observed galaxies \citep[e.g.][]{Arnouts1999,Ilbert2006,Brammer2008}. It allows physical parameters to be derived in addition to redshift probability density functions (PDF). The first neural networks for the estimation of photometric redshifts emerged in the early 2000s \citep{firth2003MNRAS.339.1195F,Tagliaferri2003, Collister2004}. Since then, machine learning progressed enormously, helped by the growing wealth of data and computing capabilities. A machine learning algorithm learns to map the multi-dimensional photometric information using labelled and/or unlabelled data, i.e. data with or without known spectroscopic redshifts (supervised and unsupervised training methods respectively, which can be combined). The accuracy of the photometric redshifts derived from such optimized mapping is much higher than via SED fitting provided the galaxies span the same parameter space as the training sample (see \citet[][]{Brescia2021} for a review and  \citet{Henghes2022} for a comparison of several such methods). Unlike SED fitting however, most machine learning methods have only provided point estimates. Few studies have had a probabilistic approach able to estimate uncertainties, \citet{Sadeh2014} and \citet{Sadeh2016} being the first to provide redshift PDFs via a classifier. \citet{Jones2023} recently proposed Bayesian neural networks assuming gaussians PDFs “as a promising way to provide accurate predictions with uncertainty estimates”.

"Deep learning" is the latest step forward in the pursuit of photometric redshifts. Thanks to the development of convolutional neural networks \citep[CNN,][]{Lecun1998}, and with the help of graphics processing units (GPUs), (regularly sampled) images may now be used directly instead of, or sometimes in addition to extracted features (magnitudes, colors, etc.), which only transmit a fraction of the available photometric information, with variable reliability. Deep neural networks were designed to handle the much larger amount of information contained in the image pixels \citep{Hoyle2016,disanto2018}. They consist of successive layers of artificial neurons, each performing a linear transformation of the input followed by a non-linear "activation function". Weights are updated as the network processes (learns from)  batches of the training data, until a suitable solution is found (a loss function is minimized).

%B16+Rarr\'io & Zarattini (2020) kNN from extracted photometry
\citet{P19} (hereafter P19) presented a CNN to estimate photometric redshifts straight from multi-band stamp images of galaxies, without any feature extraction nor color images. The network was designed as a classifier into small contiguous redshift bins, the output of which was normalised to produce PDFs. As a proof of concept, it was applied to the flux-limited, $r<17.8$, spectroscopic Main Galaxy Sample of the SDSS \citep[]{York2000}, using $ugriz$ stamp images and the galactic reddening values along the lines of sight as input data, with the spectroscopic redshifts as labels in the context of supervised learning. The weighted mean values of the so-called PDFs were found to be photometric redshits of unprecedented accuracy in the limited redshift range of interest ($z<0.4$). Other methods exploiting galaxy images have since been proposed \citep{Hayat2021,Henghes2022,Schuldt2021,Dey2021}.

Here we use a more complex CNN architecture to estimate photometric redshifts for the $\sim$14 million galaxies at $r\leq 20$ without spectroscopy in the SDSS footprint. The photometric and spectroscopic data are presented in Section \ref{sec:data} (and Appendix \ref{sec:skymaps}). The architecture, input and output of the network are described in Section \ref{sec:CNN} (and Appendix \ref{sec:architecture_apdx}). Training experiments are described in Section \ref{sec:training}. The performance of the final experiment is tested in Section \ref{sec:test}. Its inference on the photometric sample is presented in Section \ref{sec:inference} (and Appendices \ref{sec:uncleanphoto} and \ref{sec:tomographer_apx}). We conclude this work in Section \ref{sec:conclusion}. Additionally, a recipe for classifying galaxies into blue/star-forming or red/passive types is given in Appendix \ref{sec:galtypes} and alternative training strategies are explored in Appendix \ref{sec:alternatives}. The CNN input and output are available at: \url{https://deepdip.iap.fr/treyer+2023}.
%The photometric redshifts can be downloaded from:  \url{https://deepdip.iap.fr/}.
%%%%%%%%%%%% DATA %%%%%%%%%%%%%%%%%%%%%%%%%%%%%%%%%

\section{The data}
\label{sec:data}

The data detailed below are summarized in Table \ref{table:data}.

\subsection{The photometric data}
\label{subsec:photodata}

Our catalog is drawn from the SDSS data release 16 \citep[DR16,][]{SDSS_DR16}. The SDSS is a multi-band imaging and spectroscopic redshift survey that was conducted on a dedicated 2.5m telescope at Apache Point Observatory in New Mexico. It provides photometry in the $ugriz$ passbands over $\sim$11,500 deg$^2$ of the North and South galactic caps to a limiting magnitude of $r=22.5$. Via the SDSS CasJob web service, we retrieved $\sim$15.3M catalog entries of non point-like sources ({\bf type}=3) with dereddened petrosian  magnitudes $r\le 20$, of which $\sim$1.5M have spectroscopic redshifts. Thus the final number of purely photometric sources for which we infer redshifts is $\sim$13.8M. A sky map of this dataset is shown in Appendix \ref{sec:skymaps}.

Photometric redshifts by \citet{Beck2016} (hereafter B16) are available for nearly all these sources. They were computed using a $k$-nearest neighbor algorithm \citep[kNN,][]{Csabai2007} with five dimensions (the $r$-band magnitude and 4 colors: $(u-g)$, $(g-r)$, $(r-i)$, $(i-z)$). The training data included deep, high redshift spectroscopic surveys in addition to the SDSS. A 3D error map ($r, g-r, r-i$) built on the uncertainties measured for spectroscopic galaxies helps to identify insecure estimates based on the position of a galaxy in this grid. The accuracy of these photometric redshifts makes them a reference in machine learning based on photometric measurements. Last but not least, they are still the only ones available for comparison purposes.  

\subsection{The spectroscopic data} 
\label{subsec:specdata}

The $\sim$1.5M spectroscopic redshifts, used as training labels, come from the SDSS, GAMA and BOSS spectroscopic surveys described below, matched to the SDSS DR16 photometric catalog described above. Figure \ref{fig:nz_surveys} shows the redshift distributions of these three surveys and Fig. \ref{fig:magz_surveys} their respective magnitude/redshift and $(u-g)/(g-r)$ color distributions. A sky map of the spectroscopic data is shown in Appendix \ref{sec:skymaps}. 
%(Fig. \ref{fig:maps}, bottom panel). 

\subsubsection{The SDSS survey}

The SDSS spectroscopy is nearly complete to $r=17.8$, totalling $\sim$660k galaxies, but reaches fainter magnitude to much lower completeness with targeted populations, adding $\sim$162k galaxies with $17.8<r<20$. We use the specific star-formation rates derived from the SDSS spectra by \citet{Brinchmann04} in the data release 12 \citep[DR12;][]{Alam2015} to draw an empirical separation between blue, star-forming galaxies and red, passive galaxies in the redshift/$(u-g)/(g-r)$ space (Appendix \ref{sec:galtypes}). This separation is used to balance the training samples and evaluate the performance of the training experiments in the two populations. 

\subsubsection{The GAMA survey}

The GAMA survey \citep{Driver2009,Driver2011,Liske2015} is a joint European-Australian spectroscopic survey combining UV to FIR photometric data from several ground and space based programs, including SDSS. The spectroscopy was carried out using the 2dF/AAOmega multi-object spectrograph on the Anglo-Australian Telescope, building on previous spectroscopic surveys such as SDSS, the 2dF Galaxy Redshift Survey and the Millennium Galaxy Catalogue (MGC). 
We use the 4 equatorial fields (G02, G09, G12, and G15) available in the data release 3 and 4 \citep[DR4,][]{Driver2022}, covering a total of $\sim 235$ deg$^2$. The spectroscopy is 98\% complete to $r = 19.8$ \citep{Liske2015}, except for the G02 field where only the region north of Dec $\sim$ 6deg was observed to high completeness. This provides us with a sample of $\sim$210k spectroscopic galaxies at $r<20$ (90\% of which at $r>17.8$) matched to the photometric catalog. GAMA constitutes the main component of our training set as its completeness makes it most representative of the photometric dataset.  
           
\subsubsection{The BOSS survey}

We retrieved an additional 486k spectroscopic sources from, essentially (98.2\%), the BOSS survey \citep{Dawson2013}. These are dominated by Luminous Red Galaxies (LRG). We will refer to this sample as "BOSS", although a small contribution ($\sim$8538 galaxies) comes from other deep redshift surveys, namely: VVDS Wide and Deep \citep{LeFevre2013}, DEEP2 \citep[DR4,][]{Newman2013}, VIPERS \citep[DR2,][]{Scodeggio2018}, UDSz \citep[][]{McLure2013,Bradshaw2013}, zCOSMOS-bright \citep[]{Lilly2007}. 
% ['2QZ', '2SLAQ-LRG', '2SLAQ-QSO', '2dFGRS', '3DHST', '6dFGS','CANDELS', 'COMPARAT2015', 'COSMOS', 'COSMOS_followup','DEEP2', 'MGC', 'NED', 'Trichas2010', 'UDSz', 'UNKNOWN','VIPERS', 'VVDS', 'VVDS Deep', 'WiggleZ', 'eboss', 'hCOS20.6','hCOSMOS', 'zCOSMOS bright']
% the GAMA survey \citep[Data release 3,][]{Baldry2018}; 
% the WiggleZ survey  \citep[final release, ][]{Drinkwater2018}; 

\begin{table}
\centering
\caption{The photometric and spectroscopic data (Sections \ref{subsec:specdata} and \ref{subsec:photodata}).}
\label{table:summary_spectro}
\begin{tabular}{ |c|c|c|c|c|}
 \hline
 Survey & Magnitude & Size & Spectra  \\
 \hline
SDSS & $10\le r\le20$ & 13.8M & no \\  
SDSS & $r \le 17.8$ & 660k & yes \\   
SDSS & $17.8\le r\le 20$ & 162k & yes \\ 
GAMA & $r \le 20$ & 210k & yes \\
BOSS & $r \le 20$ & 486k & yes \\
\hline
\label{table:data}
\end{tabular}
\end{table}

\begin{figure}
\includegraphics[width=8.3cm]{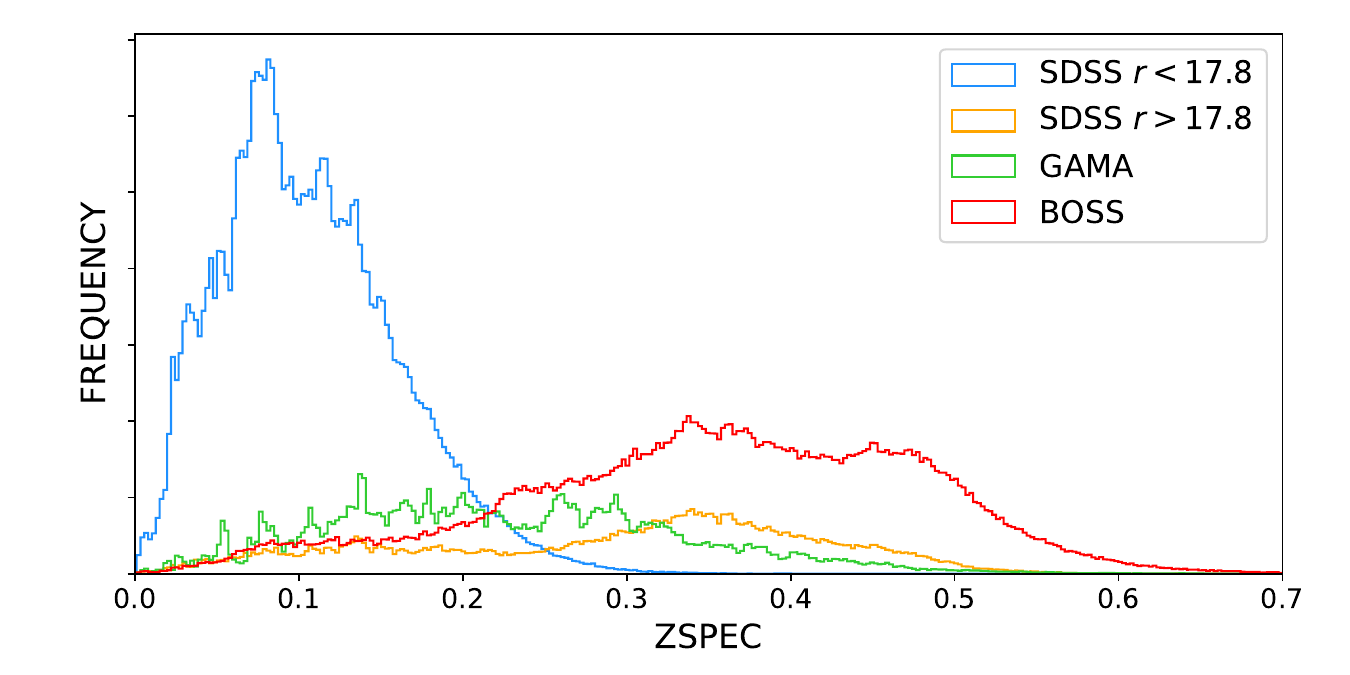}
\caption{The redshift distributions of the SDSS, GAMA and BOSS surveys presented in Section \ref{subsec:specdata}, totalling 1.5M galaxies.}
\label{fig:nz_surveys}
\end{figure}

\begin{figure}
\includegraphics[width=4.2cm]{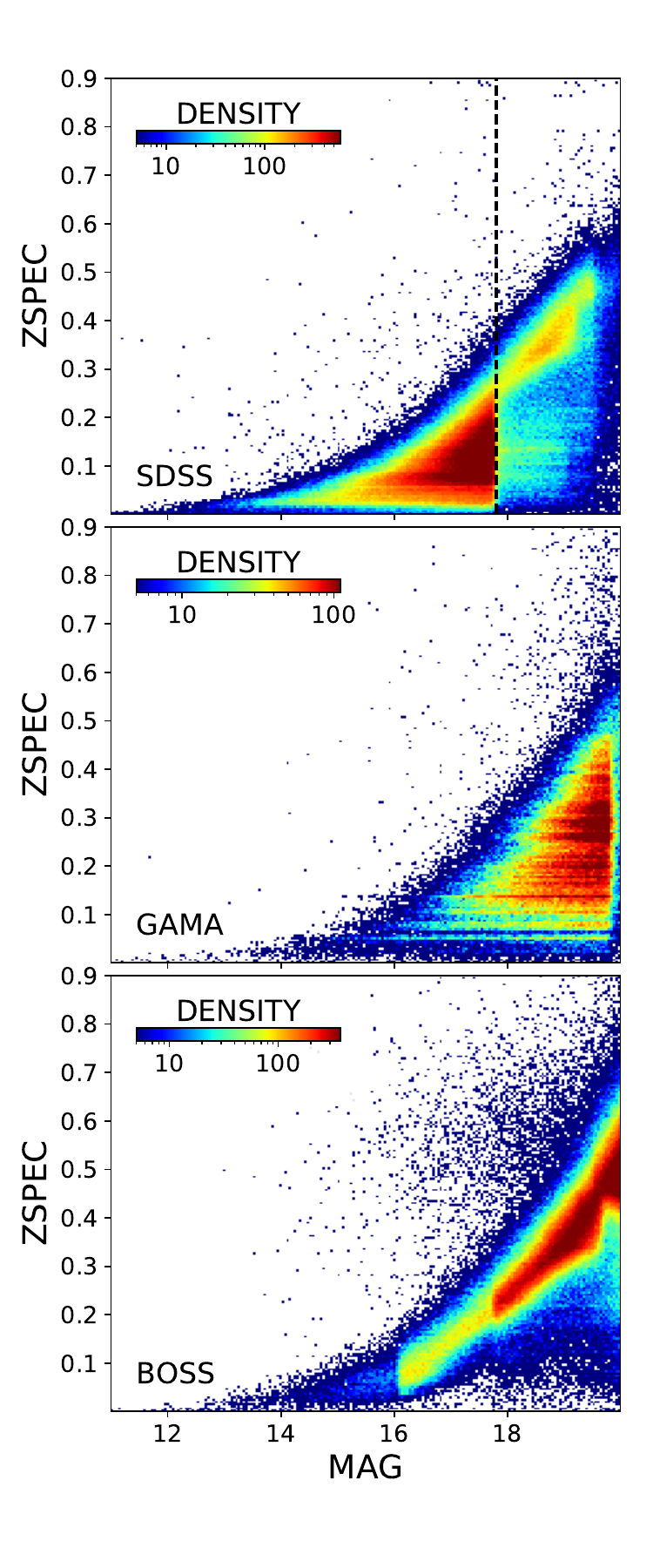}
\includegraphics[width=4.2cm]{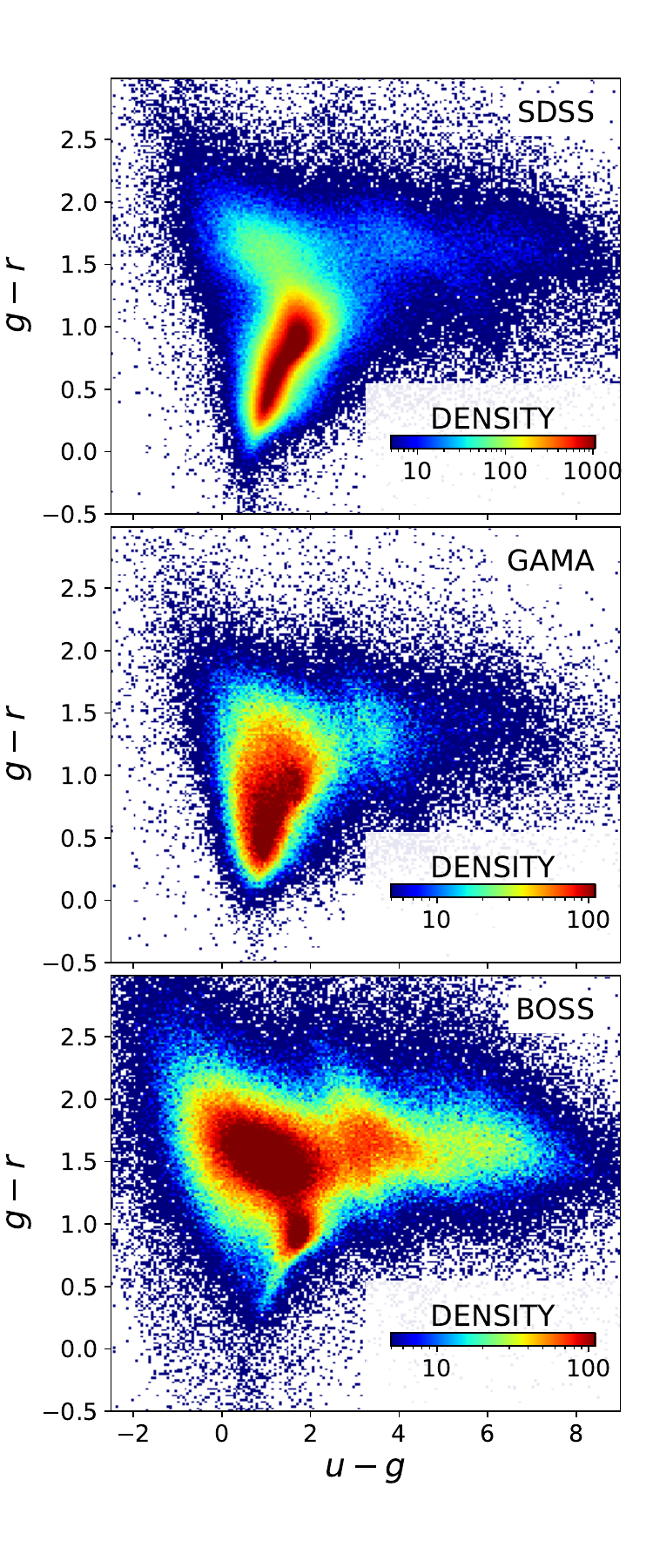}
\caption{Magnitude/redshift and $(u-g)/(g-r)$ distributions of the SDSS, GAMA and BOSS surveys (Section \ref{subsec:specdata}).}
\label{fig:magz_surveys}
\end{figure}

%%%%%%%%%%%% CNN %%%%%%%%%%%%%%%%%%%%%%%%%%%%%%%%%

\section{The CNN}
\label{sec:CNN}

\subsection{Architecture}
\label{subsec:architecture}

The present network is a more complex version of the P19 CNN, intended for a more complex dataset. Its architecture is diagrammed in Fig.~\ref{fig:cnn} and \ref{fig:inception}. Figure \ref{fig:CNN_layers} in Appendix \ref{sec:architecture_apdx} lists all the layers with their type, shape, number of parameters and the layer(s) they are connected to upstream. We refer to P19 for a pedagogical description of the role played by the different types of layers. Their Fig. 4 can also be compared to our Fig.~\ref{fig:cnn}.  

As in P19, the input data consist of images of galaxies in the five $ugriz$ bands of the SDSS surveys, with the galactic extinction along the line of sight added downstream before the fully connected layers (Section \ref{subsec:images}). The training labels are spectroscopic redshifts.

Here too, we choose to handle the redshift estimation task by means of a classification rather than, but aided by, a non-linear regression. The gain of using a classification rather than a regression or of adding a regression to the classification is negligible with a rich data set such as the bright SDSS used by P19, but it proves more significant with sparser training sets such as the present data (Appendix \ref{subsec:regression}) and even more so with high redshift data (Ait Ouahmed et al. 2023). 

The classes correspond to narrow, mutually exclusive, redshift bins, i.e. each training galaxy belongs to a single class (one-hot encoding of the spectroscopic redshifts). The classifier is trained using the softmax cross-entropy loss function \citep{Baum1987,Solla1988} and the regression with the root mean square error. The two loss functions are simply added (after trying different weighted sums). 

A softmax activation function \citep{Bridle1990} applied to the output layer of the classifier normalizes the outputs to 1. Such outputs were shown, both theoretically and experimentally, to provide good estimates of the posterior probability of classes in the input space \citep{Richard1991,Rojas1996} 
provided the network is sufficiently complex and properly trained. Whether that is the case here may be questioned but we find the classification outputs to be very useful probability density function (PDF) proxies. We will call them "PDF". 
%to avoid pompous arguments.

Compared to P19, the present network has 1 additional convolutional layer upstream, 6 inception blocks of similar complexity instead of 4 + 1 simpler one, and 3 additional convolutional layers following the last inception module. These have no padding and are followed by an average pooling layer, which reduces the number of trainable parameters to $\sim$7M compared to $\sim$27M in P19. Apart from the softmax activation function used in the last dense layer to produce the "PDFs", all but two of the non-linear activation functions (introducing non-linearity into the network) are the commonly used ReLU \citep[Rectified Linear Unit,][]{NairHinton2010}. The second convolutional layer and the last dense layer before the regression output use a hyperbolic tangent activation function, which clips the dynamic range. The network is trained with the Adam optimizer \citep{KingmaBa2015}, a stochastic gradient descent method based on the adaptive estimation of first-order and second-order moments.

We note that the results presented below are quite robust to the specifics of the CNN architecture. Replacing the inception blocks by simple convolutions while retaining the same depth and number of trainable parameters, only slightly degrades the metrics. The gain from the inception modules is of the order of that found with averaging a large number of models (Fig. \ref{fig:metrics_average25models}): it takes averaging a large number of trained networks without inception blocks to achieve results similar to no averaging using inception blocks.

\begin{figure}
\hspace{1.5cm}
\includegraphics[width=13.cm]{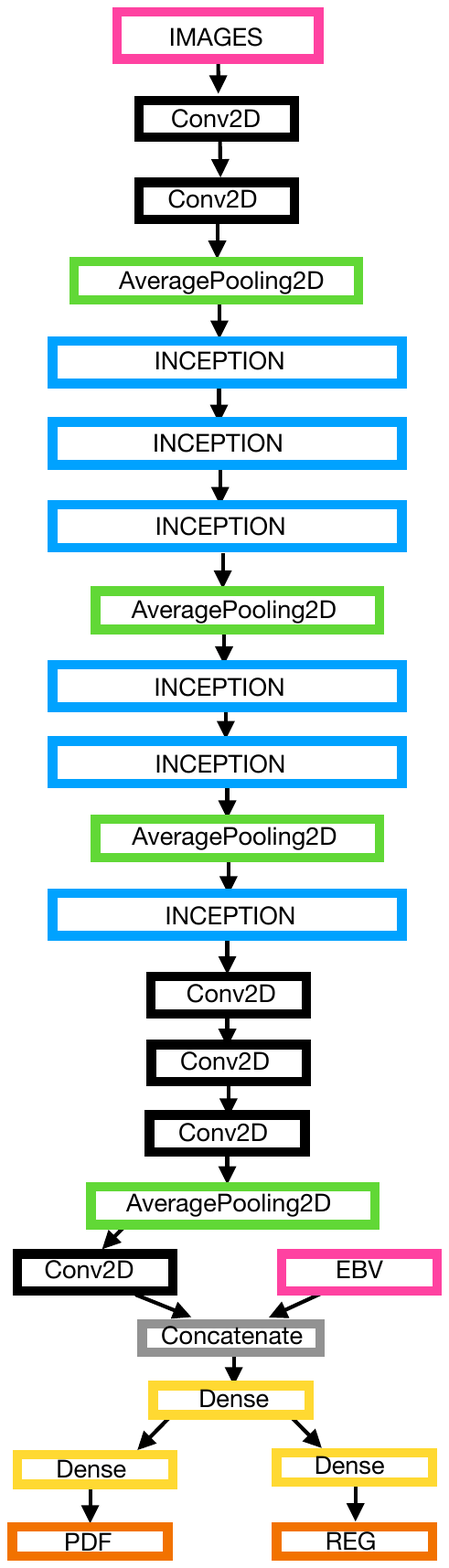}
%\vspace{-0.5cm}
\caption{The CNN architecture. The inception module is detailed in Fig. \ref{fig:inception}.}
\label{fig:cnn}
\end{figure} 

\begin{figure}
\includegraphics[width=8.5cm]{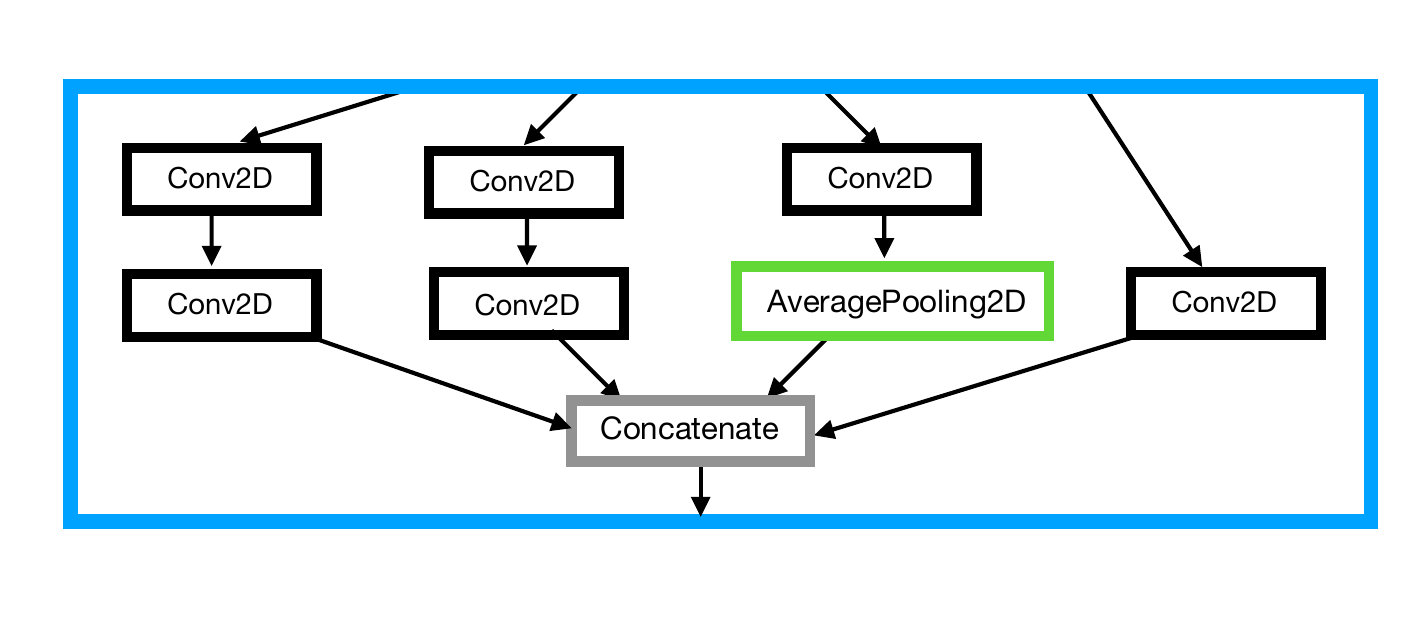}
\vspace{-0.5cm}
\caption{Inception module}
\label{fig:inception}
\end{figure}

\subsection{Training input}
\label{subsec:images}

The CNN is fed $64 \times 64$ pixel image cutouts centered on the galaxy coordinates in the five $ugriz$ SDSS bands, to which we add the \citet{Schlegel1998} galactic extinction value along the line of sight.
%% Description of the new image making procedure (EB)
The much larger number of galaxies and the different samples involved in the analysis compared to P19 requires a more efficient approach for generating the cutouts. Instead of extracting and resampling individual SDSS  frames for every galaxy, we adopt the following procedure:
\begin{itemize}
\item Using \textsc{SWarp} \citep{2002ASPC..281..228B}, we first re-project at once the whole SDSS imaging survey on a grid of $27\,070$ overlapping tiles covering the survey footprint in the five filters, relying  on the WCS parameters of the input image headers \citep{2002A&A...395.1077C} for the astrometry. The number of input frames contributing to a given output pixel ranges from one or two for "regular" SDSS images, to 64 for some of the galaxies in Stripe 82. Each output tile is $18\,192 \times 18\,192$ pixels wide ($2^\circ \times 2^\circ$ with 0.396" pixels), and is aligned with the local North-South axis using the ZEA (zenithal equal area) projection. 5 arcmin overlaps between nearest neighbors at mid-width/height guarantee that any cutout is entirely contained in at least one of the tiles.
\item Galaxy sample requests are organized by tile, and $64 \times 64$ pixel cutouts are extracted without resampling around every projected galaxy position.
\end{itemize}
While the new procedure is more than two orders of magnitude faster than that of P19, the generated cutouts are not centered as precisely (up to half-a-pixel only), and the direction to the local north is not always as perfectly aligned with the vertical axis of the pixel grid. However we did not find these changes to have a measurable impact on the quality of the inferred redshifts.
A (64,64,5) data cube is thus produced for the $\sim$15.3M sources in the DR16 photometric catalog.
%% End of the description of the new image making procedure (EB)

\iffalse
\subsection{Training protocol}
\label{subsec:protocol}

For a given training set, the database is split into 5 cross-validation samples. Each of these is used in turn for testing while the remaining 80\%, augmented with randomly flipped and rotated images\footnote{Note that these data augmentation processes can have an impact on the training if the image transfer function is not isotropic, e.g., if differential chromatic refraction is not negligible in the data.}, is used for training. This operation is repeated 5 times with randomly initialized weights. For a given cross-validation sample, the final PDFs are the average of these 5 classifications. For test galaxies, not part of the training set, the final PDF is the average of 25 classifications. The regression values are similarly averaged. The variance between the models and the gain from averaging them are illustrated in Appendix \ref{sec:training_apdx}. 
\fi 

\subsection{Output assessment}
\label{subsec:assessement}

\subsubsection{"PDF"}
\label{subsubsec:PDF}

We use several tests and quantities designed to evaluate PDFs to assess the 
%PDF-like 
behavior of our "PDFs":
\begin{itemize}
    \item The Probability Integral Transform statistic \citep[PIT,][]{dawid.2307/2981683} is based on the histogram of the cumulative probabilities at the true value (${\rm CDF}_i =\sum_{z\le z_i} {\rm PDF}_i(z)$ for galaxy $i$ at spectroscopic redshift $z_i$). A flat PIT distribution is expected from well calibrated PDFs, whereas convex or concave distributions point to over or under-confident PDFs \citep{Polsterer2016}. Indeed excessively narrow (overconfident) PDFs will miss the target too often, overproducing PIT values close to 0 or 1, whereas PDFs that are too wide (underconfident) will encompass the true redshifts more often than not, overproducing intermediate PIT values. 

\item The credibility test proposed by \citet{Wittman2016} (hereafter WBT) is based on the cumulative distribution of the "threshold credibilities", defined as the cumulative probabilities equal to or above the probability at the true value ($c_i=\sum_{{\rm PDF}_i(z) \ge {\rm PDF}_i(z_i)} {\rm PDF}_i(z)$), i.e. the smallest credible interval (CI) in which the spectroscopic redshift of a galaxy lies. With well calibrated PDFs, 1\% of the galaxies have their spectroscopic redshift within their 1\% CI, 2\% within their 2\% CI, etc., which translates into the cumulative distribution of $c_i\le c$ being equal to c.

\item The Continuous Ranked Probability Score \citep[CRPS, well known in meteorological predictions,][]{Hersbach} is a quadratic measure of the difference between the forecast cumulative distribution function (CDF) and the empirical CDF of the observation \citep{Zamo2018}, here a unit step function around the spectroscopic redshift ($CRPS_i$=$\int_{-\infty}^{z_i} CDF_i(z)^{2}dz + \int_{z_i}^{+\infty} (CDF_i(z)-1)^{2} dz$). It can be viewed as a generalization of the MAE to distributional predictions.

\item We quantify the uncertainty of the "PDFs" by the width of the 68\% central credible interval, i.e. the redshift width encompassing 68\% of the distribution after chopping off the left and right wings in equal measure.
%a width defined as the redshift interval encompassing 68\% of the distribution after chopping off the left and right wings in equal measure. 

\item Other important aspects of the redshift estimate contained in the full shape of the "PDF" can also be estimated, e.g. skewness and multi-modality.
\end{itemize}

Although passing one test does not ensure PDF quality \citep{Amaro2019} nor, for that matter, PDF status, we expect several successful tests combined with measures of the accuracy of the point estimates (next section) to provide some level of photometric redshift reliability.

%In other words the distribution can be deformed into a unimodal one by moving the CDF by at most the dip at each point, and the dip is the smallest number for which this is true.

\subsubsection{Point estimates}
\label{subsubsec:metrics}

Although redshift PDFs may be directly incorporated into Bayesian schemes in certain cosmological studies (e.g., inferring cosmological parameters), point estimates are required for many others (especially if the "PDFs" are not true PDFs). We consider the following 4 photometric redshift estimators:  
i/ the weighted mean ($z_{mean}$), ii/ the median value ($z_{med}$) and iii/ the peak value ($z_{peak}$) of the "PDF", and iv/ the regression output ($z_{reg}$). The metrics we use to assess the accuracy of these point estimates are identical to P19. We define:
\begin{itemize}
\item the \textbf{normalized residuals} $\Delta z_{\rm norm}= (z_{cnn}-z_{spec})/(1+z_{spec})$
\item the \textbf{prediction bias} $<\Delta z_{\rm norm}>$ (mean of the residuals)
\item the \textbf{deviation} $\sigma_{\rm MAD}=1.4826\times {\rm MAD}$, where MAD (Median Absolute Deviation) is the median of $|\Delta z_{\rm norm} - \textrm{Median}(\Delta z_{\rm norm})|$
\item the \textbf{fraction $\eta$  of outliers} with $|\Delta z_{\rm norm}|$>0.05
\end{itemize}

%%%%%%%%%%%%%% Section 5: TRAININGS %%%%%%%%%%%%%%%%%%%%%%

\section{Training experiments}
\label{sec:training}

\subsection{SDSS at $r\leq 17.8$}
\label{subsec:sdss}

For the purpose of comparison with P19, we first train the CNN with the same dataset: 
$\sim$510k SDSS galaxies at $r<17.8$, the same binning: 180 redshift classes in the range $0<z<0.4$ with constant width $\delta z = 0.4/180$, and the same protocol as P19: the database is split into 5 cross-validation samples, each one used in turn for testing while the remaining 80\%, augmented with randomly flipped and rotated images\footnote{Note that these data augmentation processes can have an impact on the training if the image transfer function is not isotropic, e.g., if differential chromatic refraction is not negligible in the data.}, is used for training (other training parameters are given in Appendix \ref{sec:architecture_apdx}). This operation is repeated 5 times with randomly initialized weights. The final "PDFs" are the average of the 5 classification outputs. 

\begin{figure}
\begin{center}
\includegraphics[width=8.3cm]{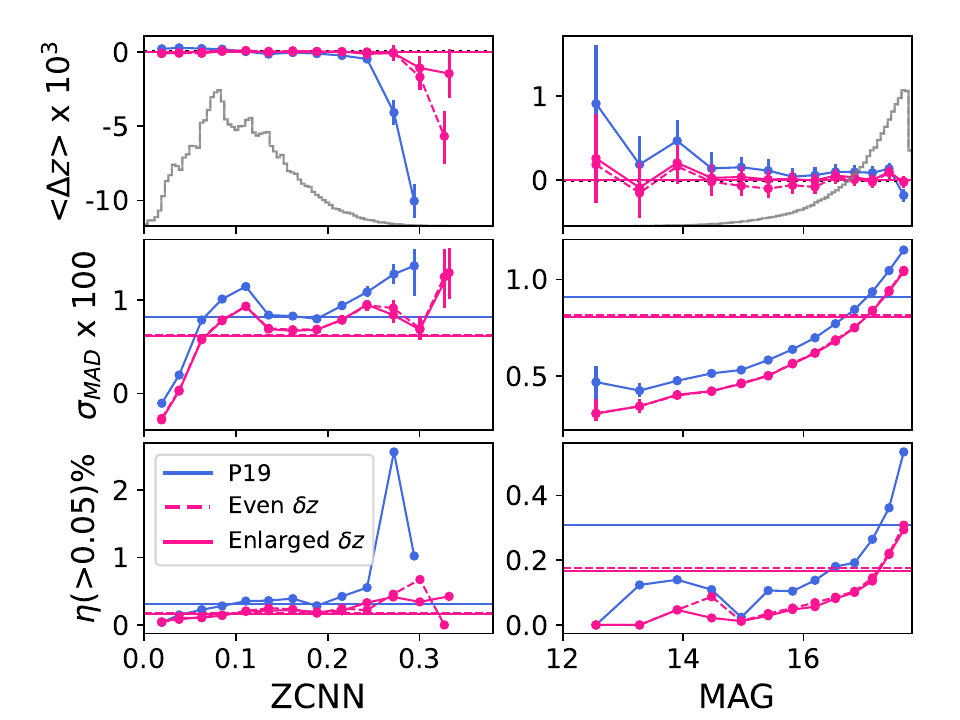}
\includegraphics[width=8.3cm]{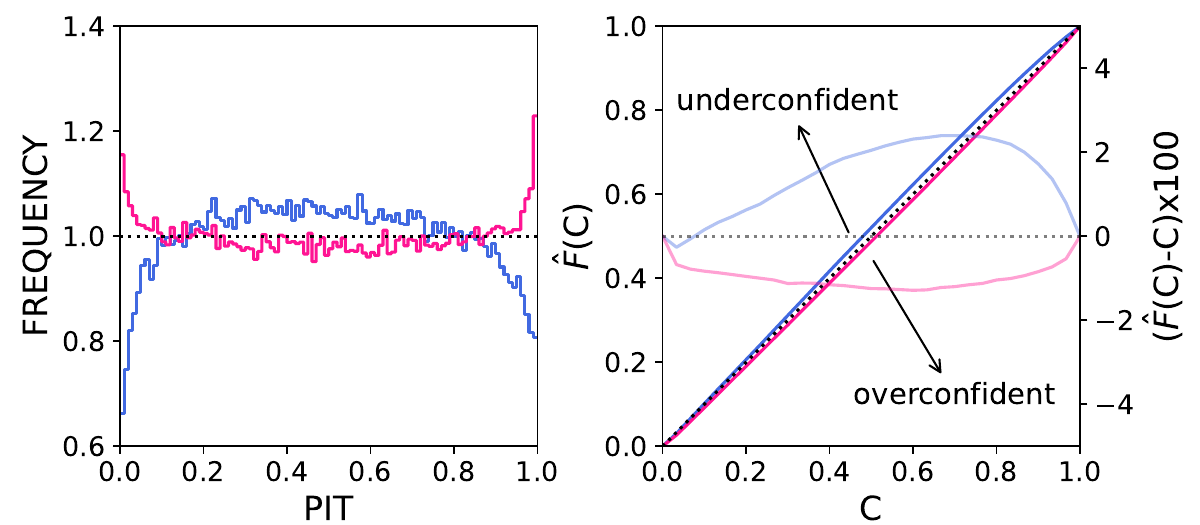}
\end{center}
\caption{Comparison between the P19 results (in blue) and the present work (in pink) for the SDSS sample at $r\le 17.8$.  {\bf Top panels:} bias, $\sigma_{\rm MAD}$ and catastrophic failures as a function of $z_{mean}$ and magnitude. The dashed and solid pink lines show the present metrics before and after expanding the binning at $z\gtrapprox 0.33$; {\bf Bottom panels:} the PIT and WBT tests. The PIT distribution is expected to be flat, the WBT test is expected to follow the unity line. The departures from the unity line are overlayed in faded colors, with units along the right hand side y-axis ($10^{-2}$).}
\label{fig:sdss_rlt178}
\end{figure}

Figure \ref{fig:sdss_rlt178} shows how this network compares with P19 (pink versus blue lines). The top panels show the bias, $\sigma_{\rm MAD}$ and outlier fraction $\eta$ as a function of CNN $z_{mean}$ and magnitude. All metrics are significantly reduced. However the dashed pink line in the top left panel shows that the new CNN is plagued with a redshift ceiling effect similar to P19, manifest as a steep drop in the bias at $z_{mean}\sim 0.3$, above which no $z_{mean}$ is predicted despite spectroscopic redshifts reaching higher values. We found that this effect could be mitigated by enlarging the bins at the highest, underpopulated redshifts, so that each class contains at least 20 training galaxies ($\delta z=0.0022$ at $z<0.33$, increasingly larger at $z>0.33$, adding up to 158 bins). The result is shown as the solid pink line. Sparse sampling may not be the only reason for this issue. The $z \gtrapprox 0.3$ tail of the SDSS data at $r<17.8$ is entirely populated by red galaxies and these are affected by a color degeneracy at this particular redshift, which distorts their CNN redshift distribution (see Section \ref{subsec:reddegeneracy}). The lower panels of Fig. \ref{fig:sdss_rlt178} show the PIT and WBT tests. According to both, the new "PDFs" are slightly over-confident where P19's were under-confident. Their mean CRPS is 0.0060, versus 0.0067 for P19.

The performance of the various redshift estimators are shown in Table ~\ref{table:point_estimates_perfs_sdss}, with the P19 results in parenthesis. P19 had used $z_{mean}$ as optimal point estimate as it minimizes the bias and the rate of catastrophic failures without significantly degrading $\sigma_{\rm MAD}$. However each point estimate has pros and cons: $z_{med}$ tends to minimize the dispersion at the expense of the bias, $z_{peak}$ also optimizes the dispersion but maximises the catastrophic failures and generates very noisy redshift distributions, while $z_{reg}$ is a slightly degraded $z_{mean}$.

\begin{table}
\centering
\caption{Performance comparison between of the 4 redshift estimators for the SDSS trainings at $r<17.8$. The P19 statistics are in parenthesis.}
\label{table:point_estimates_perfs_sdss}
\begin{tabular}{ c|ccccc}
\hline
& $z_{mean}$   & $z_{med}$ & $z_{peak}$ & $z_{reg}$ & unit \\
 \hline

\multicolumn{1}{c|}{$\sigma_{\rm MAD}$}       & 808 (908)  & 800 (902)  & 818 (918) & 810 & $10^{-5}$\\
\multicolumn{1}{c|}{<$\Delta z_{\rm norm}$>}  & 3 (4)      & -31 (-43) & -87 (-125) & -27  & $10^{-5}$\\
\multicolumn{1}{c|}{$\eta(>0.5) $}            & 0.17 (0.31)& 0.18 (0.31)& 0.27 (0.39)  & 0.17 & \% \\
\hline
\end{tabular}
\end{table}

These statistics marginally outperform the two recent attempts at improving the P19 performance. \citet{Hayat2021} proposed a self-supervised representation learning method in which a network was pre-trained with 1.2M unlabelled SDSS galaxies, then fine-tuned with the labelled data. While the great potential of such methods is undeniable, in particular at high redshifts where spectroscopy is very sparse, the present supervised training technique remains competitive with the SDSS data. \citet{Dey2021} presented a deep capsule network that jointly estimates the redshift and the basic morphological type of galaxies (spiral/elliptical). Its backbone consists of a primary convolutional layer followed by Conv-Caps ("capsule") layers, composed of multiple neurons that compute not only the presence or absence of a feature, but also its properties such as rotation, size, velocity, or color. Although this design is robust and invariant to image orientation, data augmentation techniques such as rotation and flipping were used during training. The latent space of the network has only 16 dimensions, which helps to study its interpretability. However compared to classical CNNs, capsule networks are trickier to train and can hardly be adapted to deeper architectures for more complex tasks.

\subsection{SDSS+GAMA+BOSS at $r\le 20$}
\label{subsec:s20go}

We expand the training to $r=20$ using the deeper spectroscopy (Section \ref{subsec:specdata}). 
Randomly splitting the full spectroscopic data for training/validation as was done for the SDSS at $r<17.8$ (e.g. 80\%/20\%) is not appropriate here, as the sample is not at all representative of the galaxies expected to populate the photometric sample. While SDSS and GAMA are representative sets of galaxies at $r\leq 17.8$ and $17.8\leq r\leq 19.8$, respectively, given their large redshift completeness, the bright SDSS data is over-represented compared to the faint GAMA data. Both surveys also feature strong local structures in their redshift distributions (Fig. \ref{fig:nz_surveys}). Last but not least, the large BOSS sample is over-populated with LRGs at $z_{spec}\gtrapprox 0.2$, which really are quite rare compared to "normal" red galaxies at $r<20$. Over-representing a specific population in a specific redshift interval will bias the predictions in favor of those redshifts. The results will be deceptively good on the validation sample as it matches the training sample by design, but sub-optimal on a different population, which the photometric sample is expected to be. 
These effects are quantified in Appendix \ref{subsec:LRG}. For these reasons we attempt to create training and testing samples at least roughly representative of the general population at $r<20$. 

\subsubsection{Test and training samples}
\label{subsubsec:samples}

We use SDSS and GAMA as models of the Universe in their respective magnitude range of completeness to create mock samples out of the full spectroscopic data. To do so, we model the redshift distributions of blue and red galaxies in bins of magnitude with smooth ad hoc functions; in each magnitude bin, we randomly extract from the full spectroscopic sample subsets of blue and red galaxies with redshift distributions matching these smooth distributions. The magnitude and redshift bin widths are chosen to strike a compromise between mock resemblance and size. 

\begin{figure}
\includegraphics[width=8.8cm]{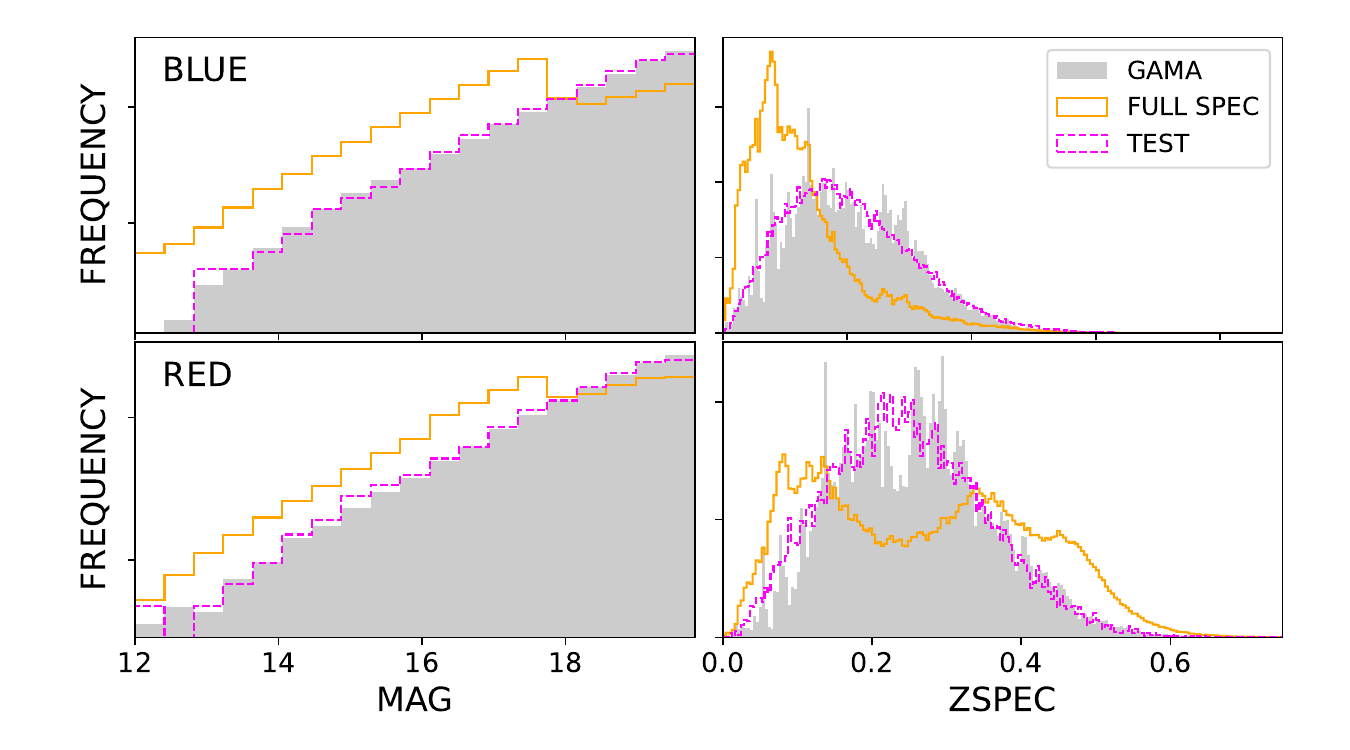} 
\caption{The normalized redshift and magnitude (log scale) distributions of the GAMA sample (gray), of the full spectroscopic sample (orange), and of the GAMA-like test sample (pink) for blue and red galaxies (top and bottom panels respectively). 
}
\label{fig:gamalikesample}
\end{figure}

\begin{figure}
\includegraphics[width=4.2cm]{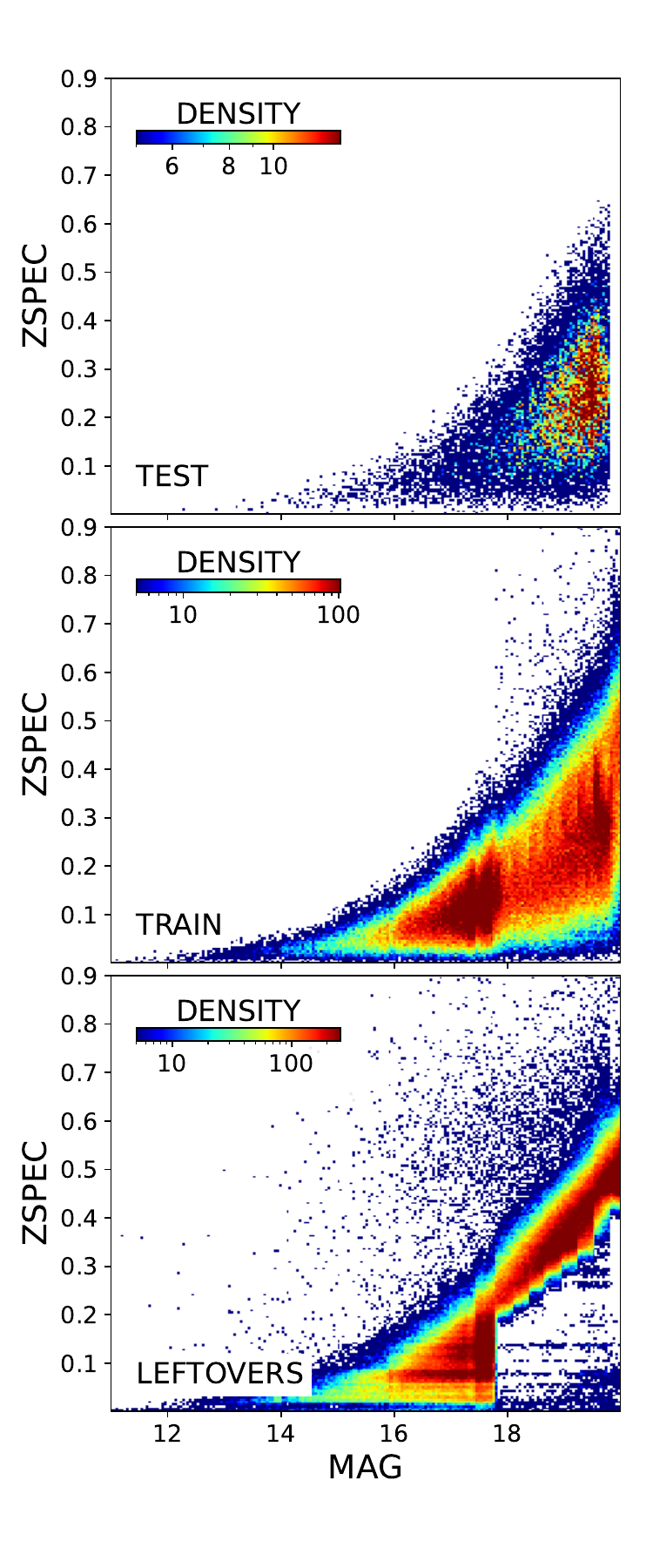}
\includegraphics[width=4.2cm]{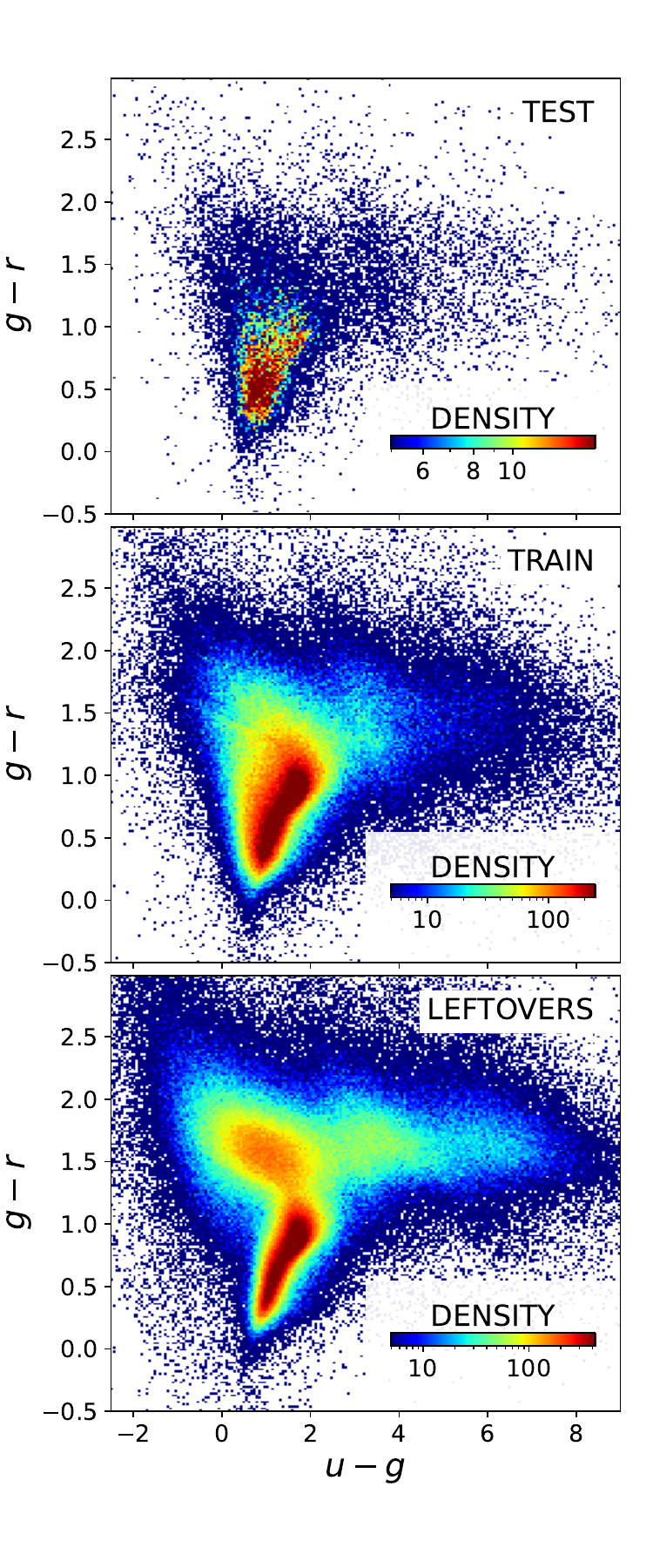}
\caption{Magnitude/redshift distributions (left panels) and $(u-g)/(g-r)$ distributions (right panels) of the test, training and leftover samples described in Section \ref{subsubsec:samples}.}
\label{fig:magz_testrain}
\end{figure}

We first create a GAMA-like test sample using the above method but avoiding GAMA itself in the random extraction in order to sample the full sky coverage. We match the blue/red ratio and the magnitude distribution of GAMA by randomly extracting from a blue and a red redshift-matched sample. The result is a test sample of 25,856 galaxies at $r\leq 19.8$ resembling GAMA but with much reduced redshift structures and spanning the entire sky. Its magnitude, redshift, and color distributions are shown in Fig.~\ref{fig:gamalikesample} and in the top panels of Fig.~\ref{fig:magz_testrain}.
Galaxy properties are not limited to magnitude, redshift and red/blue type, but more sophisticated methods taking more parameters into account (e.g. SOM) would reduce the size of the mock sample too significantly, defeating its purpose.

We subtract this test sample from the spectroscopic sample and proceed to extract training samples using the above method without excluding GAMA and adding galaxies at $19.8<r<20$, also based on GAMA but more loosely as it is less representative in this range. The SDSS and GAMA-like subsets are concatenated without matching the GAMA number counts, which would deplete the bright end too drastically. We create in this way 25 training samples of $\sim$370k galaxies, significantly smaller than 
throwing in the full GAMA and SDSS samples but doing so propagates unwanted training features in the predicted redshift distributions. The 25 samples total $\sim$910k unique spectroscopic galaxies and leave out a sample of $\sim$580k galaxies, half of them at $r<17.8$, the other half dominated by LRGs (Section \ref{subsec:leftovers}). The intersection between any 2 training samples is between 45 and 80\%. The intersection of all 25 samples amounts to 78,682 sources, most of them in the troughs of the GAMA redshift distribution, whose features reappear when combining any 2 samples. 
%By digging under the troughs of the redshift distributions, 
Figure \ref{fig:magz_testrain} shows the magnitude/redshift and $(u-g)/(g-r)$ distributions of the test, training and leftover samples. 

Another strategy, leaving a larger sample of leftovers available for testing, would be to train just a few training samples several times. Using just one training sample for instance, picked at random, leaves nearly twice as many galaxies for testing. These however, are, by design, as un-representative as the smaller sample (bright galaxies and faint red ones) and thus do not provide a more informative testing opportunity. The strategy is explored in Appendix \ref{subsec:randomness}. 

\subsubsection{Averaging}
\label{subsubsec:averaging}

\begin{figure}
\includegraphics[width=4.2cm]{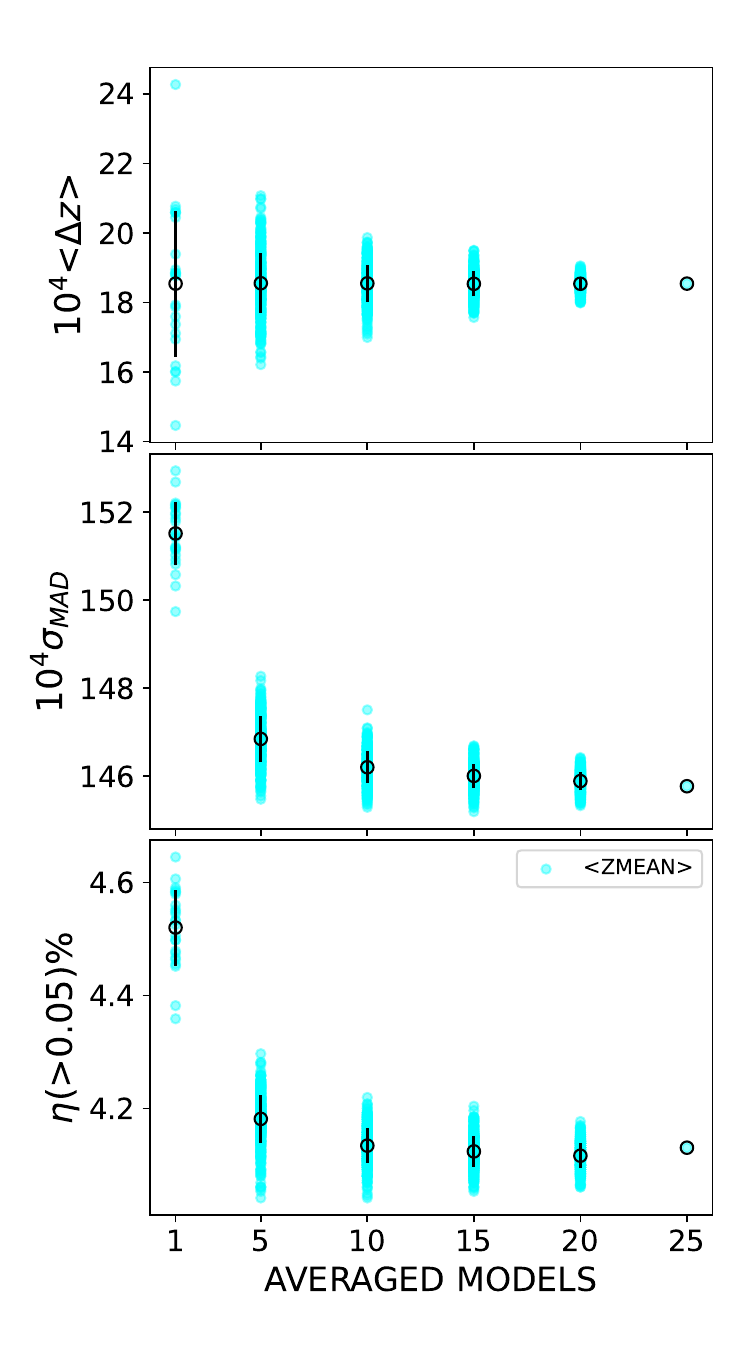}
\includegraphics[width=4.2cm]{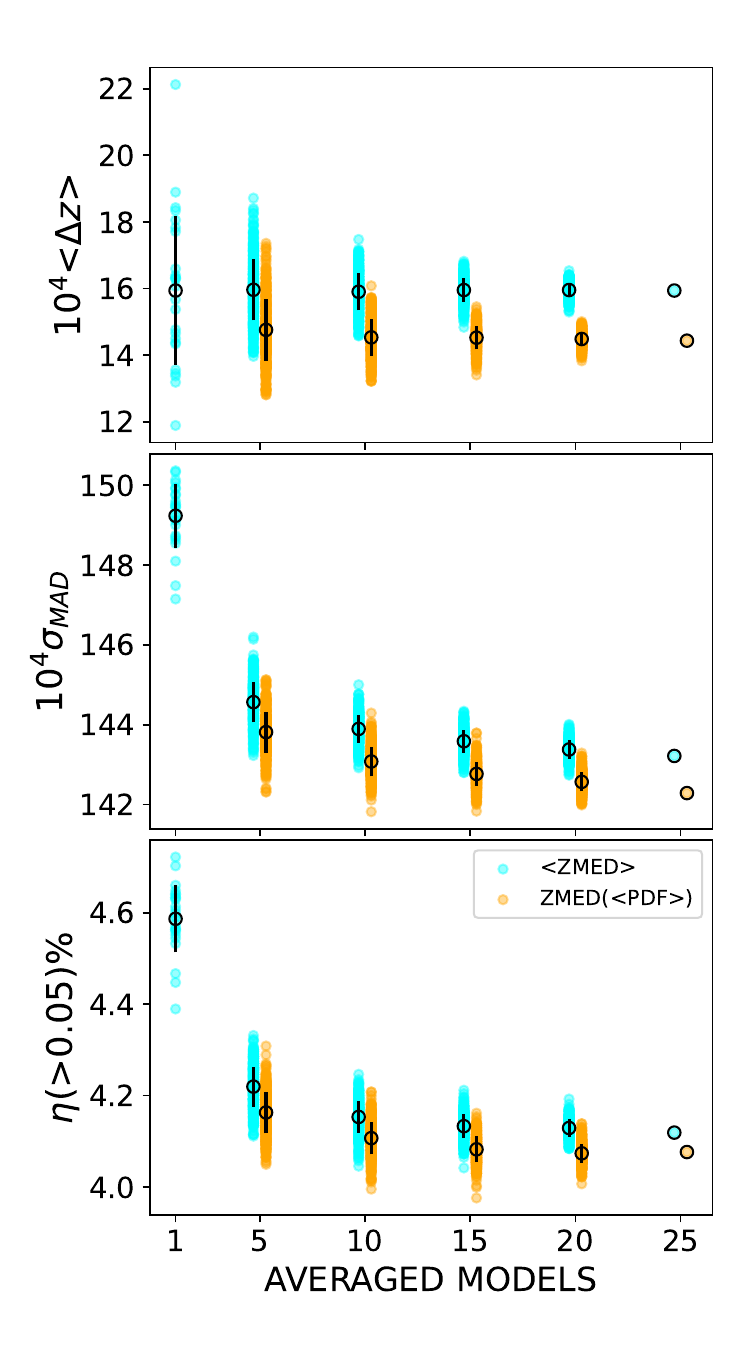}
\includegraphics[width=4.2cm]{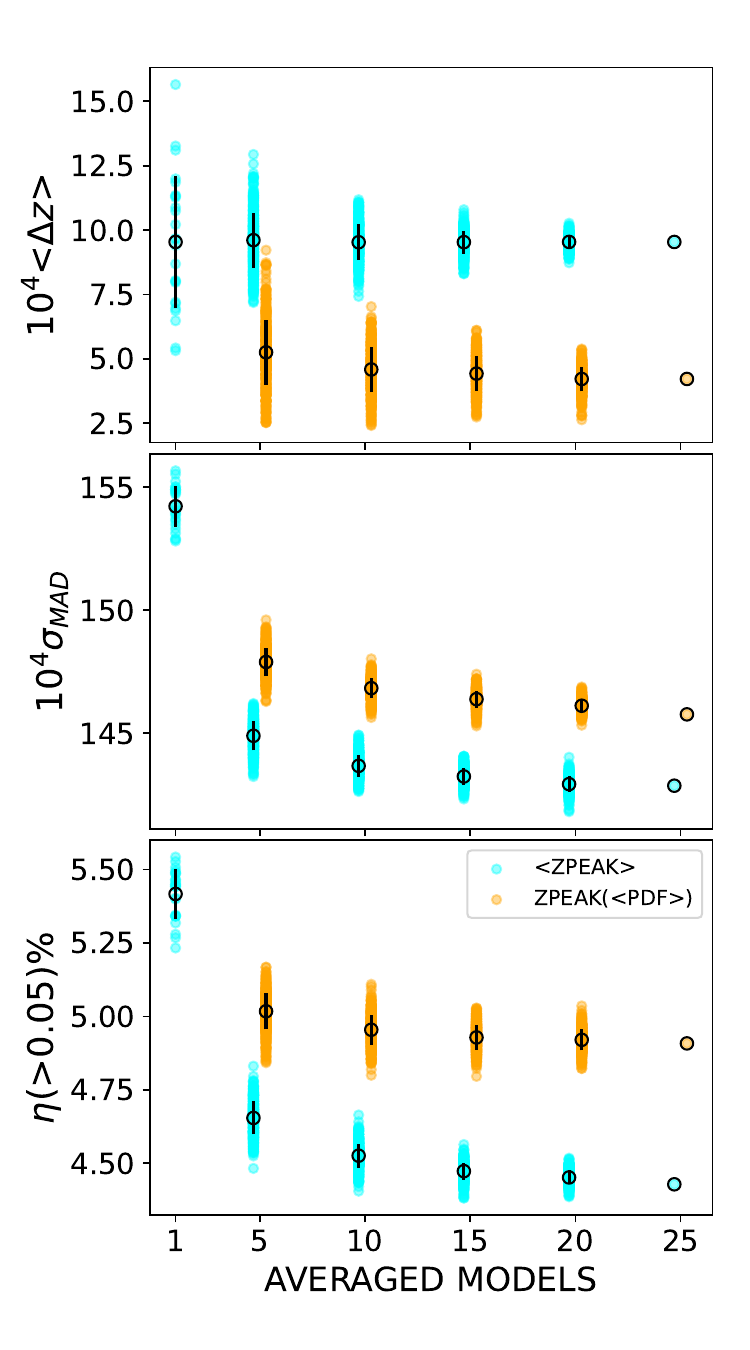}
\includegraphics[width=4.2cm]{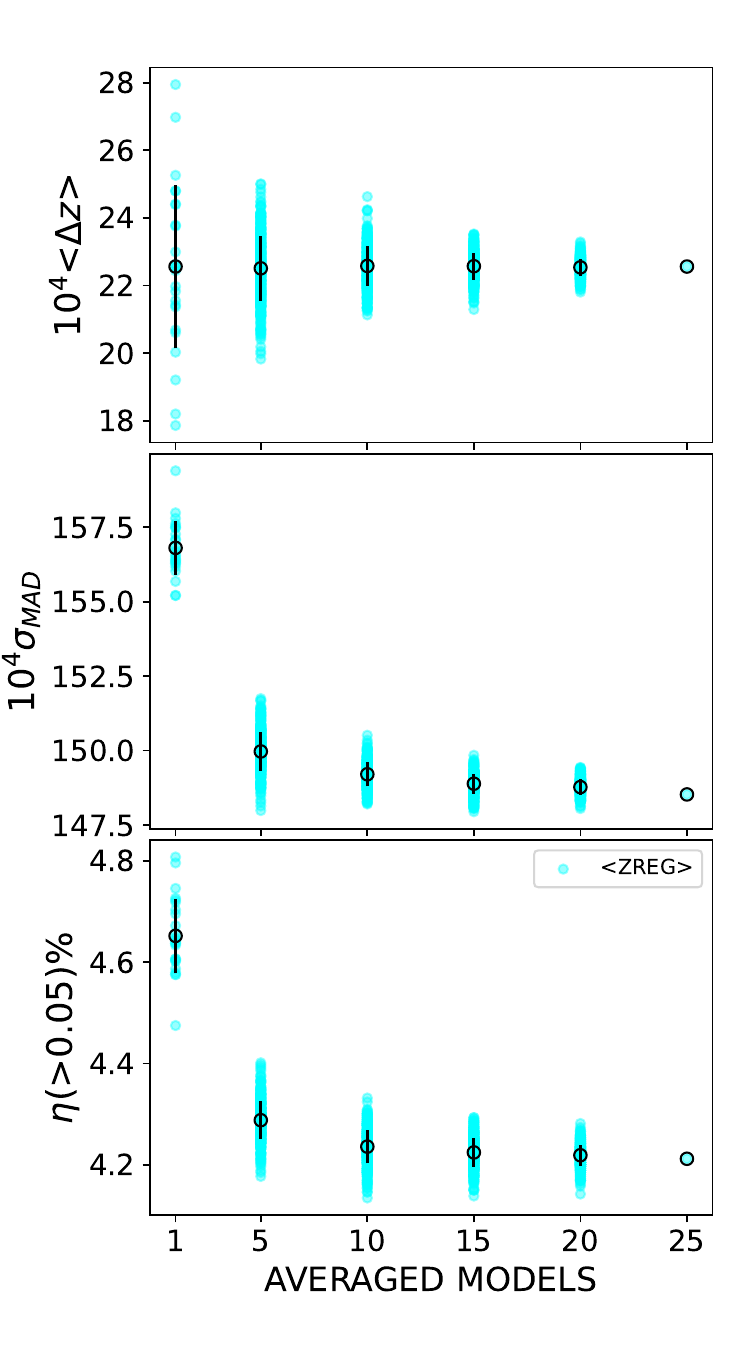}
\caption{Bias, deviation and rate of catastrophic failure of the averaged point estimates (blue) and of the corresponding point estimate of the averaged PDFs 
when relevant (orange), as a function of the number of CNN outputs being averaged for the test sample. From top left to bottom right: $z_{mean}$, $z_{med}$, $z_{peak}$ and $z_{reg}$. The black circles and vertical lines show the mean and standard deviation of the distributions.} 
\label{fig:metrics_average25models}
\end{figure}

\begin{figure}
\includegraphics[width=8.9cm]{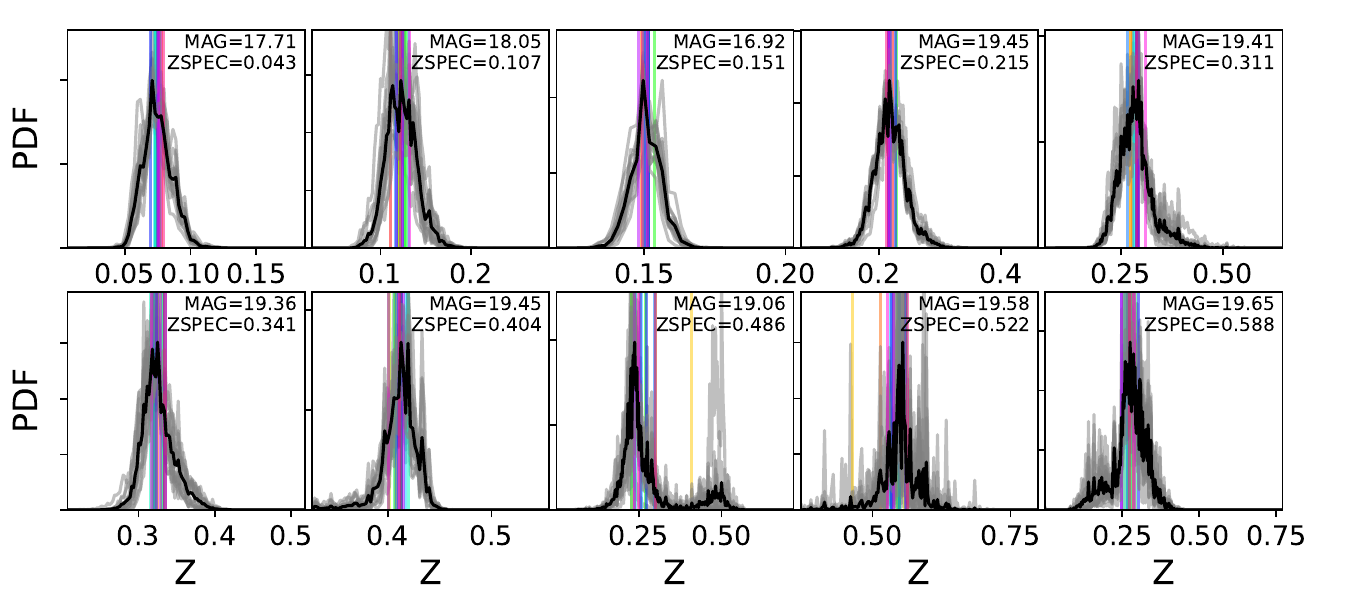}
\caption{A random sample of outputs in increasing order of spectroscopic redshift. The outputs of the 15 models are shown in gray, with their respective median values marked as colored vertical lines. The averaged outputs (the final "PDF") are shown in back.} 
\label{fig:PDF_average25models}
\end{figure}

We train the CNN with each of the 25 samples in full (without cross-correlations) using 400 redshift bins between $z=0$ and 0.9, corresponding to a bin width $\delta z=0.00225$ similar to the SDSS training at $r<17.8$. Other training parameters are given in Appendix \ref{sec:architecture_apdx}. From the 25 trained networks, we infer "PDFs" and regression values for the test sample.  

The benefits of averaging the outputs of several networks is shown in Fig. \ref{fig:metrics_average25models}, where the bias, deviation and rate of catastrophic failure of the averaged $z_{mean}$, $z_{med}$, $z_{peak}$ and $z_{reg}$ (in blue) and of the point estimates of the averaged "PDFs" when relevant (in orange), are plotted as a function of the number of models being averaged. The 25 points at N=1 illustrate the variance between the 25 models. In between these points and the final averages of the 25 models, we randomly picked 400 combinations of N different models among the tens of thousand of possibilities (several million between 10 and 15). The black circles and vertical lines show the mean and standard deviation of the colored points. 

For all the point estimates, $\sigma_{\rm MAD}$ gains the most from averaging, while the bias is quite insensitive to it. The median of the averaged "PDFs" does consistently better than the averaged $z_{med}$. On the contrary the averaged $z_{peak}$ do better for $\sigma_{\rm MAD}$ and $\eta$ than the mode of the averaged "PDFs", but a factor of 2 worse for the bias, making the latter method preferable in this case too. For all the point estimates, the gain from  using a single model to averaging 5 is significant for $\sigma_{\rm MAD}$ and $\eta$: averaging any 5 models does better than the best model among the 25. Then the overlap becomes large and it becomes possible to do better with certain combinations of, say, 10 than of more models, including the final average of 25. However the means and standard deviations show that it is unlikely, and also that 25 is an overkill. 

In all that follows, we use 15 training samples randomly selected among the 25, a reasonable compromise between performance and computing time. This choice leaves unchanged the total number of spectroscopic galaxies used for training and left over. We compute the point estimates of the averaged "PDFs" rather the average of the point estimates. The uncertainty on a given metric may be estimated from the standard deviations: they are of order $3\times 10^{-5}$ for $\sigma_{\rm MAD}$ and the bias, and of order 0.03 for $\eta$. 

The benefit of averaging "PDFs" is further illustrated in Fig. \ref{fig:PDF_average25models}, which shows random examples of outputs in increasing order of spectroscopic redshift. The 15 CNN outputs are shown in gray, with their respective $z_{med}$ marked as colored vertical lines. The average of the 15 outputs are plotted in black. We interpret the fact that these consistently provide better redshift estimations than the individual ones, which are themselves consistent with one another, as convergence towards true PDFs. At any rate, we remove the quotes to alleviate the text. 

% sigbin standard deviations sig*1.e5 bias*1.e5 eta 
% zmean 3.0 3.1 0.03
% zmed 2.9 3.0 0.03
% zpeak 4.9 3.2 0.03
% zreg 3.8  3.2 0.02

\subsubsection{Redshift binning}
\label{subssubsec:binning}

\begin{figure}
\hspace{-0.3cm}
\includegraphics[width=8.8cm]{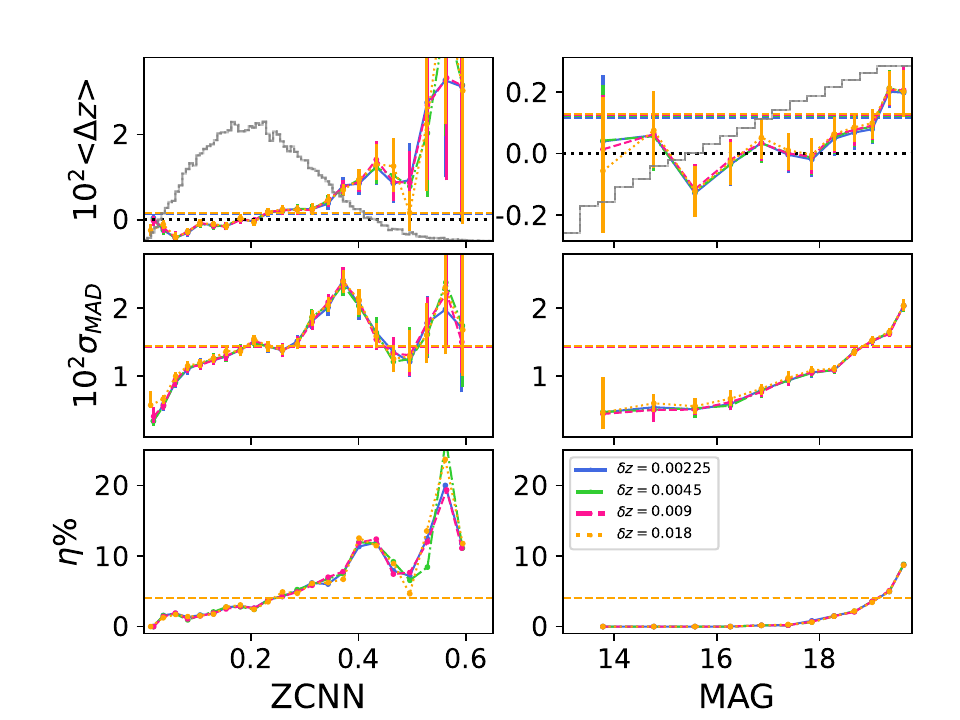}
\caption{Test sample metrics as a function of $z_{med}$ and magnitude for the 4 classification schemes. The gray histograms in the top panels are the shapes of the spectroscopic redshift and magnitude distributions, in log scale for the latter. The metrics are very little sensitive to the choice of classification binning within the explored range.}
\label{fig:dzsig_magz_bins}
\end{figure}

In addition to the binning used above with $\delta z=0.00225$, 
%for which we retained the 15 selected models, 
we train the 15 samples using 200, 100 and 50 bins between $z=0$ and 0.9, corresponding to $\delta z=0.0045$, 0.009 and 0.018. 
Figure \ref{fig:dzsig_magz_bins} shows how little sensitive the point estimates ($z_{med}$ here) are to the PDF resolution. The bias, $\sigma_{\rm MAD}$, and rate of catastrophic failure, plotted as a function of $z_{med}$ and magnitude, are undistinguishable from the smallest to the largest binning, suggesting that too fine a binning is superflous. 

\begin{figure}
\includegraphics[width=8.5cm]{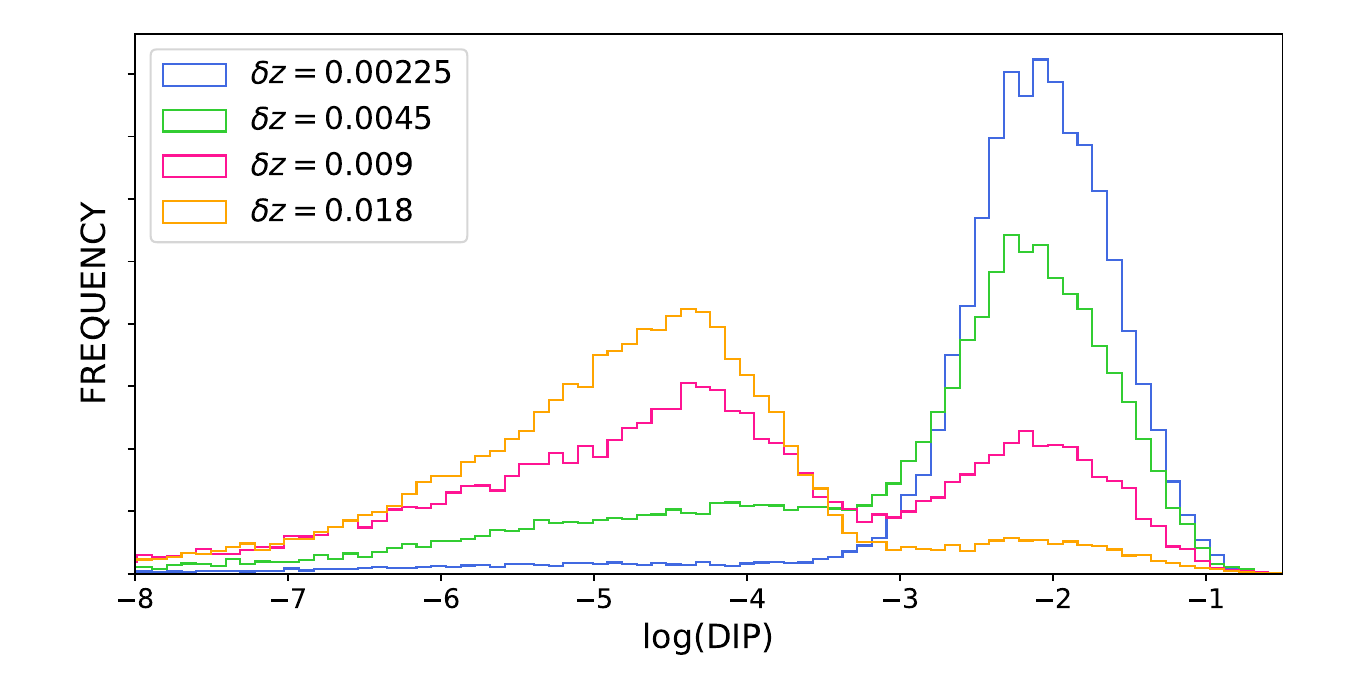}
\caption{The DIP score distribution of the test sample for the 4 binning experiments.}
\label{fig:DIPDIS_bins}
\end{figure}

We use the DIP test \citep{Hartigan85} to detect multi-modality in the different sets of PDFs. It consists of measuring the maximum distance at any point between the CDF and the closest uni-modal CDF, a uni-modal distribution having a score of 0 by definition. Figure \ref{fig:DIPDIS_bins} shows the significant impact of the bin width on the distribution of DIP scores in the test sample. Since the bin width has, on the contrary, very little impact on the point estimates in the range we tested, we conclude that the many spikes generated by the small bins do not represent meaningful multi-modalities but simply the incapacity of the CNN to classify redshifts in such fine a grid. Increasing the bin width makes the PDFs increasingly uni-modal but would eventually degrade the predictions significantly, to the point of no prediction at all in the extreme case of only 1 bin. There is therefore an optimal resolution for the classifier, given the training data.

\begin{figure}
\includegraphics[width=8.5cm]{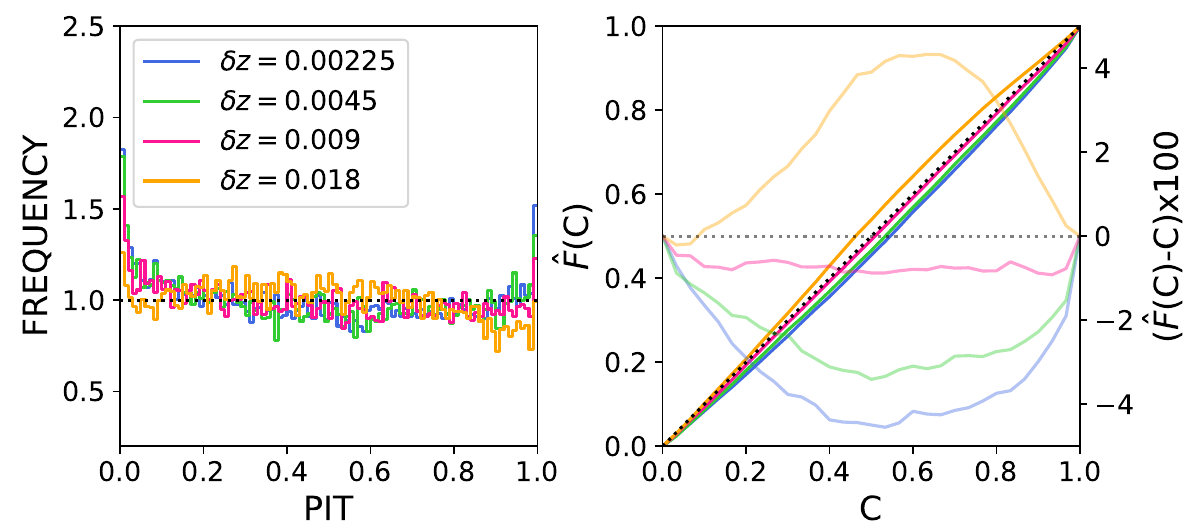}
\caption{PIT and WBT statistics for the 4 binning scenarios. The departures from the unity line in the WBT test are overlayed in faded colors, with units along the right hand side y-axis ($10^{-2}$).}
\label{fig:PIT_QQ_bins}
\end{figure}

Figure \ref{fig:PIT_QQ_bins} shows the PIT and WBT tests for the 4 binning scenarios. Both are nearly perfect but the latter is more discriminatory. It shows that $\delta z=0.009$ is close to the optimal resolution. Smaller bins produce "over-confident", overly spiky PDFs, larger bins produce "under-confident", under-informative ones. 
The $\delta z=0.009$ binning is consequently our final choice in the rest of this work. The mean CRPS for this binning and the two smaller ones is $\sim 0.013$. It is slightly higher ($\sim 0.015$) for the largest binning.

\section{Test results}
\label{sec:test}

\subsection{Color dependence}
\label{subsec:reddegeneracy}

\begin{figure}
\includegraphics[width=8.8cm]{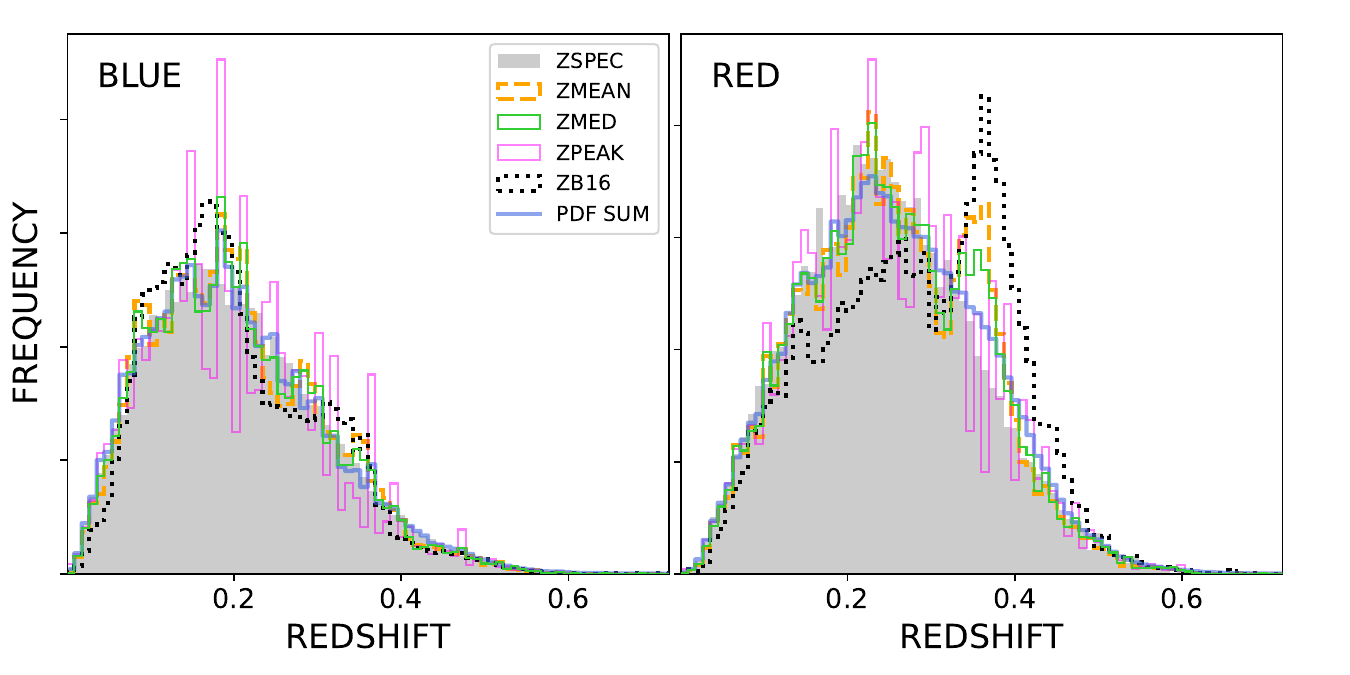}
\caption{The point estimate distributions of blue and red test galaxies, with their stacked PDFs (the missing regression value is very similar to $z_{mean}$). The red galaxy distribution is strongly distorted around $\sim 0.35$, to a variable degree depending on the point estimate. This is attributed to a color degeneracy (see Section \ref{subsec:reddegeneracy}). }
\label{fig:nz_test}
\end{figure}

Figure \ref{fig:nz_test} shows the distribution of the different point estimates for the blue and red galaxies in the test sample, as well as the stacked PDF, the spectroscopic redshift distribution and the B16 distributions (the regression value is not displayed to limit the clutter and because it is very similar to $z_{mean}$). The strong distortion around $z_{spec}\sim 0.35$ for the red galaxies  can be attributed to a degeneracy in their optical colors. It is more or less severe depending on the point estimate: $z_{mean}$ (and $z_{reg}$) generate the strongest distortion 
%(though not quite as bad as B16) 
within the smoothest distributions, $z_{peak}$ minimizes the distortion but generates the noisiest (discretized) distributions, $z_{med}$ is the best compromise considering both the blue and red populations. The stacked PDF best fits the $z_{spec}$ distribution of both blue and red galaxies. The B16 redshifts are significantly more distorted, probably because they were computed from the measured colors while the CNN captures more information from the full images. 

\begin{figure}
%\hspace{-0.4cm}
\includegraphics[height=4.2cm]{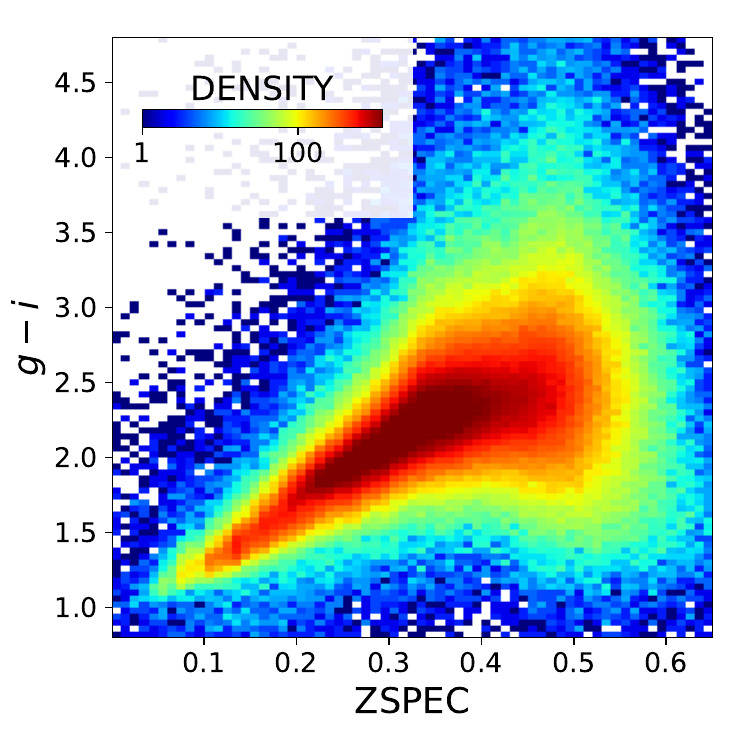} 
\includegraphics[height=4.2cm]{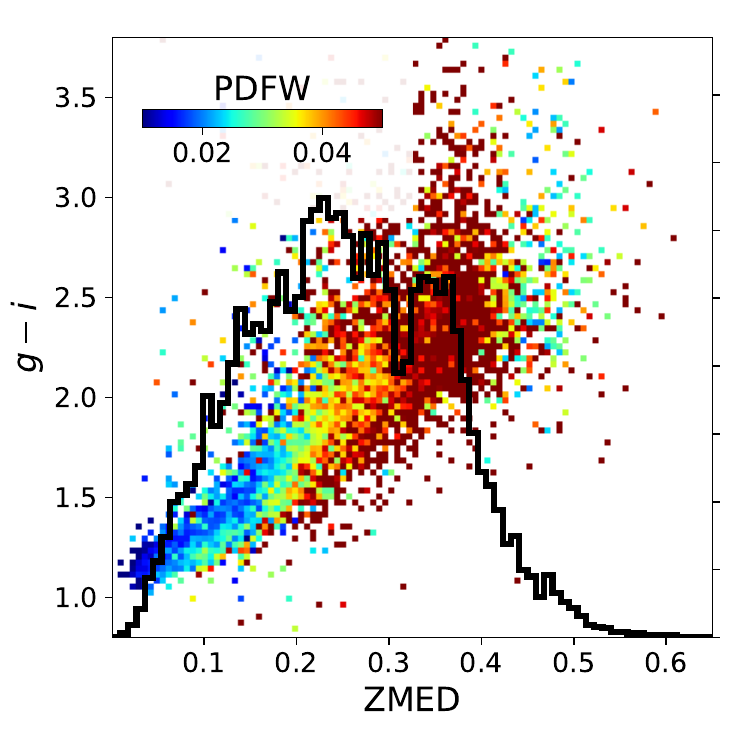}
\caption{{\bf Left:} $(g-i)$ color against $z_{spec}$ for the red galaxies in the full spectroscopic sample. The color clearly becomes indiscriminate at $z_{spec} \gtrsim 0.31$; {\bf Right:} the $(g-i)$ vs. $z_{med}$ distribution of the red galaxies in the test sample color-coded by the PDF width. The black profile is their $z_{med}$ distribution. The mislocated galaxies in the degenerate interval have wider than average PDFs, reflecting the greater uncertainty in their classification. }
\label{fig:red_degeneracy}
\end{figure}

The left panel of Fig.~\ref{fig:red_degeneracy} shows the $(g-i)$ color vs. $z_{spec}$ distribution of red galaxies in the spectroscopic sample. While color and redshift are well correlated at low redshift, the relation flattens out at $z_{spec} \gtrsim 0.31$, making it a harder task to predict redshifts in this range. The right panel displays the $(g-i)$ color vs. $z_{med}$ distribution of red galaxies in the test sample, color-coded by the mean PDF width. The overlaid black line is the shape of the $z_{med}$ distribution (also displayed in Fig.~\ref{fig:nz_test}). The strong distortion around $\sim 0.35$ coincides with increased PDF widths, i.e. increased uncertainties in the classification. As shown in Fig.~\ref{fig:nz_test}, stacking these wider PDFs nearly suppresses the point estimate distortion.

\begin{figure}
\includegraphics[width=8.5cm]{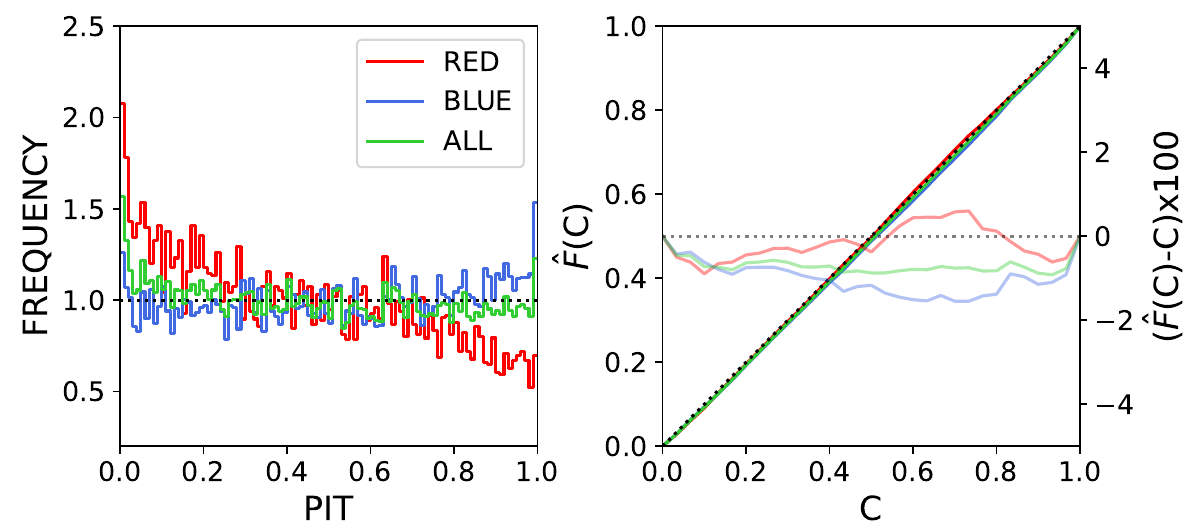}
\caption{PIT and WBT statistics for the blue and red sub-populations in the $\delta z=0.009$ scenario. The PIT distribution highligts the red population bias. However both the PIT distribution and WBT test for the full sample are near perfect. The departures from the unity line in the WBT test are overlayed in faded colors, with units along the right hand side y-axis ($10^{-2}$).}
\label{fig:PIT_QQ_bluered}
\end{figure}

Figure \ref{fig:PIT_QQ_bluered} shows the PIT and WBT tests for the blue and red sub-populations and the full sample. The PIT distribution 
%is more discriminatory in this case, it 
reveals a large positive bias for the red galaxies, whose PDFs are excessively to the right of their true redshift. This is indeed the case for the many galaxies shoved into the CNN redshift distortion from lower $z_{spec}$. The WBT test is very close to the unity line for both galaxy types, and closer still for the whole population. For comparison with traditional 1$\sigma$ coverage tests: 67.3\% of the test galaxies 
%(exactly 68\% at $z<0.3$) 
have their smallest credible interval defined by their spectroscopic redshift smaller than or equal to 68\%, while 64.6\% of the galaxies have their spectroscopic redshift within the 68\% central credible interval defining the PDF width independently of the spectroscopic redshift. The difference points to the fact that the PDFs are non gaussian.

\subsection{Magnitude dependence}
\label{subsec:threshold}

\begin{figure}
\hspace{-0.3cm}
\includegraphics[width=8.8cm]{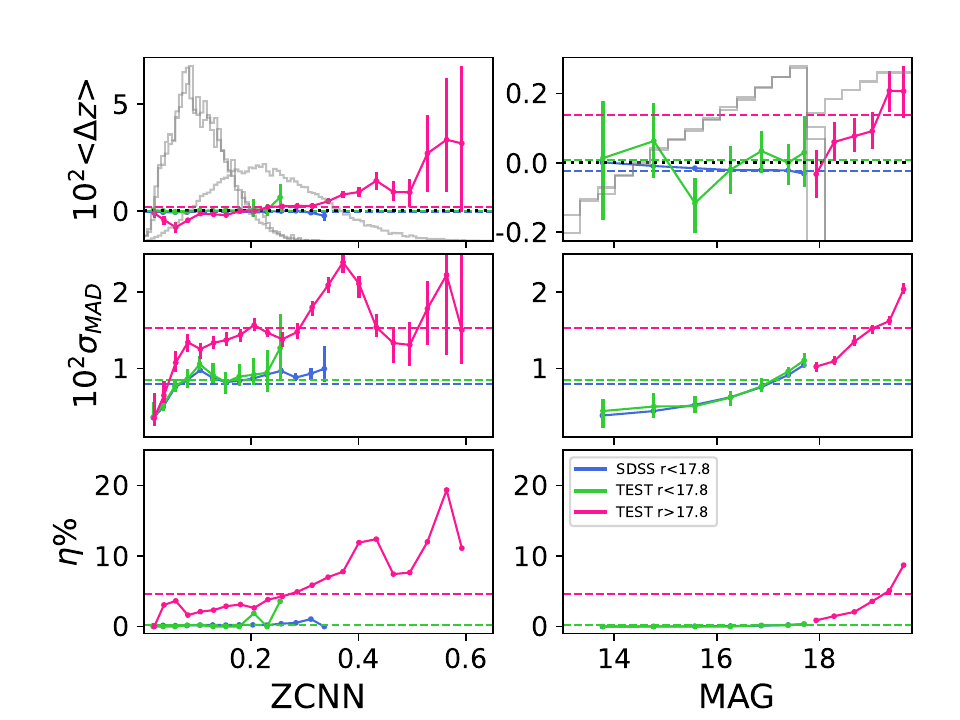}
\caption{Test sample metrics as a function of $z_{med}$ and magnitude at $r<17.8$ and $r>17.8$, with the SDSS MGS training performance for comparison (Section \ref{subsec:sdss}). The performance degrades significantly from bright to faint. The bright regime is similar to the SDSS MGS training despite the much lower number of training galaxies.}
\label{fig:dzsig_magz_test}
\end{figure}

Figure \ref{fig:dzsig_magz_test} shows how the performance on the test sample degrades, predictably, from $r<17.8$ to $r>17.8$. The $r<17.8$ regime remains competitive with the bright SDSS training described in Section \ref{subsec:sdss}, despite the much lower number of galaxies in this magnitude range in the present training samples ($\sim 132$k vs. $\sim 414$k). This is expected from P19 who found that the performance remained virtually unchanged when the training sample size was reduced to $\sim 100$k.

\begin{figure}
\includegraphics[width=8.5cm]{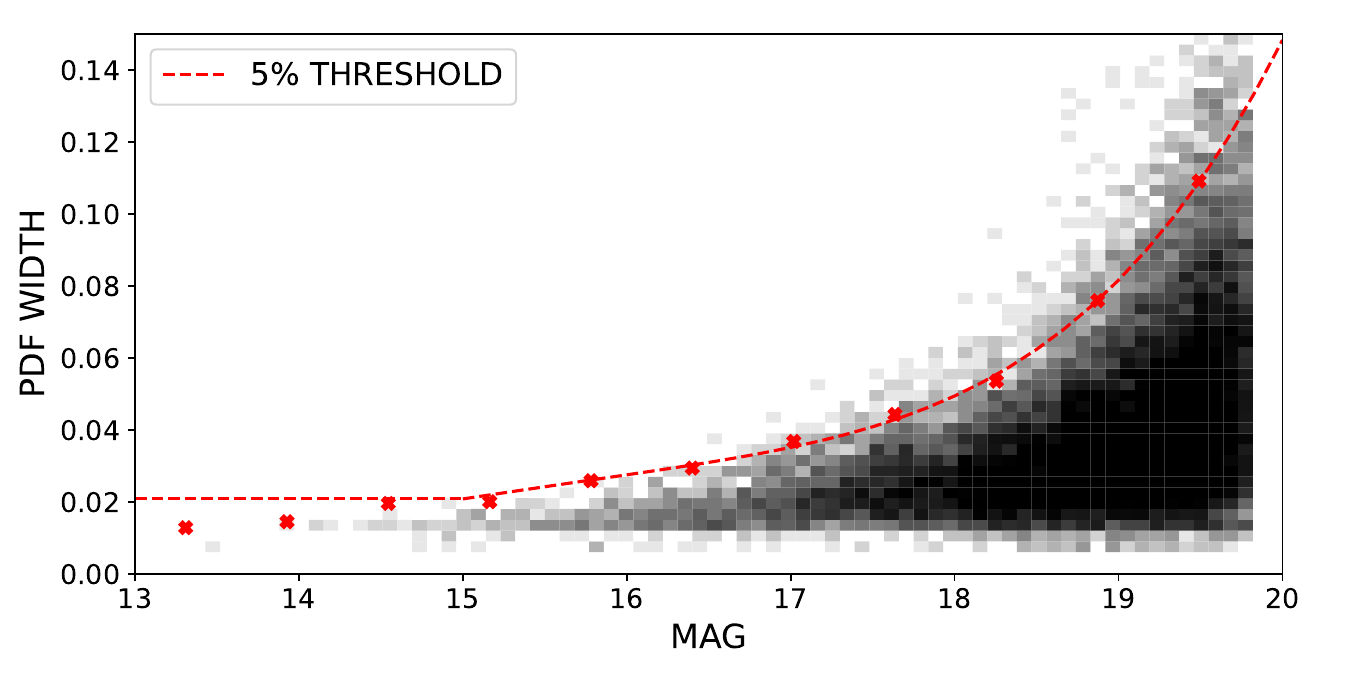}
\caption{Magnitude-dependent threshold designed to exclude the largest 5\% of PDF widths in the test sample (red crosses). The dashed line is the analytical fit (Eq. \ref{eq:pdfwmax_5pc}), plateaued out below $r=15$ at a PDF width of $\sim 0.02$.}
\label{fig:pdfw_cut}
\end{figure}

P19 also found that performance could be improved by discarding sources with the largest PDF widths, reporting the result of rejecting the largest 10\% and 20\%.
%Discarding sources with the largest PDF widths will improve the accuracy of the point estimates, as shown by P19 who reported the result of rejecting the largest 10 or 20\%. 
Here we design a more conservative, magnitude-dependent cut in PDF width meant to exclude the worst 5\% of PDF widths at a given $r-$band magnitude in the test sample. The computed threshold is shown in Fig. \ref{fig:pdfw_cut} as red crosses. We fit the trend with a 4$^{th}$ degree polynomial, $\sum_{i=0}^{4} a_{5\%} [i]~r^i$, where:
\begin{multline} \label{eq:pdfwmax_5pc} 
a_{5\%} = [ 1.26761997\times{\rm 10^1}, -3.30587750,  3.22381178\times{\rm 10^{-1}},\\ -1.39350947\times{\rm 10^{-2}}, 2.25736799\times{\rm 10^{-4}}]
\end{multline}
capping it at bright magnitude at the 
$r=15$ value of $\sim 0.02$. The final threshold is the red dashed lines in Fig. \ref{fig:pdfw_cut}. As shown in Table \ref{table:stats_gamalike} (see next section), excluding galaxies with PDF width above this threshold improves redshift quality at a minor cost (4.6\% of the test sample), while allowing for the expected increase in uncertainty with magnitude. Also expected is the underlying dependence with specroscopic redshift. High redshift galaxies not only tend to be fainter, they are also sparse in the training samples, hence their poorer outcome. The fraction of galaxies excluded by the threshold increases from 2 to 16\% between $z=0$ and 0.6.

We also note that, despite our attempt at matching representative magnitude-dependent redshift distributions, this limit inevitably carries biases. But by enveloping the spread of PDF widths of a sample of spectroscopic galaxies whose image quality is on average more reliable than in the full photometric SDSS survey (Section \ref{sec:inference}), it is nevertheless a tool of quality control. 

\subsection{Point estimate summary}
\label{subsec:perf}

\begin{figure}
\includegraphics[width=8.5cm]{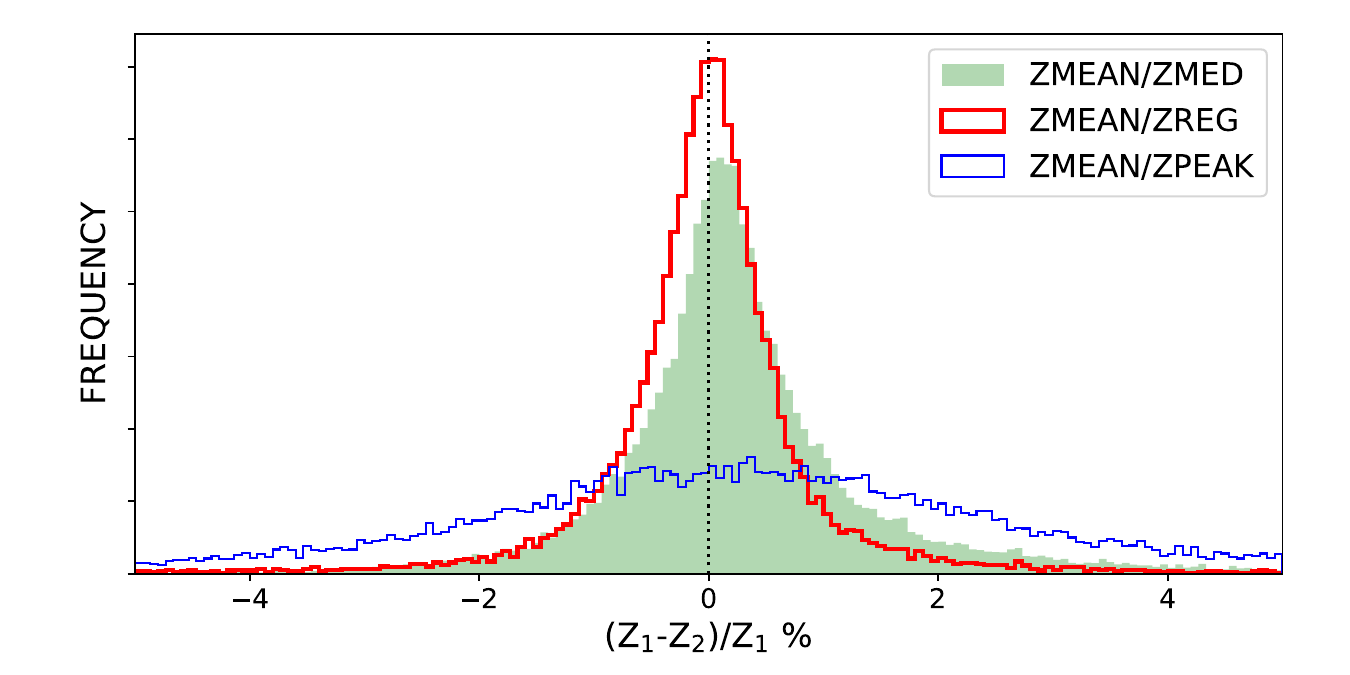}
\caption{Comparison of the point estimates (in percentages): $z_{mean}$ is very similar to the regression value but positively skewed with respect to $z_{med}$. The mean difference between any 2 point estimates is less than 0.5\%.}
\label{fig:skewness_test}
\end{figure}

Figure \ref{fig:skewness_test} shows the distributions of the discrepancies between different point estimates. The PDF mean tends to be larger than the mode, which tends to be larger than the median (positive skewness), however the mean difference between any 2 point estimates is less than 0.5\%. The regression value is very similar to the mean value. 

The performance of the different point estimates for the blue, red and combined galaxies in the test sample are reported in Table \ref{table:stats_gamalike}, with the best scores in bold face. The metrics resulting from applying the PDF width threshold are reported in the second lines in parenthesis. The best dispersions are consistently achieved with $z_{med}$.
The mean bias is a more shifting quantity since it may take negative values. Good scores are of the order of $10^{-4}$, poor ones above $10^{-3}$, as is the case for red galaxies. The rate of catastrophic failures remains below 5\% in all cases. Compared to B16, the precision is improved by a factor of $\sim 2$ or more, the red galaxy bias and the rate of catastrophic failures by a factor of $\sim 5$.

\begin{table*}
\centering
\caption{The CNN performance on the GAMA-like test sample, in units of $10^{-5}$ for $\sigma_{\rm MAD}$ and <$\Delta z_{\rm norm}$>, and in \% for $\eta(>0.05)$. The B16 performance is reported under $z_{\rm B16}$. The numbers and second lines in parenthesis are the results of excluding galaxies above the PDF width threshold (Eq. \ref{eq:pdfwmax_5pc}).
} 
\label{table:stats_gamalike}
\begin{tabular}{ c|ccccc|ccccc|ccccc|}
\hline
 & \multicolumn{5}{|c|}{Blue} & \multicolumn{5}{|c|}{Red} & \multicolumn{5}{|c|}{All} \\
 \hline
N & \multicolumn{5}{|c|}{15120 (14463)}  & \multicolumn{5}{|c|}{10736 (10192)} & \multicolumn{5}{|c}{25856 (24655)} \\
 \hline
 $z_{phot}$ & {$z_{mean}$} & {$z_{med}$} & {$z_{peak}$} & {$z_{reg}$} & {$z_{\rm B16}$} & {$z_{mean}$} & {$z_{med}$}  & {$z_{peak}$} & {$z_{reg}$} & {$z_{\rm B16}$}& {$z_{mean}$} & {$z_{med}$}  & {$z_{peak}$} & {$z_{reg}$} & {$z_{\rm B16}$} \\

\hline

{$\sigma_{\rm MAD}$} & 1446 & {\bf 1403} &1455 &1474 &3058 &1474 & {\bf 1443} & 1466 &1503 &2853 &1466 &\bf{1421} &1460 &1481 &3062 \\

& (1392 &{\bf 1355} & 1403 &1413 &3002 &1400 & \bf{1382} &1415 &1421 &2734 &1402 &{\bf 1367} &1406 &1422 & 2987 ) \\

{<$\Delta z$>} & -46 & -78 &-163  &{\bf -22} & -59 & 559 & 479 & {\bf 339} & 581& 2706  &205 & 153 &  \bf{45} &  228& 1089 \\

& (-77 & -99 & -128 & -66 & {\bf -59} & 443 & 373 & {\bf 281} & 459 & 2572  & 138 &  97 &  {\bf 41} & 151& 1031) \\

{$\eta (\%)$} & {\bf 3.7} & {\bf 3.7} &  4.76 & 3.81 & 15.97 & 4.66 & {\bf 4.41} & 4.96 & 4.81 &21.36 & 4.1 & {\bf 3.99} &  4.84 & 4.22 & 18.2 \\

& ({\bf 2.75} & 2.8 &  3.56 & 2.77 & 15.3 & {\bf 3.} &   3.01 & 3.61 & 3.08 & 19.85 & {\bf 2.86} & 2.88 & 3.58 & 2.9 & 17.18 )\\
\hline
\end{tabular}
\end{table*}

\subsection{Outliers}

\begin{figure}
\includegraphics[width=8.5cm]{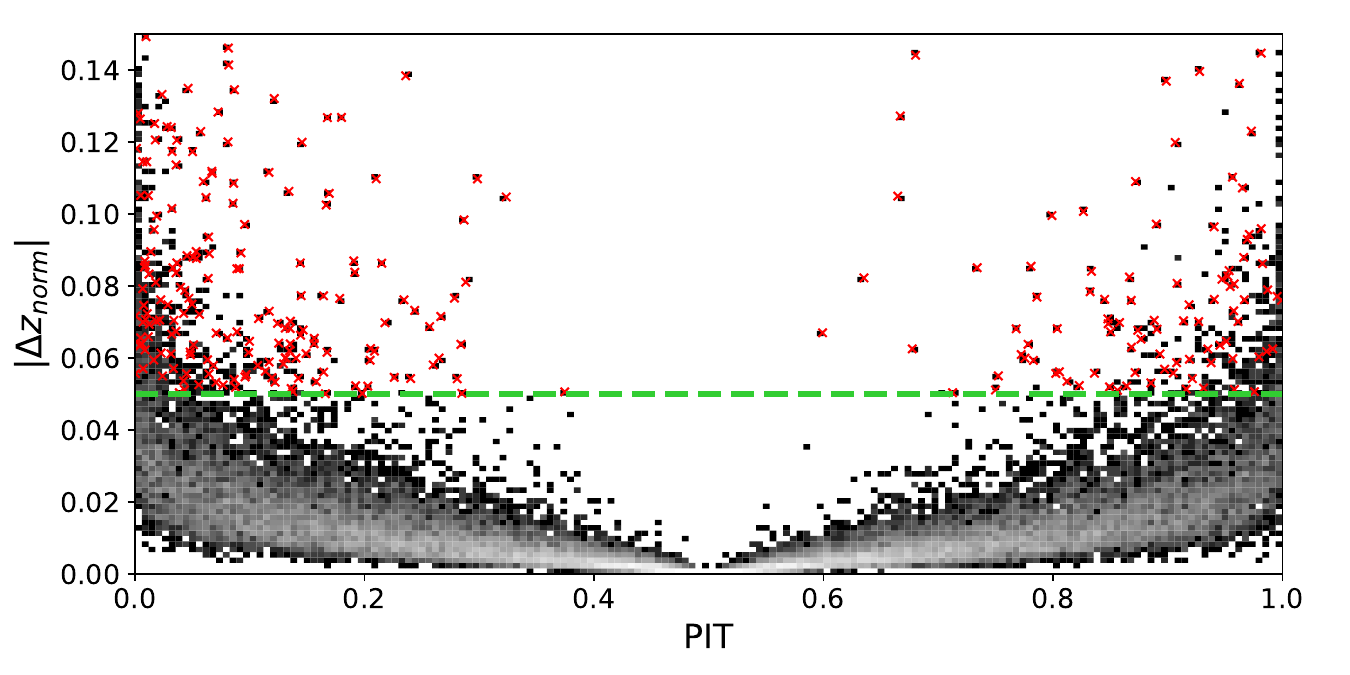}
\caption{PIT intervals against $|\Delta z_{\rm norm}|$ assuming $z_{med}$ (the convergence at PIT=0.5 and $\Delta z_{norm}=0$ arises from the very definition of $z_{med}$). The gray scale is inverted to highlight the low density outlier region above the green dashed line ($|\Delta z_{norm}|>0.05$). Outliers excluded by the PDF width threshold are crossed in red. All the others have PIT values close to 0 and 1, i.e. narrow PDFs missing the spectroscopic redshifts.}
\label{fig:outliers}
\end{figure}

Figure \ref{fig:dzsig_magz_test} shows that the fraction of catastrophic failures increases with magnitude and predicted redshift (it increases similarly with spectroscopic redshift), like all metrics. Table \ref{table:stats_gamalike} shows that applying the PDF width threshold significantly reduces their fraction, however about two thirds remain within the range of other galaxies. Figure \ref{fig:outliers}, which displays the PIT intervals against $|\Delta z_{\rm norm}|$, assuming $z_{med}$, reveals that most outliers, defined as being above the green dashed line ($|\Delta z_{\rm norm}|>0.05$), tend to have PIT values close to 0 and 1 (equivalently, WBT credibility intervals close to 1). This means that their spectroscopic redshift tends to lie to the left or right of their PDF. Those with intermediate PIT values would be largely excluded by the PDF width threshold (excluded outliers are crossed in red). Let's note that most cases of extreme PIT value are not outliers, they are expected from a flat PIT distribution.  Outliers with narrow PDFs missing the spectroscopic redshifts cannot be easily identified. Their visual inspection in the 5 bands does not reveal any specific photometric defect, nor is there anything noteworthy in their spatial distribution. They simply are the tail of a continuous $|\Delta z_{\rm norm}|$ degradation occurring with increasing redshift and magnitude. Reducing it is likely to be difficult without a richer training set at high redshift.

%They have larger than average DIP scores, with most falling in the right most bump in Fig. \ref{fig:DIPDIS_bins}.   

\subsection{Spectrocopic leftovers}
\label{subsec:leftovers}

\begin{figure}
\includegraphics[width=8.5cm]{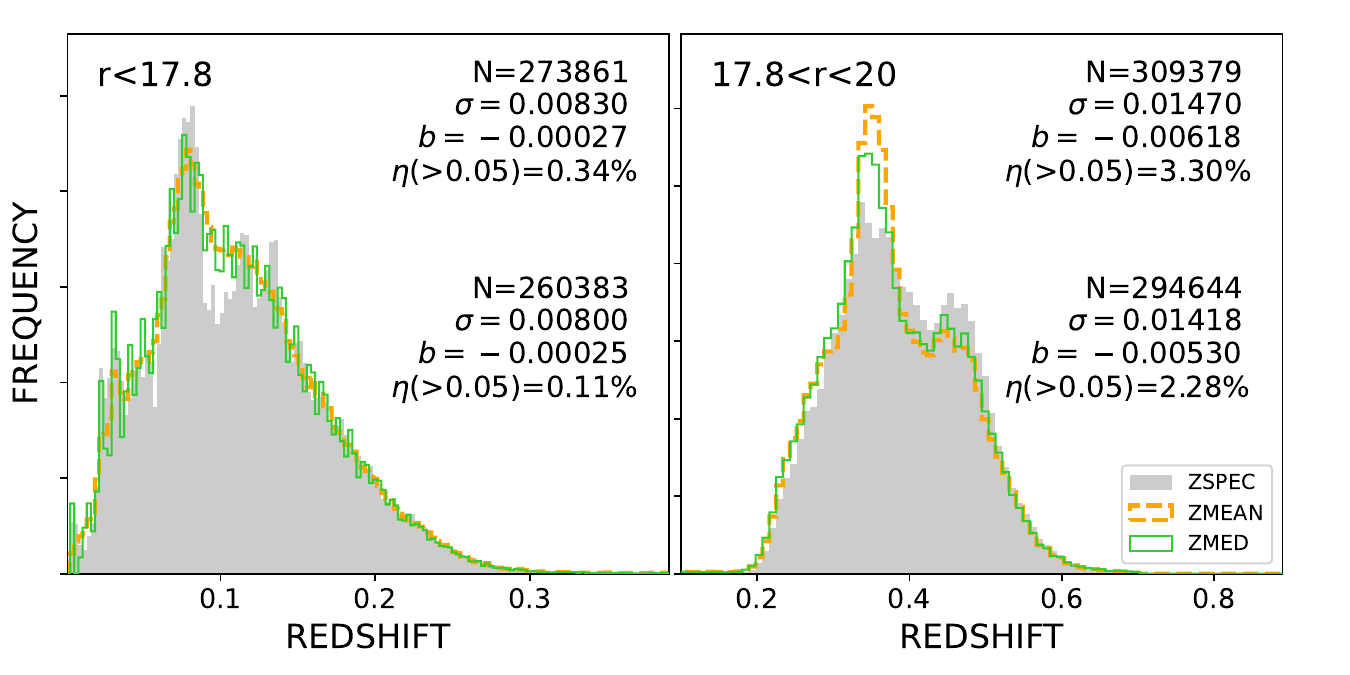}
\caption{The redshift distributions of leftover galaxies at $r<17.8$ and $17.8<r<20$ (left and right panel respectively). The gray shaded histograms are the spectroscopic redshift distributions, $z_{mean}$ and $z_{med}$ are in orange and green respectively. The faint interval is dominated by LRGs in the BOSS sample, hence the strong CNN redshift distortion around $\sim 0.35$, here emphasizing an existing feature. The metrics reported in the top right corners are for $z_{med}$.}  
\label{fig:nz_leftover}
\end{figure}

The $\sim$583k galaxies in the spectroscopic sample that belong neither to the test sample nor to any of the 15 training samples have a very bimodal distribution (see Fig \ref{fig:magz_testrain}): one half at $r<17.8$ is a mix of red and blue galaxies left over from, mostly, the bright SDSS catalog, the other half at $17.8<r<20$ is mainly red galaxies. Figure \ref{fig:nz_leftover} shows the redshift distributions of these 2 subsets, with the $z_{med}$ metrics before and after applying the PDF width threshold. 
In the bright interval, the excess of galaxies around $z_{spec}\sim 0.08$ is significantly larger than in the full SDSS catalog as the training samples were designed to avoid it. The CNN predictions are overly smooth compared to the true distribution, probably a counter bias from smoothing the training redshift distribution in order to avoid biases from such local structures. %One can never win on all fronts. 
Nevertheless, the metrics remain close to the values derived in Table \ref{table:point_estimates_perfs_sdss}.

In the faint magnitude panel, the situation is less favorable. The sample is dominated by high redshift LRGs that we chose to avoid in the training samples, and by red galaxies in the redshift interval of the color degeneracy (Section \ref{subsec:reddegeneracy}). The distortion induced for these galaxies around $\sim 0.35$ happens to emphasize an actual feature, also a leftover from the creation of the smooth training catalogs. Although this population largely differs from the test sample, the deviation and rate of catastrophic failures remain within the range of values derived in Table \ref{table:stats_gamalike}. However the bias is much larger, and negative. The redshifts are visibly under-estimated.
The training samples were designed to represent "normal" red galaxies within the magnitude and redshift range of the GAMA sample. 
The performance is poorer for these galaxies due to the color degeneracy, but poorer still for LRGs that are purposely under-represented in the training samples compared to their overwhelming presence in the spectroscopic sample. 
Adding high redshift LRGs to the training samples reduces the present bias but at the cost of increasing it in the test sample.
%Again, we couldn't win on all fronts. 
This is shown in Appendix \ref{subsec:LRG}. 

\section{Inference results}
\label{sec:inference}

We infer the PDFs of the $\sim$13.8M sources without spectroscopic redshift in the photometric sample (Section \ref{subsec:photodata}), split in two according to the SDSS keyword {\bf "clean"} referring to photometric quality.  
 %and 74.5\% have clean photometry and are outside masked regions ({\bf insideMask}=0). 
The results are presented below for the "clean" ({\bf clean}=1) sources ($\sim$81.7\%), and in Appendix \ref{sec:uncleanphoto} for the "dirty" ({\bf clean}=0) sources. 

\subsection{The "clean" sample}
\label{subsec:cleanphoto}

\begin{figure}
\includegraphics[width=4.2cm]{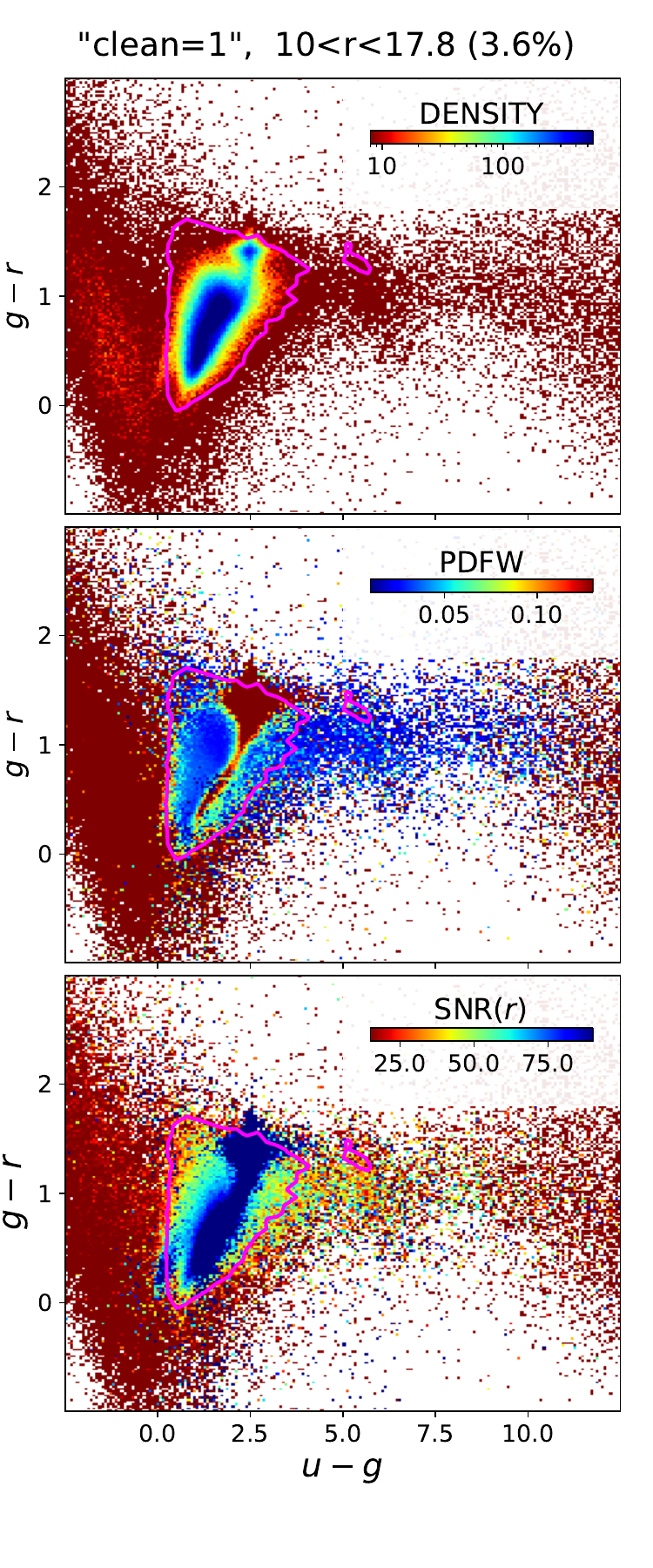}
\includegraphics[width=4.2cm]{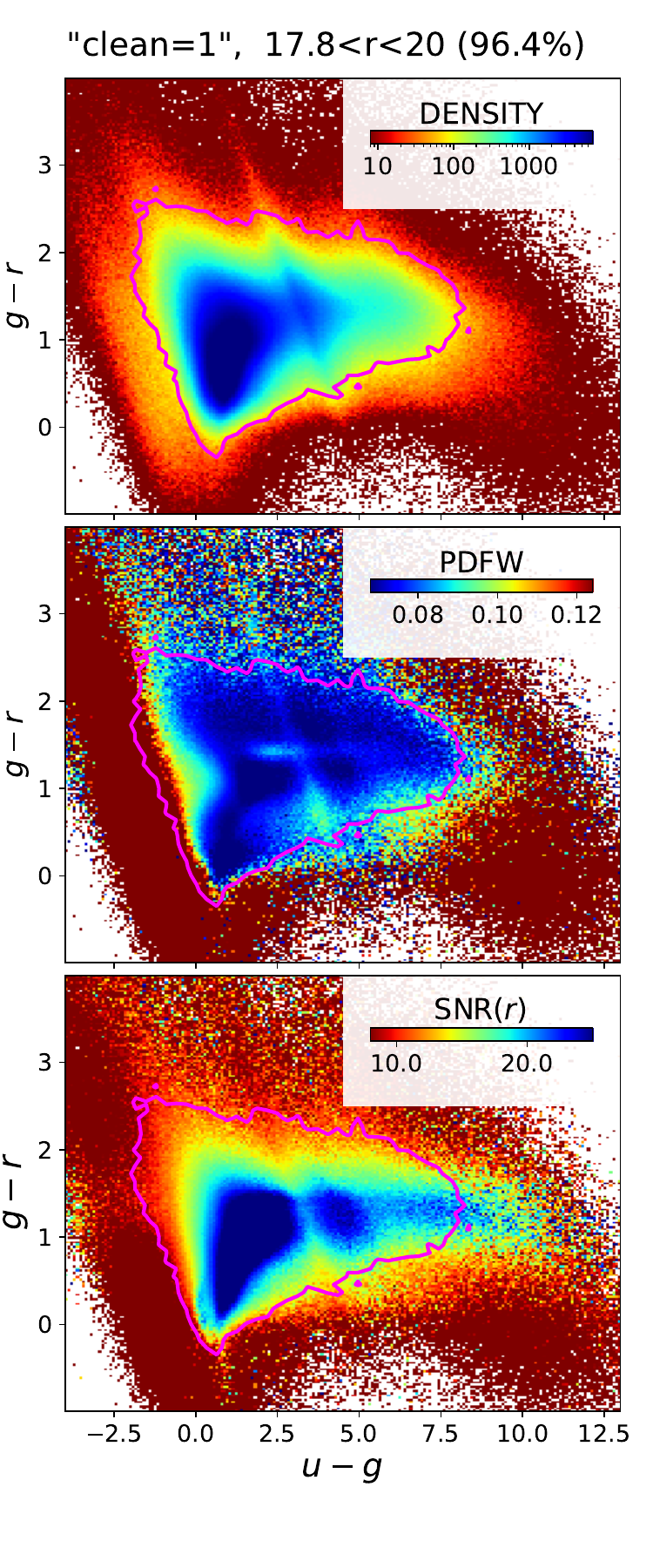}
\caption{The $(u-g)/(g-r)$ color distributions of the "{\bf clean}=1" photometric sample at $r<17.8$ and $r>17.8$ in the top left and right panels respectively. The distribution are color-coded by the mean PDF width in the middle panels and by the mean $r$-band signal-to-noise ratio (SNR) in the bottom panels.  A bright magnitude, the star sequence stands out with the poorest PDFs and the highest SNR. The PDFs are predictably very inconclusive outside of the training contours (shown in pink), where the SNRs are also very poor.}
\label{fig:colcol_inference_clean}
\end{figure}

Figure \ref{fig:colcol_inference_clean} shows the $(u-g)/(g-r)$ color distributions of the "clean" sources at $10<r<17.8$ and $17.8<r<20$ (left and right panels respectively). 
%The "insideMask>0" sources, which represent 7.5\% of this sub-sample, are not removed. 
The color-code indicates the density in the top panels, the mean PDF width in the middle panels and the mean $r-$band signal-to-noise ratio (SNR) in the bottom panels. The star sequence is conspicuous at $r<17.8$ with very poor PDFs, unsurprisingly since stars were not included in the training. Also unsurprisingly, the PDFs are very inconclusive in regions of the color/color plots not or ill represented in the training samples, the contours of which are shown in pink. These regions devoid of spectroscopy also have very poor SNR.
%in the $r$-band (and in the other 4 bands as well).
They could be the locus of bona fide galaxy populations that were systematically missed as spectroscopic targets due to their low optical SNR but it seems more likely that their colors are wrong and their PDFs useless due to poor image quality. In any case, whether the CNN infers in uncharted territories where machine learning techniques are unable to perform or whether the input data are flawed, the PDF widths clearly signal worthless predictions. We use the threshold introduced in the previous section (Eq. \ref{eq:pdfwmax_5pc}) to discard them. 

\begin{figure}
\includegraphics[width=8.5cm]{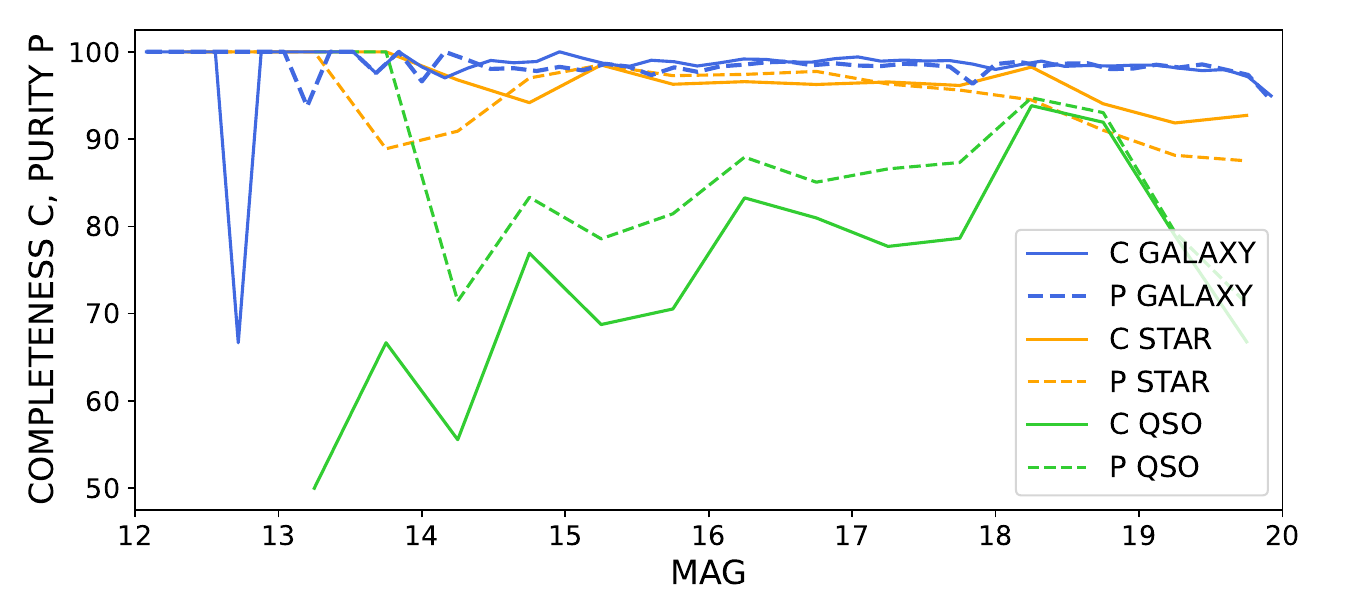}
\caption{Completeness and purity scores of the galaxy/star/QSO classifier as a function of magnitude in the concatenated validation samples. Only type 3 sources are considered.}
\label{fig:reda_class}
\end{figure}

We also build a galaxy/star/QSO classifier. This is a CNN similar to the redshift classifier, with the same type of input data, into which we insert, at the output of the last convolution, two successive layers of 96 neurons and a final 3 neuron layer for the triple classification. 
%(the regression output was removed). 
We train it with 80,000 sources in each class, randomly extracted from the SDSS spectroscopic catalog (type 3 and 6 objects were included for stars and QSOs). The results are cross-validated 5 times, with 80\% of the data used for training and 20\% for validation. Predicted classes assigned according to the highest probability yield completeness and purity scores\footnote{Completeness is defined as the fraction of galaxies (stars or QSOs) that are correctly classified, purity as the fraction of classified galaxies (stars or QSOs) that really are galaxies (stars or QSOs).} above 98\% for galaxies at all magnitudes (save for a glitch at $r\sim 12.7$) as shown in Fig. \ref{fig:reda_class}. The two scores are actually above 99\% for red galaxies, and 97.6\% and 96\% respectively for blue galaxies. Only type 3 sources are considered here, as in the photometric catalog, which leaves only 6194 stars and 7775 QSOs. (The scores are higher for type 6 stars and QSOs, except for the small fraction of bright, $r<18$ QSOs 
%($\sim 6\%)$ 
whose purity remains low). 

\begin{figure}
\includegraphics[width=8.5cm]{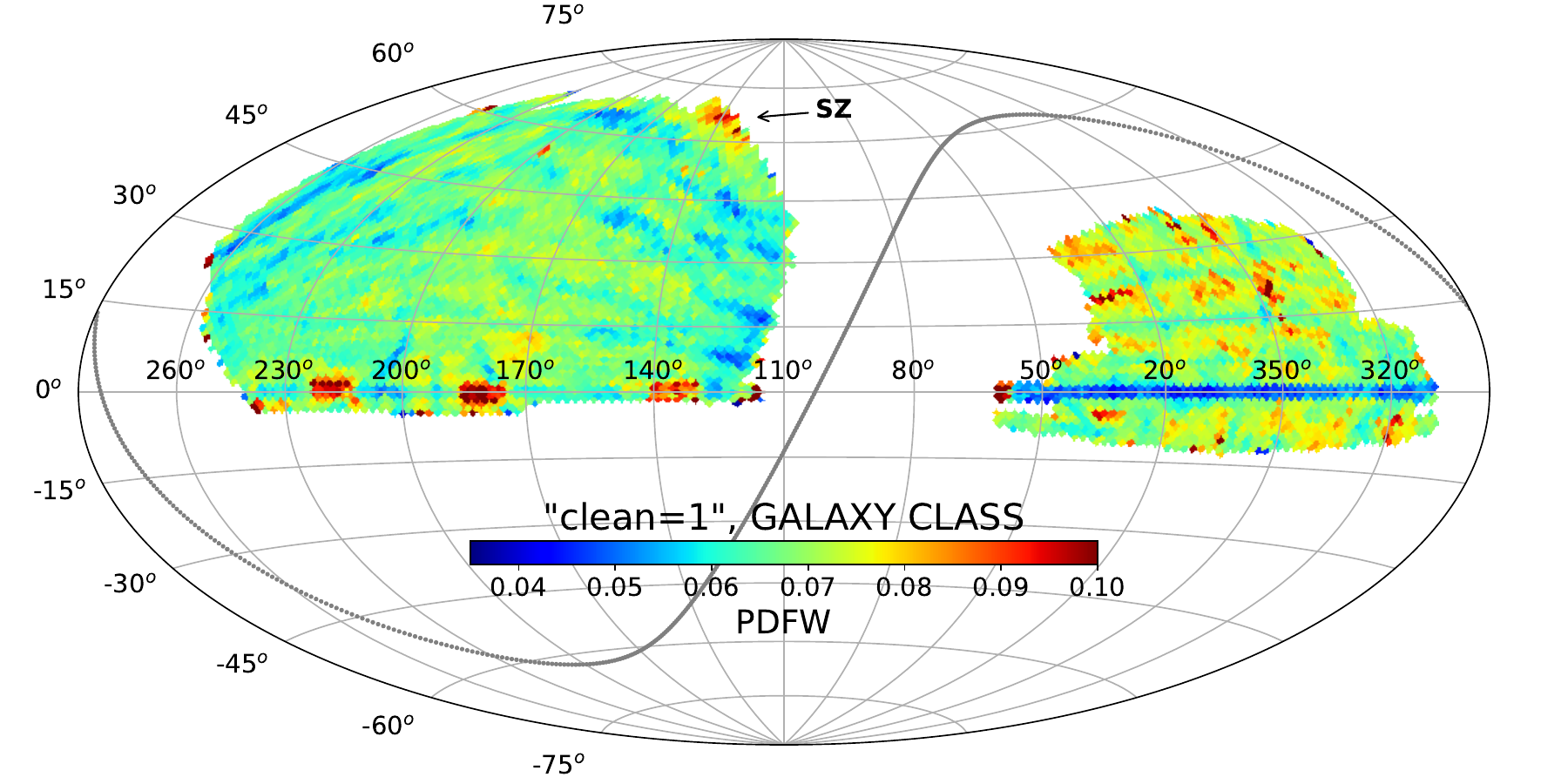}
\caption{PDF width-colored map of the sources classified as galaxies in the "clean=1" photometric samples. The "suspect zone" identified by P19 in the SDSS at $r<17.8$ in the northern region is marked "sz". Several similarly degraded patches also show in the southern region. The larger than average PDF widths in the  equatorial GAMA regions are due to their deeper than average magnitude. The lower than average PDF widths in Stripe 82 (equatorial blue stripe) are due to better image quality. }
\label{fig:map_clean_pdfw}
\end{figure}

Figure \ref{fig:map_clean_pdfw} shows the PDF width-colored sky map of the "clean" photometric sources classified as galaxies. Redshift prediction quality is unevenly distributed. The "suspect zone" (marked "sz") identified by P19 in the SDSS at $r<17.8$ is visible in the northern region. Other similarly degraded patches also show in the southern region (many more are visible in the "{\bf clean}=0" sub-sample shown in Fig. \ref{fig:map_notclean_pdfw}). On the other hand, the larger than average PDF widths in the GAMA regions are due to their deeper than average magnitude (nearly all GAMA sources with $r<19.8$ being part of the spectroscopic sample). The lower than average PDF widths in Stripe 82 (blue equatorial stripe in the South) are due to better image quality (P19).

\begin{figure}
\includegraphics[width=4.2cm]{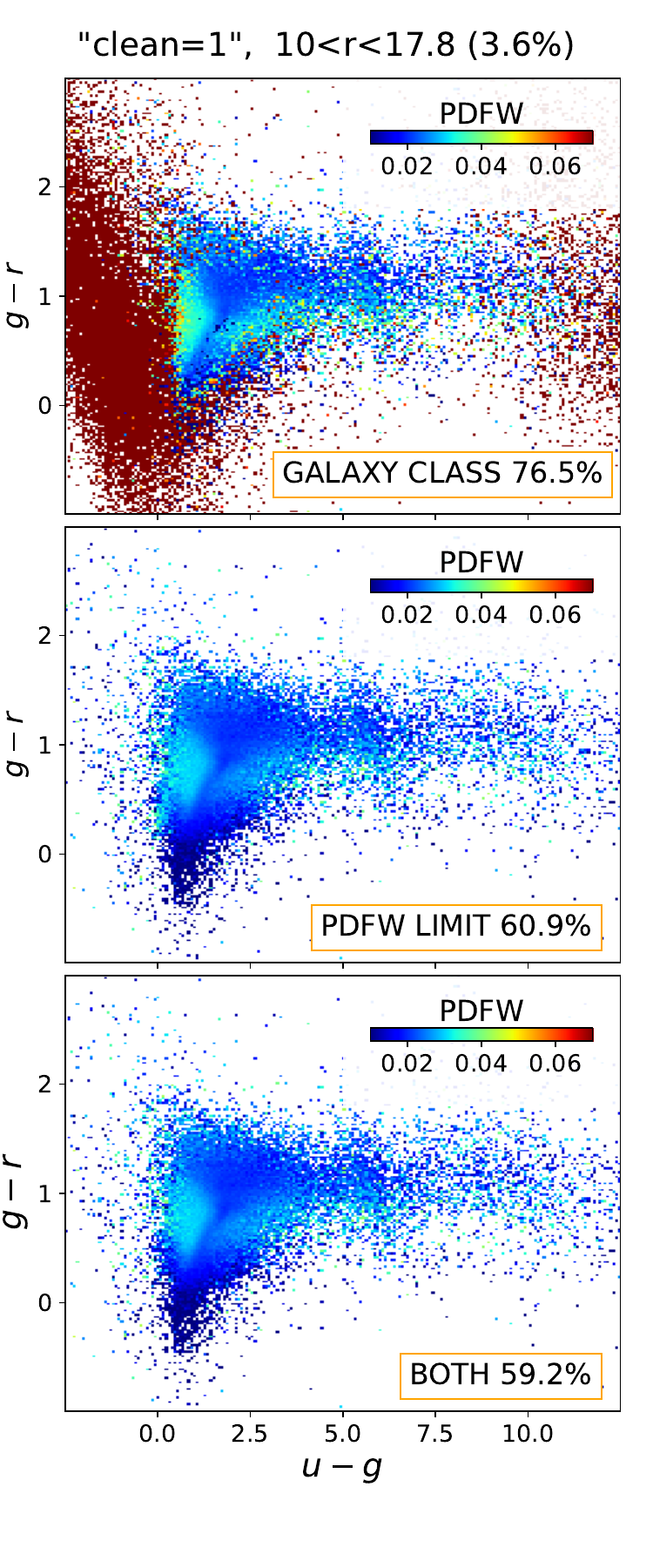}
\includegraphics[width=4.2cm]{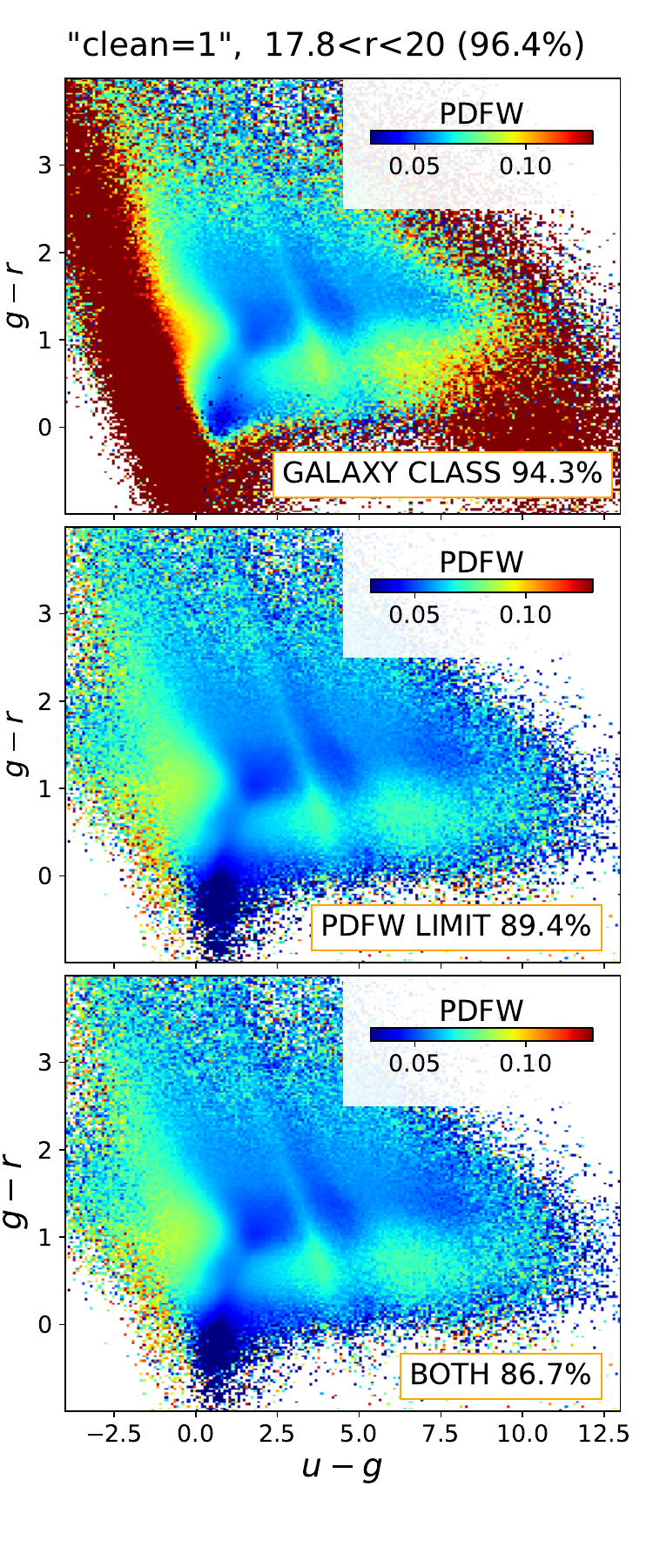}
\caption{The PDF width-colored $(u-g)/(g-r)$ distribution of the "clean=1" photometric sample at $r<17.8$ (left panels) and $r>17.8$ (right panels), following three cleansing procedures: the top panels restrict the sample to sources classified as galaxies, the middle panels applies the PDF width threshold (Eq. \ref{eq:pdfwmax_5pc}), and the bottom panels use both constraints. The PDF width limit alone efficiently screens both the faulty color regions and most of the classified stars and QSOs. The equivalent distributions are shown in Fig. \ref{fig:colcol_inference_notclean_cleaned} for the "{\bf clean}=0" sample.
}
\label{fig:colcol_inference_clean_cleaned}
\end{figure}

Figure \ref{fig:colcol_inference_clean_cleaned} shows the $(u-g)/(g-r)$ color distributions of the sample color-coded by the mean PDF width and split into the bright and faint magnitude intervals, following three cleansing procedures: the top panels are restricted to sources classified as galaxies (76.5\% at $r<17.8$, 94.3\% at $r>17.8$), the middle panels to sources with PDF widths below the threshold (60.9\% at $r<17.8$, 89.4\% at $r>17.8$), and finally the bottom panels to classified galaxies with PDF widths below the threshold (59.2\% at $r<17.8$, 86.7\% at $r>17.8$). This final procedure rejects 14.3\% of the initial data (6.4\% classified as stars or QSOs, 11.6\% with PDF widths above the threshold). 

The PDF width limit efficiently screens both the offending color regions and most of the classified stars and QSOs. 
%in both the "clean=1" and "clean=0" samples. 
It also clears the red, "suspect zone"-like blotches in the sky maps (Fig. \ref{fig:map_clean_pdfw} and \ref{fig:map_notclean_pdfw}). Expectedly, the waste is much greater than in the test sample, and much worse still for the "{\bf clean}=0" sources, but the procedure greatly and homogeneously improves the quality of the photometric redshifts in the two photometric sub-samples, which can thus be combined.

\subsection{The full photometric sample}
\label{subsec:fullphoto}

\begin{figure}
\includegraphics[width=8.5cm]{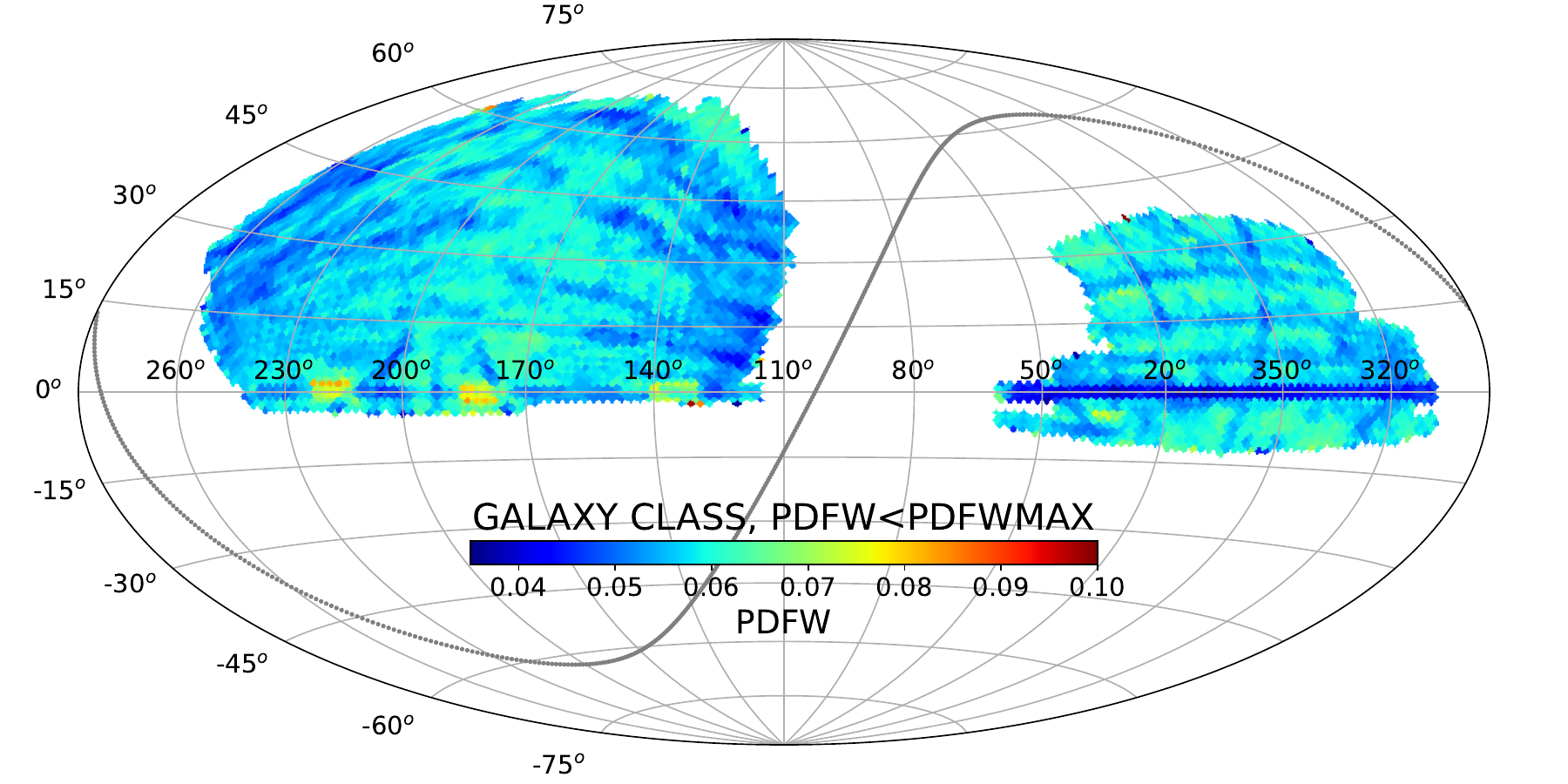}
\caption{PDF width-colored map of sources classified as galaxies with PDF width below the threshold in the full photometric samples ($\sim$10.75 million sources). The color code is the same as in Fig. \ref{fig:map_clean_pdfw} and \ref{fig:map_notclean_pdfw} to highlight the effect of the applied threshold.}
\label{fig:map_allsky_pdfw}
\end{figure}

The combined "{\bf clean}=0"+"{\bf clean}=1" photometric sample contains $\sim$11M sources classified as galaxies with PDF widths below the threshold. We simply refer to them as "galaxies" in the rest of this work. Figure \ref{fig:map_allsky_pdfw} shows the PDF width-colored sky distribution of these galaxies, with the same color code as in Fig. \ref{fig:map_clean_pdfw} and \ref{fig:map_notclean_pdfw} to highlight the effect of the applied threshold. The cleansing procedure rejects $20\%$ of the initial data ($\sim$10\% are classified as stars or QSOs, $\sim$17\% have PDF widths above the threshold). However it is not drastic enough to wash away the stripy disparities due to the SDSS observing conditions.

\begin{figure}
\includegraphics[width=8.5cm]{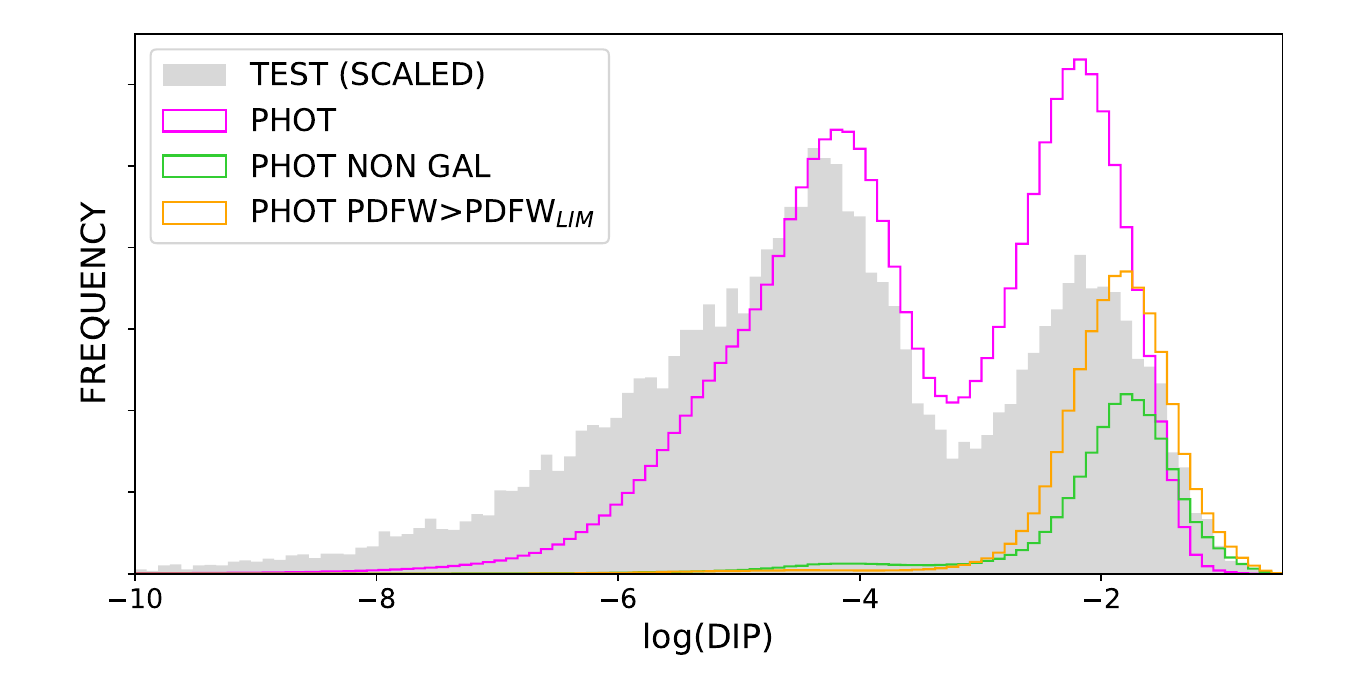}
\caption{The DIP score distribution of galaxies in the photometric sample (pink histogram) compared with that of the test sample (gray histogram, accordingly scaled). The green and orange histograms are the DIP score distributions of sources classified as stars or QSOs and of sources with PDF width above the threshold, respectively.}
\label{fig:DIPDIS_photo}
\end{figure}

\begin{figure}
\includegraphics[width=8.9cm]{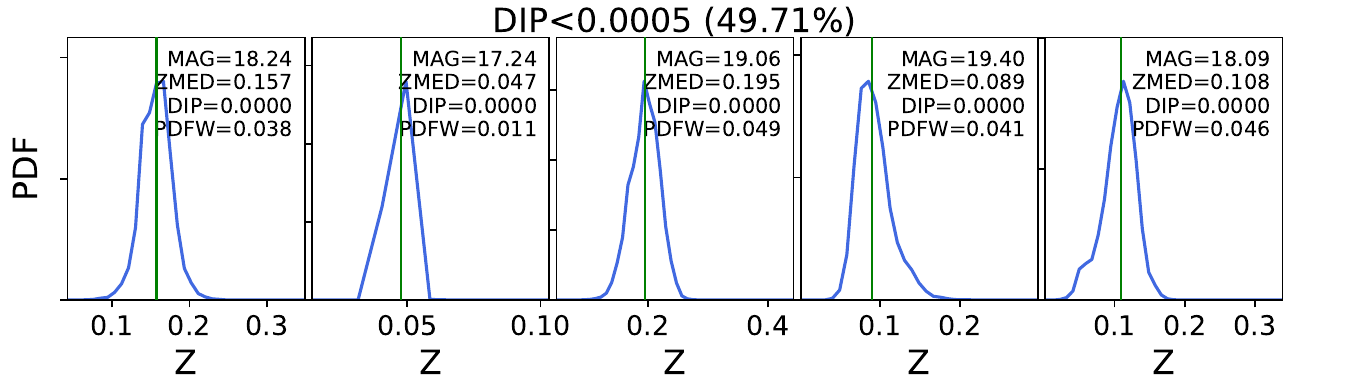}
\includegraphics[width=8.9cm]{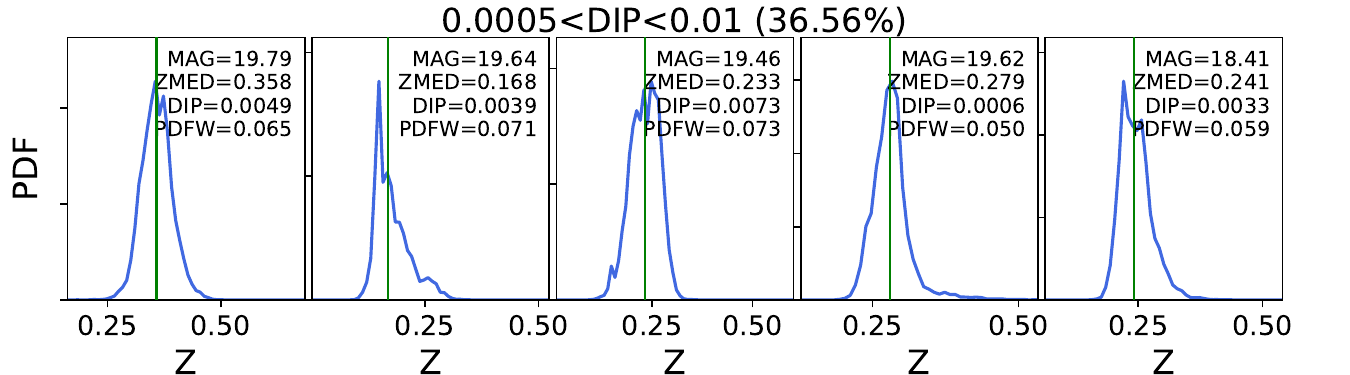}
\includegraphics[width=8.9cm]{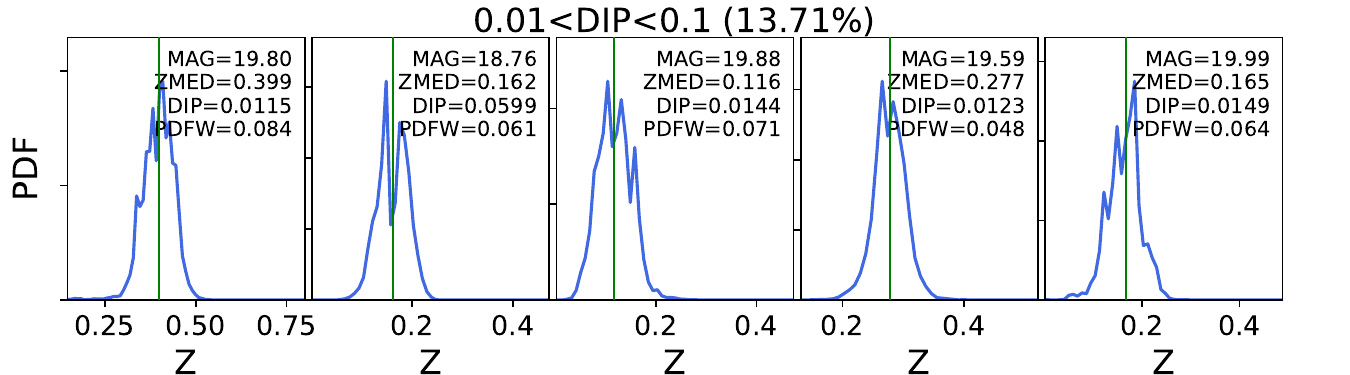}
\includegraphics[width=8.9cm]{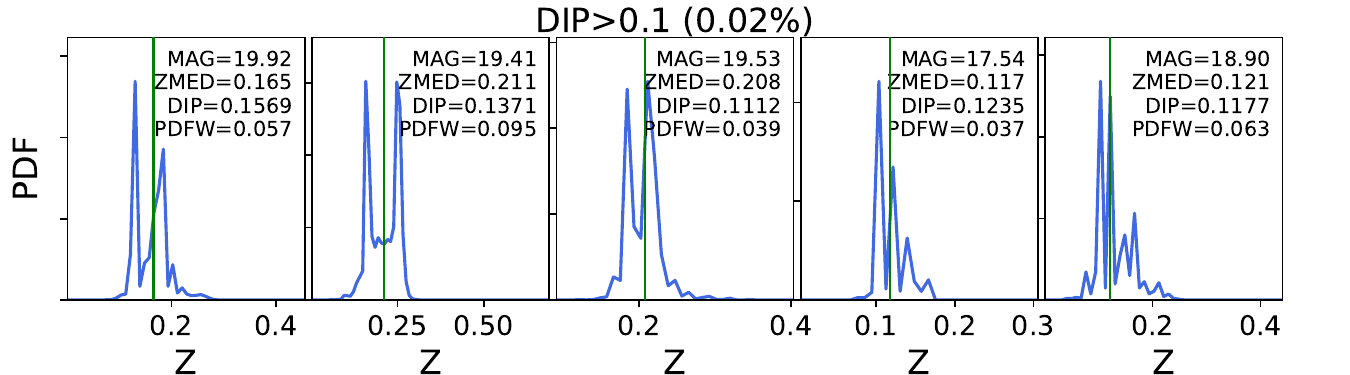}
\caption{Randomly selected PDFs with DIP<0.0005 (49.7\%), 0.0005<DIP<0.01 (36.5\%), 0.01<DIP<0.1 (13.7\%) and DIP>0.1 (0.02\%), from top to bottom.}
\label{fig:pdf_dip}
\end{figure}

Figure \ref{fig:DIPDIS_photo} shows the DIP score distribution of the photometric galaxies compared to the scaled test sample. Also shown for comparison are the DIP score distributions of sources classified as stars or QSOs and of sources with PDF width above the threshold. Figure \ref{fig:pdf_dip} shows samples of PDFs of photometric galaxies with DIP scores in the intervals: DIP $< 0.0005$ (49.7\%), $0.0005<\rm DIP< 0.01$ (36.5\%), $0.01<\rm DIP<0.1$ (13.7\%) and $\rm DIP>0.1$ (0.02\%). The green vertical lines mark the median point estimates. PDFs with DIP $<0.01$ (86\%) look very close to uni-modal. Of the remaining 14\%, only a very small minority with DIP $>0.1$ look compellingly multi-modal with probabilities dropping very low between peaks. %For comparison, 60\% of the sources classified as star or QSO and 53\% of the galaxy class with PDF width above the threshold belong in last 2 intervals. 

\iffalse
Figure \ref{fig:skewness_photo} shows the histograms of the discrepancies between the various point estimates. The wings are more extended than in the test sample (Fig. \ref{fig:skewness_test}) but the mean discrepancy between $z_{mean}$ and $z_{med}$ remains less than 0.5\%. 
\begin{figure}
\includegraphics[width=8.5cm]{skewness_photo.pdf}
\caption{Comparison of the point estimates (in percentages): as in the test sample, $z_{mean}$ is very similar to the regression value and positively skewed with respect to $z_{med}$. The wings are more extended but the mean difference remains less than 0.5\%.}
\label{fig:skewness_photo}
\end{figure}
\fi

\subsection{Redshift distributions}
\label{subsec:tomographer}

We compare the shapes of the CNN redshift distributions to those derived from the clustering redshift technique \citep[]{Menard2013,Rahman2015} using the online platform "Tomographer"\footnote{\url{https://tomographer.org/}}. This tool is perfectly suited to our case since it relies on the SDSS-BOSS spectroscopic population (Main Galaxy Sample, LRG and quasar samples), covering the north and south galactic caps as well as the redshift range of our bright, $r\le 20$ sample \citep[]{Chiang2019a,Chiang2019b}. The technique consists in spatially cross-correlating the spectroscopic population in bins of redshift with the sky positions of a test sample. The clustering amplitude is directly related to the redshift distribution of the test population, $dN/dz$, scaled by its bias $b(z)$ with respect to the underlying dark matter density field. In a narrow interval of magnitude, the redshift evolution of the bias may be neglected and the reconstructed distribution compared to the test distribution with a constant scaling factor \citep[e.g. ][]{Menard2013}.

We first run Tomographer on the spectroscopic sample in 12 intervals of magnitude from $r=17.6$ to 20 ($\Delta \rm mag=0.2$) to test the level of accuracy of the reconstruction. The results are shown in Appendix \ref{sec:tomographer_apx}. A noisy high redshift tail to $z\sim 3$ is present in all the magnitude bins, which we choose to ignore in the normalization to allow for a satisfactory, though far from perfect agreement between the observed and reconstructed redshift distributions in the redshift ranges of interest. This comparison gauges the accuracy we may expect for unknown distributions.

\begin{figure}
\includegraphics[width=8.7cm]{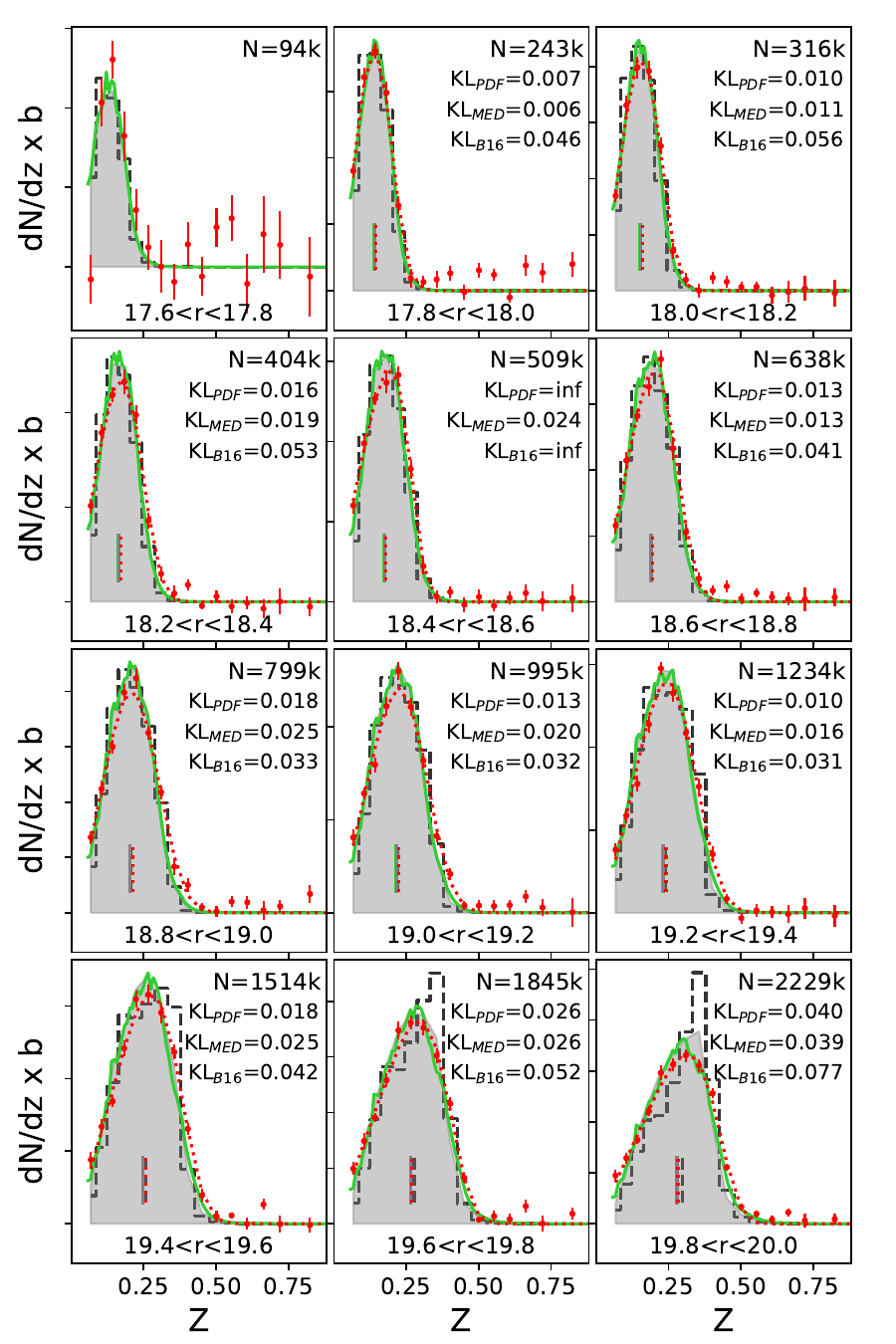}
\caption{Normalized CNN redshift distributions in intervals of magnitude compared to the Tomographer-derived distributions (red dots, fitted by dotted lines). The PDF sums (in green) are shown in their native binning while $z_{med}$ (gray shaded histogram) and $z_{B16}$ (dashed black histogram) use the Tomographer binning. Noted on each subplot are the KL divergences between the 3 photometric redshift distributions and Tomographer. The vertical segments mark the mean redshift of each distribution in their respective color. }
\label{fig:tomo_photo}
\end{figure}

In Fig. \ref{fig:tomo_photo}, we compare the redshift distributions of the photometric galaxies with the outputs of Tomographer in the 12 intervals of magnitude. 
%The PDF sums are shown in their native redshift binning while the $z_{med}$ and $z_{B16}$ histograms use the binning of Tomographer. 
The high redshift tails are not shown for clarity but as in the spectroscopic case, they need to be excluded in the normalization. Here we neglect them by fitting the outputs with the following ad-hoc, 4 free parameter function $F(z)= a z^b e^{-(z/c)^d}$. This operation does away with the oscillations around zero and the necessity to cherry-pick the last bin of interest to normalize each distribution and compute Kullback-Leibler divergences ($dN_{\rm CNN}/dz || dN_{\rm tomo}/dz$). These are reported in each panel, except in the first panel where no fit could be found. The $z_{med}$ KL (KL$_{\it MED}$) range from 0.006 to 0.039 from the brightest to the faintest bin. The KL of the stacked PDFs tend to be smaller. The B16 distributions are significantly more discrepant. They are plagued with a growing feature at $z\sim 0.35$ as magnitude increases, presumably related to the red galaxy degeneracy (Section \ref{subsec:reddegeneracy}). The mean redshift values, indicated for the three distributions with vertical lines of the appropriate color, all agree within $\sim$ 5\%. 
%The agreement between the CNN and Tomographer in all the magnitude bins substantiate the reliability of the PDFs across the whole survey area and magnitude range, at the resolution of the Tomographer's redshift binning.  

\subsection{Cluster membership}
\label{subsec:redmapper}

Another test of the CNN redshifts is provided by the redMaPPer cluster catalog \citep[]{Rykoff2014}\footnote{We use the 6.3 version of the redMaPPer catalogs based on the SDSS DR8 and available at  \url{https://cdsarc.cds.unistra.fr/ftp/J/ApJ/785/104/}}. The redMaPPer Cluster Finder is a red sequence cluster finder combining a calibration of the red sequence with spectroscopic redshifts and a matched-filter technique to find the clusters. For each cluster candidate, a redshift and a richness are assigned as well as a list of member galaxies at $r\lesssim 22$ with their membership probability. The catalog contains $\sim$26k cluster candidates at $z \lesssim 0.6$ over the two SDSS galactic caps. Based on the central galaxies with spectroscopic redshifts, the redshift uncertainty of the clusters is estimated to be lower than $\sigma=0.01$ at $z\le0.3$ and $\sigma=0.015$ at $z\le0.5$, with a systematic offset $\Delta z_{\rm norm}<0.003$ over the whole redshift range. 
 
\begin{figure}
\includegraphics[width=8.4cm]{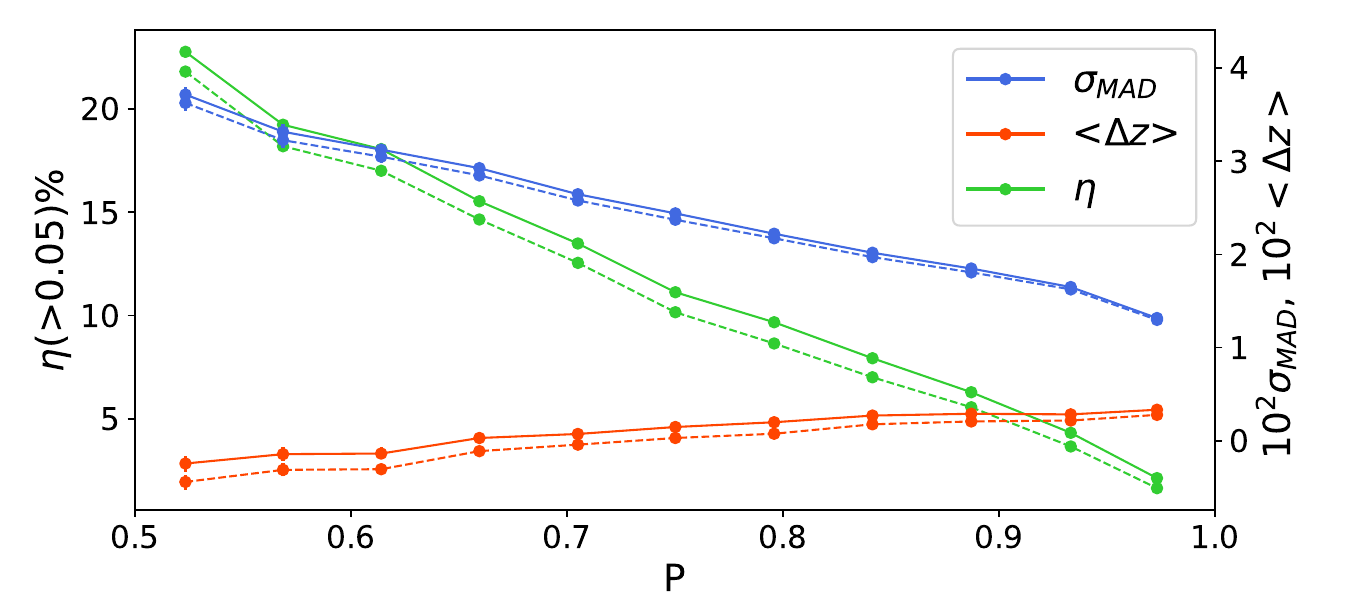}
\includegraphics[width=4.3cm]{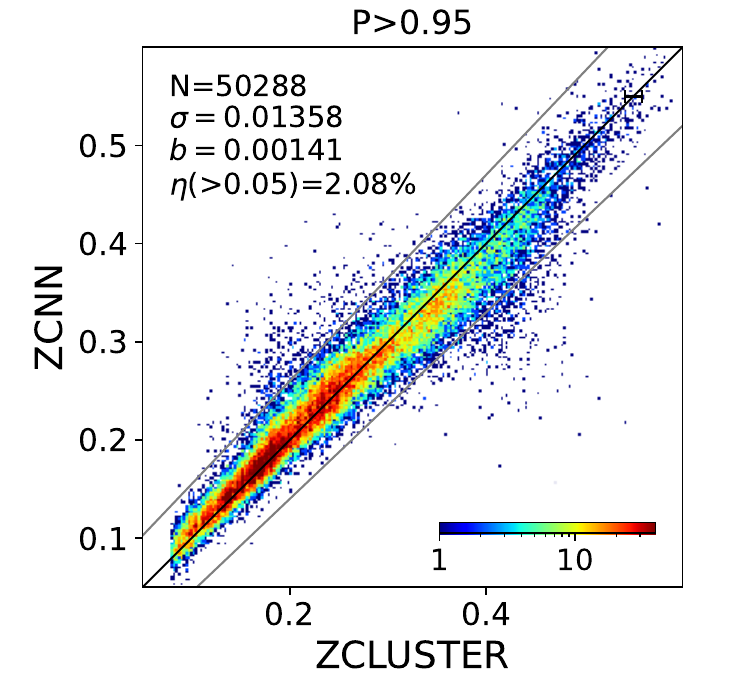}
\includegraphics[width=4.3cm]{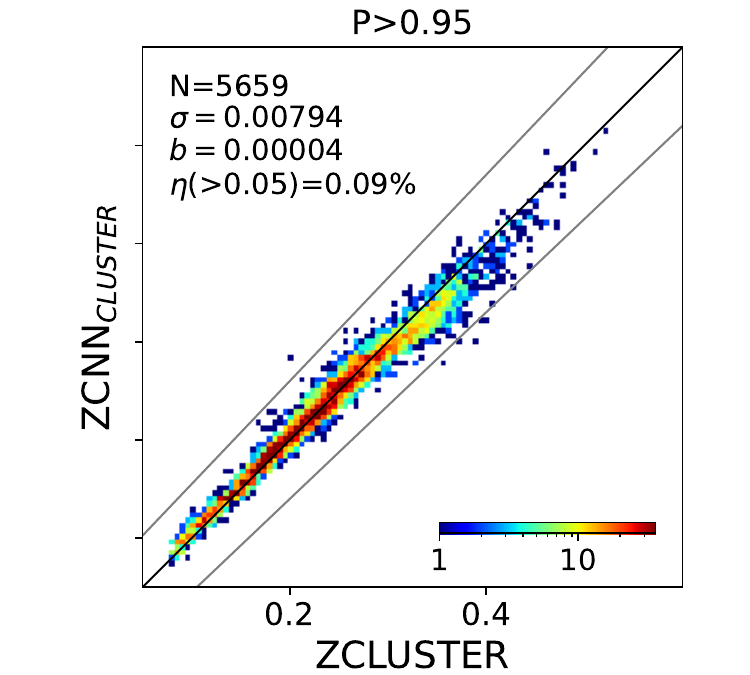}
\caption{Comparison between the photometric redshifts of rich redMaPPer galaxy clusters and the CNN $z_{med}$ of probable cluster members in the photometric sample. {\bf Top panel:} $\sigma_{\rm MAD}$, bias and fraction of catastrophic failures as a function of galaxy membership probability $P$. The dashed lines correspond to galaxies with PDF width below the threshold, which we assume in the bottom panels; {\bf Bottom left:} the redMaPPer redshifts of the clusters versus the CNN redshifts of the galaxies assigned to them with a membership probability $P>0.95$. The straight lines are the catastrophic failure borders and the identity line; {\bf Bottom right:} the redMaPPer redshifts of the clusters compared to the 
%median value of the combined PDFs of their members. 
mean value of the PDF product of their members.
Only clusters with at least 3 members are retained.}
\label{fig:zredMap}
\end{figure}

Among the $\sim$1.7M galaxies in the redMaPPer catalog of probable cluster members, $\sim$377k belong to our photometric sample of sources classified as galaxies. We assign them the redshift of their associated cluster, $z_{\rm CL}$ (the velocity dispersion within such a system rarely exceeding $\sigma_V=1000$km/s \citep[]{Clerc2016}, corresponding to an individual redshift uncertainty $\sigma_z<0.004$) and define $\Delta z = (z_{\rm CNN} - z_{\rm CL} )/(1 + z_{\rm CL})$ using $z_{med}$ as CNN point estimate. The top panel of Fig. \ref{fig:zredMap} shows how the deviation, bias and rate of catastrophic failures ($|\Delta z|>0.05$) evolve with the membership probability. The dashed lines correspond to galaxies with PDF width below our quality threshold ($\sim$360k). The accuracy gradually improves from $\sigma_{\rm MAD} \sim 0.035$ for galaxies with $0.50<P<0.55$ to $\sigma_{\rm MAD} \sim 0.013$ for galaxies with $0.95<P<1.0$, while the catastrophic fraction decreases from 20\% to 2\%.

The bottom left panel compares the redMaPPer redshift of the clusters to the CNN redshifts of the galaxies attributed to them with a probability $P>0.95$, and with PDF width below the threshold (N=50,288). The $\sigma_{\rm MAD}$, mean bias and rate of catastrophic failures are comparable to  those measured in the test sample (Table \ref{table:stats_gamalike}). In the bottom right panel, we compare the redMaPPer redshifts of the clusters to the 
%median value of the combined PDFs of their members (which is slightly more favorable than averaging the $z_{med}$), 
weighted mean value of the PDF product of their members, restricting the sample to clusters with at least 3 members with $P>0.95$ and PDF width below the threshold
(N=5659, with only 4 very rich clusters not surviving the PDF product).
%(N=5663). 
The catastrophic fraction is negligible and $\sigma_{\rm MAD}$ drops to 0.00794. 
%0.00825. It reaches $0.00633$ 
It reaches 0.00628
if we further restrict the sample to clusters with at least 5 members 
(N=3274, with the same 4 clusters having inconsistent members).
%(N=3278). 
This very good agreement mutually confirms the completely independent redshift quality of the CNN  and of the red sequence clusters, also supported by the SPIDERS cluster follow-up \citep[]{Clerc2016}.

%\section{Other tests}

%Batch size : 128 instead of 32 on sdss x1 -> reduces training time by a lot but also the performance a little (especially for zmean)
%Image size: 46x46 and 90x90 with old cfhtls catalog \& TF1 x1
%Activation functions: cfhtls x4 PReLU -> increases training time by a lot and the performance just a little; cfhtls x4 ELU -> in progress more bands: with HSC

%%%%%%%%%%%%%%%%%%%%%Section 6: CONCLUSION %%%%%%%%%%%%%%%%%%%%%

\section{Conclusion}
\label{sec:conclusion}

Inferring from a CNN trained with $ugriz$ stamp images of galaxies from the SDSS, GAMA and BOSS spectroscopic surveys, we estimated redshifts for the $\sim$14 million sources at $r\leq 20$ without spectroscopic redshift in the 11,500 deg$^2$ of the SDSS north and south footprints. The redshifts extend to $\sim$0.7. 
To limit biases, particularly those resulting from the overpopulation of Luminous Red Galaxies in the BOSS data, we extracted from the full spectroscopic catalog several training samples approximately representative of the general galaxy population at $r<20$, using GAMA as a model. The CNN was built to classify redshifts into narrow, contiguous bins. 
The classification outputs offer all the benefits of well-behaved bona fide PDFs, passing several statistical tests and efficiently flagging, via their widths, unreliable estimates due to poor photometry or stellar sources. Based on a representative spectroscopic test sample, the point estimates are more than twice as accurate as the photometric redshifts currently published for the SDSS at $r<20$.

We designed a magnitude-dependent PDF width threshold and a galaxy/star/QSO classifier to clean the inference sample, leaving $\sim$11M sources whose CNN redshift quality we deem reliable and homogeneous over the whole footprint. These redshifts are in very good agreement with the independently derived photometric redshifts of the redMaPPer galaxy clusters for the probable cluster members among them. Their distributions in narrow bins of magnitudes also match the results of tomographic analyses satisfactorily. 

 Pending the release of spectroscopic redshifts by the Dark Energy Spectroscopic Instrument (DESI) for their Bright Galaxy Sample in a few years %\citep{DESI2022}, 
these photometric redshifts are of interest for a variety of statistical analyses. They are accurate enough to improve cluster membership and reveal the cosmic web in thin redshift slices \citep[e.g.][]{Laigle2018}, allowing us to extend the spectroscopic analyses probing its impact on galaxy properties \citep[]{Kraljic2018, Malavasi2017}. They may also improve the cosmological information retrieved from cross-correlating sparse spectroscopic samples with photometric data, e.g. by reducing the shot noise at the baryon acoustic oscillation scale \citep[]{Patej2018}. They can be used to measure, e.g.: the connectivity of groups and clusters to study the properties of their member galaxies as a function of group mass and assembly history %\citep[as traced by the connectivity, ][]
\citep{DarraghFord2019}; 
the evolution of the mean connectivity with redshift, which depends on the cosmological model and on the nature of the dark energy \citep[]{Codis2018}; the one-point distribution of the cosmic density field in cylinder at a given radius in a way complementary to standard power spectrum analysis \citep[albeit with different biases and sensitivity to cosmology, ][]{Ulhemann2018}. 

%While 
However the range of the present CNN redshifts is limited and their precision degrade as they and magnitude increase. Going down in magnitude is a big challenge, especially as spectroscopic data become very sparse. More complex deep learning techniques are necessary for deeper on-going surveys such as HSC-CLAUDS \citep[]{Sawicki2019} and future ones such as LSST \citep[]{Ivezic2019}. Work is underway that promises to reach high photometric redshift accuracy to $z\sim 1.5$ (Ait-Ouahmed et al. in prep.),  allowing us to extend the scope of the above cosmological investigations. The challenge of satisfying the LSST science requirements to $z=3$ is yet to be met.

\section*{Data availability}
The CNN input (5-band image cutouts and associated spectroscopic and photometric catalogs) and output (PDFs and derived quantities) for the $\sim$14M sources in the SDSS DR16 at $r<20$ are available at: \url{https://deepdip.iap.fr/treyer+2023}.

%The CNN input and output are available at https://dx.doi.org/[TBD]. 
 
%%%%%%%%%%%%%%%%%%%%%THANKS%%%%%%%%%%%%%%%%%%%%%

\section*{Acknowledgements}
This work was granted access to the HPC resources of IDRIS under the allocation 2023-AD010414147 made by GENCI. It was carried out thanks to the support of the DEEPDIP ANR project (ANR-19-CE31-0023, \url{http://deepdip.net}) and  the Programme National Cosmologie et Galaxies (PNCG) of CNRS/INSU with INP and IN2P3, co-funded by CEA and CNES. This publication makes use of Sloan Digital Sky Survey (SDSS) data. Funding for SDSS-III has been provided by the Alfred P. Sloan Foundation, the Participating Institutions, the National Science Foundation, and the U.S. Department of Energy Office of Science. The SDSS-III web site is \url{http://www.sdss3.org/}.
SDSS-III is managed by the Astrophysical Research Consortium for the Participating Institutions of the SDSS-III Collaboration including the University of Arizona, the Brazilian Participation Group, Brookhaven National Laboratory, Carnegie Mellon University, University of Florida, the French Participation Group, the German Participation Group, Harvard University, the Instituto de Astrofisica de Canarias, the Michigan State/Notre Dame/JINA Participation Group, Johns Hopkins University, Lawrence Berkeley National Laboratory, Max Planck Institute for Astrophysics, Max Planck Institute for Extraterrestrial Physics, New Mexico State University, New York University, Ohio State University, Pennsylvania State University, University of Portsmouth, Princeton University, the Spanish Participation Group, University of Tokyo, University of Utah, Vanderbilt University, University of Virginia, University of Washington, and Yale University.

\bibliographystyle{mnras}
\bibliography{aa.bib}

\begin{thebibliography}{}
\makeatletter
\relax
\def\mn@urlcharsother{\let\do\@makeother \do\$\do\&\do\#\do\^\do\_\do\%\do\~}
\def\mn@doi{\begingroup\mn@urlcharsother \@ifnextchar [ {\mn@doi@} {\mn@doi@[]}}
\def\mn@doi@[#1]#2{\def\@tempa{#1}\ifx\@tempa\@empty \href {http://dx.doi.org/#2} {doi:#2}\else \href {http://dx.doi.org/#2} {#1}\fi \endgroup}
\def\mn@eprint#1#2{\mn@eprint@#1:#2::\@nil}
\def\mn@eprint@arXiv#1{\href {http://arxiv.org/abs/#1} {{\tt arXiv:#1}}}
\def\mn@eprint@dblp#1{\href {http://dblp.uni-trier.de/rec/bibtex/#1.xml} {dblp:#1}}
\def\mn@eprint@#1:#2:#3:#4\@nil{\def\@tempa {#1}\def\@tempb {#2}\def\@tempc {#3}\ifx \@tempc \@empty \let \@tempc \@tempb \let \@tempb \@tempa \fi \ifx \@tempb \@empty \def\@tempb {arXiv}\fi \@ifundefined {mn@eprint@\@tempb}{\@tempb:\@tempc}{\expandafter \expandafter \csname mn@eprint@\@tempb\endcsname \expandafter{\@tempc}}}

\bibitem[\protect\citeauthoryear{{Ahumada} et~al.,}{{Ahumada} et~al.}{2020}]{SDSS_DR16}
{Ahumada} R.,  et~al., 2020, \mn@doi [\apjs] {10.3847/1538-4365/ab929e}, \href {https://ui-adsabs-harvard-edu.insu.bib.cnrs.fr/abs/2020ApJS..249....3A} {249, 3}

\bibitem[\protect\citeauthoryear{{Alam} et~al.,}{{Alam} et~al.}{2015}]{Alam2015}
{Alam} S.,  et~al., 2015, \mn@doi [\apjs] {10.1088/0067-0049/219/1/12}, \href {http://adsabs.harvard.edu/abs/2015ApJS..219...12A} {219, 12}

\bibitem[\protect\citeauthoryear{{Amaro} et~al.,}{{Amaro} et~al.}{2019}]{Amaro2019}
{Amaro} V.,  et~al., 2019, \mn@doi [\mnras] {10.1093/mnras/sty2922}, \href {https://ui-adsabs-harvard-edu.insu.bib.cnrs.fr/abs/2019MNRAS.482.3116A} {482, 3116}

\bibitem[\protect\citeauthoryear{{Arnouts}, {Cristiani}, {Moscardini}, {Matarrese}, {Lucchin}, {Fontana}  \& {Giallongo}}{{Arnouts} et~al.}{1999}]{Arnouts1999}
{Arnouts} S.,  {Cristiani} S.,  {Moscardini} L.,  {Matarrese} S.,  {Lucchin} F.,  {Fontana} A.,   {Giallongo} E.,  1999, \mn@doi [\mnras] {10.1046/j.1365-8711.1999.02978.x}, \href {http://adsabs.harvard.edu/abs/1999MNRAS.310..540A} {310, 540}

\bibitem[\protect\citeauthoryear{Baum \& Wilczek}{Baum \& Wilczek}{1987}]{Baum1987}
Baum E.~B.,  Wilczek F.,  1987, in Proceedings of the 1987 International Conference on Neural Information Processing Systems. NIPS'87.
MIT Press, Cambridge, MA, USA, pp 52--61, \url {http://dl.acm.org/citation.cfm?id=2969644.2969650}

\bibitem[\protect\citeauthoryear{{Beck}, {Dobos}, {Budav{\'a}ri}, {Szalay}  \& {Csabai}}{{Beck} et~al.}{2016}]{Beck2016}
{Beck} R.,  {Dobos} L.,  {Budav{\'a}ri} T.,  {Szalay} A.~S.,   {Csabai} I.,  2016, \mn@doi [\mnras] {10.1093/mnras/stw1009}, \href {http://adsabs.harvard.edu/abs/2016MNRAS.460.1371B} {460, 1371}

\bibitem[\protect\citeauthoryear{{Bertin}, {Mellier}, {Radovich}, {Missonnier}, {Didelon}  \& {Morin}}{{Bertin} et~al.}{2002}]{2002ASPC..281..228B}
{Bertin} E.,  {Mellier} Y.,  {Radovich} M.,  {Missonnier} G.,  {Didelon} P.,   {Morin} B.,  2002, in {Bohlender} D.~A.,  {Durand} D.,   {Handley} T.~H.,  eds,  Astronomical Society of the Pacific Conference Series Vol. 281, Astronomical Data Analysis Software and Systems XI. p.~228

\bibitem[\protect\citeauthoryear{{Bradshaw} et~al.,}{{Bradshaw} et~al.}{2013}]{Bradshaw2013}
{Bradshaw} E.~J.,  et~al., 2013, \mn@doi [\mnras] {10.1093/mnras/stt715}, \href {https://ui.adsabs.harvard.edu/abs/2013MNRAS.433..194B} {433, 194}

\bibitem[\protect\citeauthoryear{{Brammer}, {van Dokkum}  \& {Coppi}}{{Brammer} et~al.}{2008}]{Brammer2008}
{Brammer} G.~B.,  {van Dokkum} P.~G.,   {Coppi} P.,  2008, \mn@doi [\apj] {10.1086/591786}, \href {http://adsabs.harvard.edu/abs/2008ApJ...686.1503B} {686, 1503}

\bibitem[\protect\citeauthoryear{{Brescia}, {Cavuoti}, {Razim}, {Amaro}, {Riccio}  \& {Longo}}{{Brescia} et~al.}{2021}]{Brescia2021}
{Brescia} M.,  {Cavuoti} S.,  {Razim} O.,  {Amaro} V.,  {Riccio} G.,   {Longo} G.,  2021, \mn@doi [Frontiers in Astronomy and Space Sciences] {10.3389/fspas.2021.658229}, \href {https://ui.adsabs.harvard.edu/abs/2021FrASS...8...70B} {8, 70}

\bibitem[\protect\citeauthoryear{Bridle}{Bridle}{1990}]{Bridle1990}
Bridle J.~S.,  1990, in Souli{\'e} F.~F.,  H{\'e}rault J.,  eds, Neurocomputing. Springer Berlin Heidelberg, Berlin, Heidelberg, pp 227--236

\bibitem[\protect\citeauthoryear{{Brinchmann}, {Charlot}, {White}, {Tremonti}, {Kauffmann}, {Heckman}  \& {Brinkmann}}{{Brinchmann} et~al.}{2004}]{Brinchmann04}
{Brinchmann} J.,  {Charlot} S.,  {White} S.~D.~M.,  {Tremonti} C.,  {Kauffmann} G.,  {Heckman} T.,   {Brinkmann} J.,  2004, \mn@doi [\mnras] {10.1111/j.1365-2966.2004.07881.x}, \href {http://adsabs.harvard.edu/cgi-bin/nph-bib_query?bibcode=2004MNRAS.351.1151B&db_key=AST} {351, 1151}

\bibitem[\protect\citeauthoryear{{Calabretta} \& {Greisen}}{{Calabretta} \& {Greisen}}{2002}]{2002A&A...395.1077C}
{Calabretta} M.~R.,  {Greisen} E.~W.,  2002, \mn@doi [\aap] {10.1051/0004-6361:20021327}, \href {http://adsabs.harvard.edu/abs/2002A%26A...395.1077C} {395, 1077}

\bibitem[\protect\citeauthoryear{{Chiang} \& {M{\'e}nard}}{{Chiang} \& {M{\'e}nard}}{2019}]{Chiang2019a}
{Chiang} Y.-K.,  {M{\'e}nard} B.,  2019, \mn@doi [\apj] {10.3847/1538-4357/aaf4f6}, \href {https://ui-adsabs-harvard-edu.insu.bib.cnrs.fr/abs/2019ApJ...870..120C} {870, 120}

\bibitem[\protect\citeauthoryear{{Chiang}, {M{\'e}nard}  \& {Schiminovich}}{{Chiang} et~al.}{2019}]{Chiang2019b}
{Chiang} Y.-K.,  {M{\'e}nard} B.,   {Schiminovich} D.,  2019, \mn@doi [\apj] {10.3847/1538-4357/ab1b35}, \href {https://ui-adsabs-harvard-edu.insu.bib.cnrs.fr/abs/2019ApJ...877..150C} {877, 150}

\bibitem[\protect\citeauthoryear{{Clerc} et~al.,}{{Clerc} et~al.}{2016}]{Clerc2016}
{Clerc} N.,  et~al., 2016, \mn@doi [\mnras] {10.1093/mnras/stw2214}, \href {https://ui-adsabs-harvard-edu.insu.bib.cnrs.fr/abs/2016MNRAS.463.4490C} {463, 4490}

\bibitem[\protect\citeauthoryear{{Codis}, {Pogosyan}  \& {Pichon}}{{Codis} et~al.}{2018}]{Codis2018}
{Codis} S.,  {Pogosyan} D.,   {Pichon} C.,  2018, \mn@doi [\mnras] {10.1093/mnras/sty1643}, \href {https://ui-adsabs-harvard-edu.insu.bib.cnrs.fr/abs/2018MNRAS.479..973C} {479, 973}

\bibitem[\protect\citeauthoryear{{Collister} \& {Lahav}}{{Collister} \& {Lahav}}{2004}]{Collister2004}
{Collister} A.~A.,  {Lahav} O.,  2004, \mn@doi [\pasp] {10.1086/383254}, \href {http://adsabs.harvard.edu/abs/2004PASP..116..345C} {116, 345}

\bibitem[\protect\citeauthoryear{{Csabai}, {Dobos}, {Trencs{\'e}ni}, {Herczegh}, {J{\'o}zsa}, {Purger}, {Budav{\'a}ri}  \& {Szalay}}{{Csabai} et~al.}{2007}]{Csabai2007}
{Csabai} I.,  {Dobos} L.,  {Trencs{\'e}ni} M.,  {Herczegh} G.,  {J{\'o}zsa} P.,  {Purger} N.,  {Budav{\'a}ri} T.,   {Szalay} A.~S.,  2007, \mn@doi [Astronomische Nachrichten] {10.1002/asna.200710817}, \href {http://adsabs.harvard.edu/abs/2007AN....328..852C} {328, 852}

\bibitem[\protect\citeauthoryear{{D'Isanto} \& {Polsterer}}{{D'Isanto} \& {Polsterer}}{2018}]{disanto2018}
{D'Isanto} A.,  {Polsterer} K.~L.,  2018, \mn@doi [\aap] {10.1051/0004-6361/201731326}, \href {http://adsabs.harvard.edu/abs/2018A%26A...609A.111D} {609, A111}

\bibitem[\protect\citeauthoryear{{Dark Energy Survey Collaboration} et~al.,}{{Dark Energy Survey Collaboration} et~al.}{2016}]{DES2016}
{Dark Energy Survey Collaboration} et~al., 2016, \mn@doi [\mnras] {10.1093/mnras/stw641}, \href {https://ui.adsabs.harvard.edu/abs/2016MNRAS.460.1270D} {460, 1270}

\bibitem[\protect\citeauthoryear{{Darragh Ford} et~al.,}{{Darragh Ford} et~al.}{2019}]{DarraghFord2019}
{Darragh Ford} E.,  et~al., 2019, \mn@doi [\mnras] {10.1093/mnras/stz2490}, \href {https://ui-adsabs-harvard-edu.insu.bib.cnrs.fr/abs/2019MNRAS.489.5695D} {489, 5695}

\bibitem[\protect\citeauthoryear{Dawid}{Dawid}{1984}]{dawid.2307/2981683}
Dawid A.~P.,  1984, Journal of the Royal Statistical Society. Series A (General), 147, 278

\bibitem[\protect\citeauthoryear{{Dawson} et~al.,}{{Dawson} et~al.}{2013}]{Dawson2013}
{Dawson} K.~S.,  et~al., 2013, \mn@doi [\aj] {10.1088/0004-6256/145/1/10}, \href {https://ui.adsabs.harvard.edu/abs/2013AJ....145...10D} {145, 10}

\bibitem[\protect\citeauthoryear{{Dey}, {Andrews}, {Newman}, {Mao}, {Rau}  \& {Zhou}}{{Dey} et~al.}{2021}]{Dey2021}
{Dey} B.,  {Andrews} B.~H.,  {Newman} J.~A.,  {Mao} Y.-Y.,  {Rau} M.~M.,   {Zhou} R.,  2021, arXiv e-prints, \href {https://ui-adsabs-harvard-edu.insu.bib.cnrs.fr/abs/2021arXiv211203939D} {p. arXiv:2112.03939}

\bibitem[\protect\citeauthoryear{{Driver} et~al.,}{{Driver} et~al.}{2009}]{Driver2009}
{Driver} S.~P.,  et~al., 2009, \mn@doi [Astronomy and Geophysics] {10.1111/j.1468-4004.2009.50512.x}, \href {https://ui.adsabs.harvard.edu/abs/2009A&G....50e..12D} {50, 5.12}

\bibitem[\protect\citeauthoryear{{Driver} et~al.,}{{Driver} et~al.}{2011}]{Driver2011}
{Driver} S.~P.,  et~al., 2011, \mn@doi [\mnras] {10.1111/j.1365-2966.2010.18188.x}, \href {https://ui.adsabs.harvard.edu/abs/2011MNRAS.413..971D} {413, 971}

\bibitem[\protect\citeauthoryear{{Driver} et~al.,}{{Driver} et~al.}{2022}]{Driver2022}
{Driver} S.~P.,  et~al., 2022, \mn@doi [\mnras] {10.1093/mnras/stac472}, \href {https://ui-adsabs-harvard-edu.insu.bib.cnrs.fr/abs/2022MNRAS.513..439D} {513, 439}

\bibitem[\protect\citeauthoryear{{Firth}, {Lahav}  \& {Somerville}}{{Firth} et~al.}{2003}]{firth2003MNRAS.339.1195F}
{Firth} A.~E.,  {Lahav} O.,   {Somerville} R.~S.,  2003, \mn@doi [\mnras] {10.1046/j.1365-8711.2003.06271.x}, \href {http://adsabs.harvard.edu/abs/2003MNRAS.339.1195F} {339, 1195}

\bibitem[\protect\citeauthoryear{Hartigan \& Hartigan}{Hartigan \& Hartigan}{1985}]{Hartigan85}
Hartigan J.~A.,  Hartigan P.~M.,  1985, \mn@doi [The Annals of Statistics] {10.1214/aos/1176346577}, 13, 70

\bibitem[\protect\citeauthoryear{{Hayat}, {Stein}, {Harrington}, {Luki{\'c}}  \& {Mustafa}}{{Hayat} et~al.}{2021}]{Hayat2021}
{Hayat} M.~A.,  {Stein} G.,  {Harrington} P.,  {Luki{\'c}} Z.,   {Mustafa} M.,  2021, \mn@doi [\apjl] {10.3847/2041-8213/abf2c7}, \href {https://ui-adsabs-harvard-edu.insu.bib.cnrs.fr/abs/2021ApJ...911L..33H} {911, L33}

\bibitem[\protect\citeauthoryear{{Henghes}, {Thiyagalingam}, {Pettitt}, {Hey}  \& {Lahav}}{{Henghes} et~al.}{2022}]{Henghes2022}
{Henghes} B.,  {Thiyagalingam} J.,  {Pettitt} C.,  {Hey} T.,   {Lahav} O.,  2022, \mn@doi [\mnras] {10.1093/mnras/stac480}, \href {https://ui-adsabs-harvard-edu.insu.bib.cnrs.fr/abs/2022MNRAS.512.1696H} {512, 1696}

\bibitem[\protect\citeauthoryear{Hersbach}{Hersbach}{2000}]{Hersbach}
Hersbach H.,  2000, \mn@doi [Weather and Forecasting] {10.1175/1520-0434(2000)015<0559:DOTCRP>2.0.CO;2}, 15, 559

\bibitem[\protect\citeauthoryear{{Hoyle}}{{Hoyle}}{2016}]{Hoyle2016}
{Hoyle} B.,  2016, \mn@doi [Astronomy and Computing] {10.1016/j.ascom.2016.03.006}, \href {http://adsabs.harvard.edu/abs/2016A%26C....16...34H} {16, 34}

\bibitem[\protect\citeauthoryear{{Ilbert} et~al.,}{{Ilbert} et~al.}{2006}]{Ilbert2006}
{Ilbert} O.,  et~al., 2006, \mn@doi [\aap] {10.1051/0004-6361:20065138}, \href {http://adsabs.harvard.edu/abs/2006A%26A...457..841I} {457, 841}

\bibitem[\protect\citeauthoryear{{Ivezi{\'c}} et~al.,}{{Ivezi{\'c}} et~al.}{2019}]{Ivezic2019}
{Ivezi{\'c}} {\v{Z}}.,  et~al., 2019, \mn@doi [\apj] {10.3847/1538-4357/ab042c}, \href {https://ui.adsabs.harvard.edu/abs/2019ApJ...873..111I} {873, 111}

\bibitem[\protect\citeauthoryear{{Jones}, {Do}, {Boscoe}, {Singal}, {Wan}  \& {Nguyen}}{{Jones} et~al.}{2023}]{Jones2023}
{Jones} E.,  {Do} T.,  {Boscoe} B.,  {Singal} J.,  {Wan} Y.,   {Nguyen} Z.,  2023, \mn@doi [arXiv e-prints] {10.48550/arXiv.2306.13179}, \href {https://ui-adsabs-harvard-edu.insu.bib.cnrs.fr/abs/2023arXiv230613179J} {p. arXiv:2306.13179}

\bibitem[\protect\citeauthoryear{Kingma \& Ba}{Kingma \& Ba}{2015}]{KingmaBa2015}
Kingma D.~P.,  Ba J.,  2015, in Bengio Y.,  LeCun Y.,  eds, 3rd International Conference on Learning Representations, {ICLR} 2015, San Diego, CA, USA, May 7-9, 2015, Conference Track Proceedings. \url {http://arxiv.org/abs/1412.6980}

\bibitem[\protect\citeauthoryear{{Kraljic} et~al.,}{{Kraljic} et~al.}{2018}]{Kraljic2018}
{Kraljic} K.,  et~al., 2018, \mn@doi [\mnras] {10.1093/mnras/stx2638}, \href {https://ui-adsabs-harvard-edu.insu.bib.cnrs.fr/abs/2018MNRAS.474..547K} {474, 547}

\bibitem[\protect\citeauthoryear{{Laigle} et~al.,}{{Laigle} et~al.}{2018}]{Laigle2018}
{Laigle} C.,  et~al., 2018, \mn@doi [\mnras] {10.1093/mnras/stx3055}, \href {https://ui-adsabs-harvard-edu.insu.bib.cnrs.fr/abs/2018MNRAS.474.5437L} {474, 5437}

\bibitem[\protect\citeauthoryear{{Laureijs} et~al.,}{{Laureijs} et~al.}{2011}]{Laureijs2011}
{Laureijs} R.,  et~al., 2011, arXiv e-prints, \href {https://ui.adsabs.harvard.edu/abs/2011arXiv1110.3193L} {p. arXiv:1110.3193}

\bibitem[\protect\citeauthoryear{{Le F{\`e}vre} et~al.,}{{Le F{\`e}vre} et~al.}{2013}]{LeFevre2013}
{Le F{\`e}vre} O.,  et~al., 2013, \mn@doi [\aap] {10.1051/0004-6361/201322179}, \href {https://ui.adsabs.harvard.edu/abs/2013A&A...559A..14L} {559, A14}

\bibitem[\protect\citeauthoryear{LeCun, Bottou, Bengio  \& Haffner}{LeCun et~al.}{1998}]{Lecun1998}
LeCun Y.,  Bottou L.,  Bengio Y.,   Haffner P.,  1998, Proceedings of the IEEE, 86, 2278

\bibitem[\protect\citeauthoryear{{Lilly} et~al.,}{{Lilly} et~al.}{2007}]{Lilly2007}
{Lilly} S.~J.,  et~al., 2007, \mn@doi [\apjs] {10.1086/516589}, \href {https://ui.adsabs.harvard.edu/abs/2007ApJS..172...70L} {172, 70}

\bibitem[\protect\citeauthoryear{{Liske} et~al.,}{{Liske} et~al.}{2015}]{Liske2015}
{Liske} J.,  et~al., 2015, \mn@doi [\mnras] {10.1093/mnras/stv1436}, \href {http://adsabs.harvard.edu/abs/2015MNRAS.452.2087L} {452, 2087}

\bibitem[\protect\citeauthoryear{{Malavasi} et~al.,}{{Malavasi} et~al.}{2017}]{Malavasi2017}
{Malavasi} N.,  et~al., 2017, \mn@doi [\mnras] {10.1093/mnras/stw2864}, \href {https://ui-adsabs-harvard-edu.insu.bib.cnrs.fr/abs/2017MNRAS.465.3817M} {465, 3817}

\bibitem[\protect\citeauthoryear{{McLure} et~al.,}{{McLure} et~al.}{2013}]{McLure2013}
{McLure} R.~J.,  et~al., 2013, \mn@doi [\mnras] {10.1093/mnras/sts092}, \href {https://ui.adsabs.harvard.edu/abs/2013MNRAS.428.1088M} {428, 1088}

\bibitem[\protect\citeauthoryear{{M{\'e}nard}, {Scranton}, {Schmidt}, {Morrison}, {Jeong}, {Budavari}  \& {Rahman}}{{M{\'e}nard} et~al.}{2013}]{Menard2013}
{M{\'e}nard} B.,  {Scranton} R.,  {Schmidt} S.,  {Morrison} C.,  {Jeong} D.,  {Budavari} T.,   {Rahman} M.,  2013, arXiv e-prints, \href {https://ui-adsabs-harvard-edu.insu.bib.cnrs.fr/abs/2013arXiv1303.4722M} {p. arXiv:1303.4722}

\bibitem[\protect\citeauthoryear{Nair \& Hinton}{Nair \& Hinton}{2010}]{NairHinton2010}
Nair V.,  Hinton G.,  2010, Proceedings of the 27th international conference on machine learning (ICML-10)

\bibitem[\protect\citeauthoryear{{Newman} et~al.,}{{Newman} et~al.}{2013}]{Newman2013}
{Newman} J.~A.,  et~al., 2013, \mn@doi [\apjs] {10.1088/0067-0049/208/1/5}, \href {https://ui.adsabs.harvard.edu/abs/2013ApJS..208....5N} {208, 5}

\bibitem[\protect\citeauthoryear{{Pasquet}, {Bertin}, {Treyer}, {Arnouts}  \& {Fouchez}}{{Pasquet} et~al.}{2019}]{P19}
{Pasquet} J.,  {Bertin} E.,  {Treyer} M.,  {Arnouts} S.,   {Fouchez} D.,  2019, \mn@doi [\aap] {10.1051/0004-6361/201833617}, \href {https://ui-adsabs-harvard-edu.insu.bib.cnrs.fr/abs/2019A&A...621A..26P} {621, A26}

\bibitem[\protect\citeauthoryear{{Patej} \& {Eisenstein}}{{Patej} \& {Eisenstein}}{2018}]{Patej2018}
{Patej} A.,  {Eisenstein} D.~J.,  2018, \mn@doi [\mnras] {10.1093/mnras/sty870}, \href {https://ui-adsabs-harvard-edu.insu.bib.cnrs.fr/abs/2018MNRAS.477.5090P} {477, 5090}

\bibitem[\protect\citeauthoryear{{Polsterer}, {D'Isanto}  \& {Gieseke}}{{Polsterer} et~al.}{2016}]{Polsterer2016}
{Polsterer} K.~L.,  {D'Isanto} A.,   {Gieseke} F.,  2016, preprint, \href {http://adsabs.harvard.edu/abs/2016arXiv160808016P} {} (\mn@eprint {arXiv} {1608.08016})

\bibitem[\protect\citeauthoryear{{Rahman}, {M{\'e}nard}, {Scranton}, {Schmidt}  \& {Morrison}}{{Rahman} et~al.}{2015}]{Rahman2015}
{Rahman} M.,  {M{\'e}nard} B.,  {Scranton} R.,  {Schmidt} S.~J.,   {Morrison} C.~B.,  2015, \mn@doi [\mnras] {10.1093/mnras/stu2636}, \href {https://ui-adsabs-harvard-edu.insu.bib.cnrs.fr/abs/2015MNRAS.447.3500R} {447, 3500}

\bibitem[\protect\citeauthoryear{Richard \& Lippmann}{Richard \& Lippmann}{1991}]{Richard1991}
Richard M.,  Lippmann R.,  1991, Neural Computation, 3, 461

\bibitem[\protect\citeauthoryear{Rojas}{Rojas}{1996}]{Rojas1996}
Rojas R.,  1996, \mn@doi [Neural Computation] {10.1162/neco.1996.8.1.41}, 8, 41

\bibitem[\protect\citeauthoryear{{Rykoff} et~al.,}{{Rykoff} et~al.}{2014}]{Rykoff2014}
{Rykoff} E.~S.,  et~al., 2014, \mn@doi [\apj] {10.1088/0004-637X/785/2/104}, \href {https://ui-adsabs-harvard-edu.insu.bib.cnrs.fr/abs/2014ApJ...785..104R} {785, 104}

\bibitem[\protect\citeauthoryear{{Sadeh}}{{Sadeh}}{2014}]{Sadeh2014}
{Sadeh} I.,  2014, in {Heavens} A.,  {Starck} J.-L.,   {Krone-Martins} A.,  eds,  IAU Symposium Vol. 306, Statistical Challenges in 21st Century Cosmology. pp 316--318, \mn@doi{10.1017/S1743921314010849}

\bibitem[\protect\citeauthoryear{{Sadeh}, {Abdalla}  \& {Lahav}}{{Sadeh} et~al.}{2016}]{Sadeh2016}
{Sadeh} I.,  {Abdalla} F.~B.,   {Lahav} O.,  2016, \mn@doi [\pasp] {10.1088/1538-3873/128/968/104502}, \href {https://ui-adsabs-harvard-edu.insu.bib.cnrs.fr/abs/2016PASP..128j4502S} {128, 104502}

\bibitem[\protect\citeauthoryear{{Sawicki} et~al.,}{{Sawicki} et~al.}{2019}]{Sawicki2019}
{Sawicki} M.,  et~al., 2019, \mn@doi [\mnras] {10.1093/mnras/stz2522}, \href {https://ui-adsabs-harvard-edu.insu.bib.cnrs.fr/abs/2019MNRAS.489.5202S} {489, 5202}

\bibitem[\protect\citeauthoryear{{Schlegel}, {Finkbeiner}  \& {Davis}}{{Schlegel} et~al.}{1998}]{Schlegel1998}
{Schlegel} D.~J.,  {Finkbeiner} D.~P.,   {Davis} M.,  1998, \mn@doi [\apj] {10.1086/305772}, \href {http://adsabs.harvard.edu/abs/1998ApJ...500..525S} {500, 525}

\bibitem[\protect\citeauthoryear{{Schuldt}, {Suyu}, {Ca{\~n}ameras}, {Taubenberger}, {Meinhardt}, {Leal-Taix{\'e}}  \& {Hsieh}}{{Schuldt} et~al.}{2021}]{Schuldt2021}
{Schuldt} S.,  {Suyu} S.~H.,  {Ca{\~n}ameras} R.,  {Taubenberger} S.,  {Meinhardt} T.,  {Leal-Taix{\'e}} L.,   {Hsieh} B.~C.,  2021, \mn@doi [\aap] {10.1051/0004-6361/202039945}, \href {https://ui-adsabs-harvard-edu.insu.bib.cnrs.fr/abs/2021A&A...651A..55S} {651, A55}

\bibitem[\protect\citeauthoryear{{Scodeggio} et~al.,}{{Scodeggio} et~al.}{2018}]{Scodeggio2018}
{Scodeggio} M.,  et~al., 2018, \mn@doi [\aap] {10.1051/0004-6361/201630114}, \href {https://ui.adsabs.harvard.edu/abs/2018A&A...609A..84S} {609, A84}

\bibitem[\protect\citeauthoryear{Solla, Levin  \& Fleisher}{Solla et~al.}{1988}]{Solla1988}
Solla S.,  Levin E.,   Fleisher M.,  1988, Complex Syst., 2, 625

\bibitem[\protect\citeauthoryear{{Tagliaferri}, {Longo}, {Andreon}, {Capozziello}, {Donalek}  \& {Giordano}}{{Tagliaferri} et~al.}{2003}]{Tagliaferri2003}
{Tagliaferri} R.,  {Longo} G.,  {Andreon} S.,  {Capozziello} S.,  {Donalek} C.,   {Giordano} G.,  2003, in , Vol.~2859, Lecture Notes in Computer Science.
pp 226--234, \mn@doi{10.1007/978-3-540-45216-4_26}

\bibitem[\protect\citeauthoryear{{Treyer} et~al.,}{{Treyer} et~al.}{2018}]{Treyer2018}
{Treyer} M.,  et~al., 2018, \mn@doi [\mnras] {10.1093/mnras/sty769}, \href {https://ui-adsabs-harvard-edu.insu.bib.cnrs.fr/abs/2018MNRAS.477.2684T} {477, 2684}

\bibitem[\protect\citeauthoryear{{Uhlemann}, {Pichon}, {Codis}, {L'Huillier}, {Kim}, {Bernardeau}, {Park}  \& {Prunet}}{{Uhlemann} et~al.}{2018}]{Ulhemann2018}
{Uhlemann} C.,  {Pichon} C.,  {Codis} S.,  {L'Huillier} B.,  {Kim} J.,  {Bernardeau} F.,  {Park} C.,   {Prunet} S.,  2018, \mn@doi [\mnras] {10.1093/mnras/sty664}, \href {https://ui-adsabs-harvard-edu.insu.bib.cnrs.fr/abs/2018MNRAS.477.2772U} {477, 2772}

\bibitem[\protect\citeauthoryear{{Wittman}, {Bhaskar}  \& {Tobin}}{{Wittman} et~al.}{2016}]{Wittman2016}
{Wittman} D.,  {Bhaskar} R.,   {Tobin} R.,  2016, \mn@doi [\mnras] {10.1093/mnras/stw261}, \href {https://ui-adsabs-harvard-edu.insu.bib.cnrs.fr/abs/2016MNRAS.457.4005W} {457, 4005}

\bibitem[\protect\citeauthoryear{{York} et~al.,}{{York} et~al.}{2000}]{York2000}
{York} D.~G.,  et~al., 2000, \mn@doi [\aj] {10.1086/301513}, \href {http://adsabs.harvard.edu/abs/2000AJ....120.1579Y} {120, 1579}

\bibitem[\protect\citeauthoryear{{Zamo} \& {Naveau}}{{Zamo} \& {Naveau}}{2018}]{Zamo2018}
{Zamo} M.,  {Naveau} P.,  2018, \mn@doi [Mathematical Geosciences] {10.1007/s11004-017-9709-7}, 50, 209

\bibitem[\protect\citeauthoryear{{de Jong}, {Verdoes Kleijn}, {Kuijken}  \& {Valentijn}}{{de Jong} et~al.}{2013}]{deJong2013}
{de Jong} J. T.~A.,  {Verdoes Kleijn} G.~A.,  {Kuijken} K.~H.,   {Valentijn} E.~A.,  2013, \mn@doi [Experimental Astronomy] {10.1007/s10686-012-9306-1}, \href {https://ui.adsabs.harvard.edu/abs/2013ExA....35...25D} {35, 25}

\makeatother
\end{thebibliography}

\appendix

\section{Data sky distribution} 
\label{sec:skymaps}
 
\begin{figure}
\centering
\includegraphics[width=8.5cm]{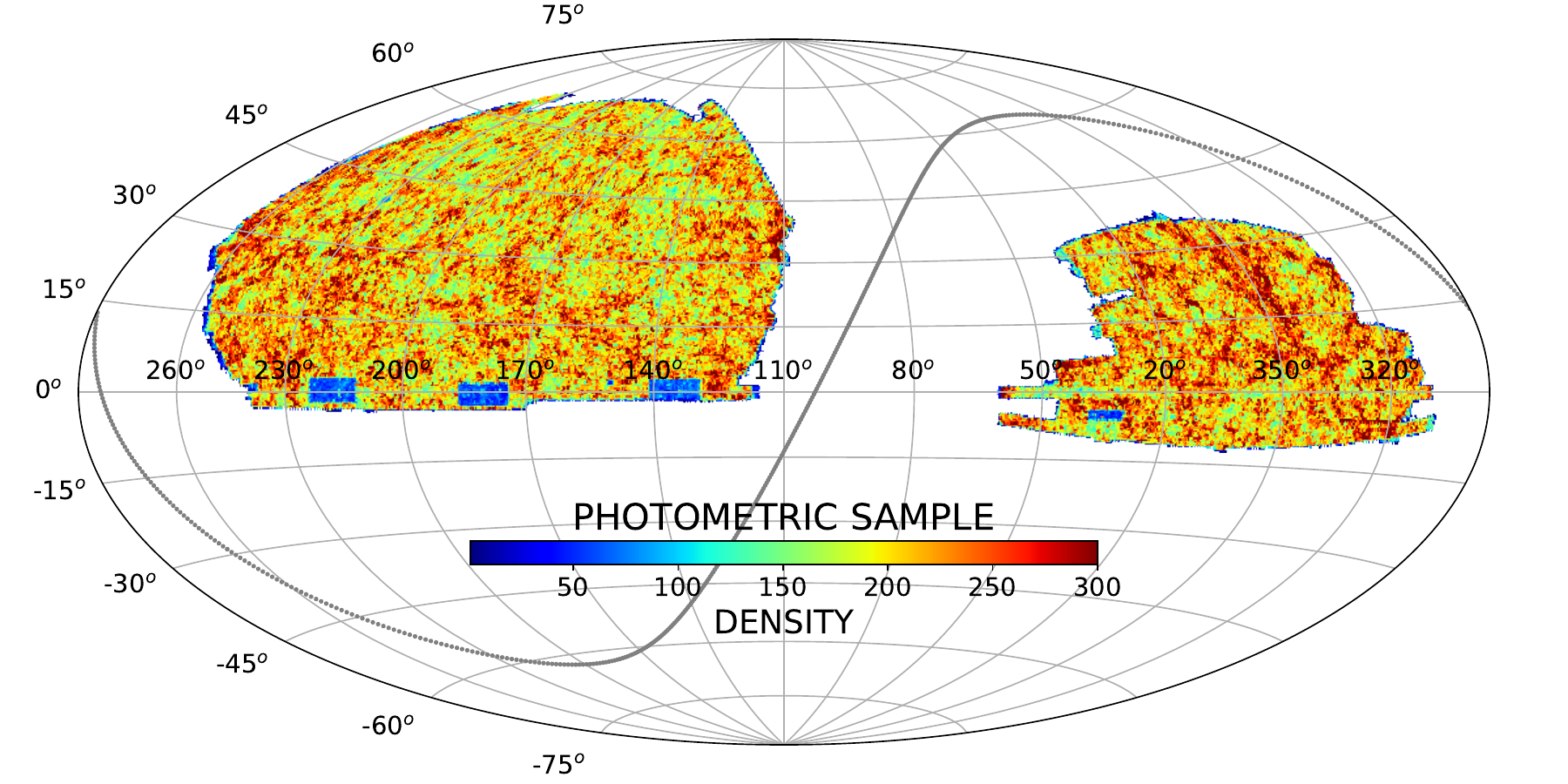}
\includegraphics[width=8.5cm]{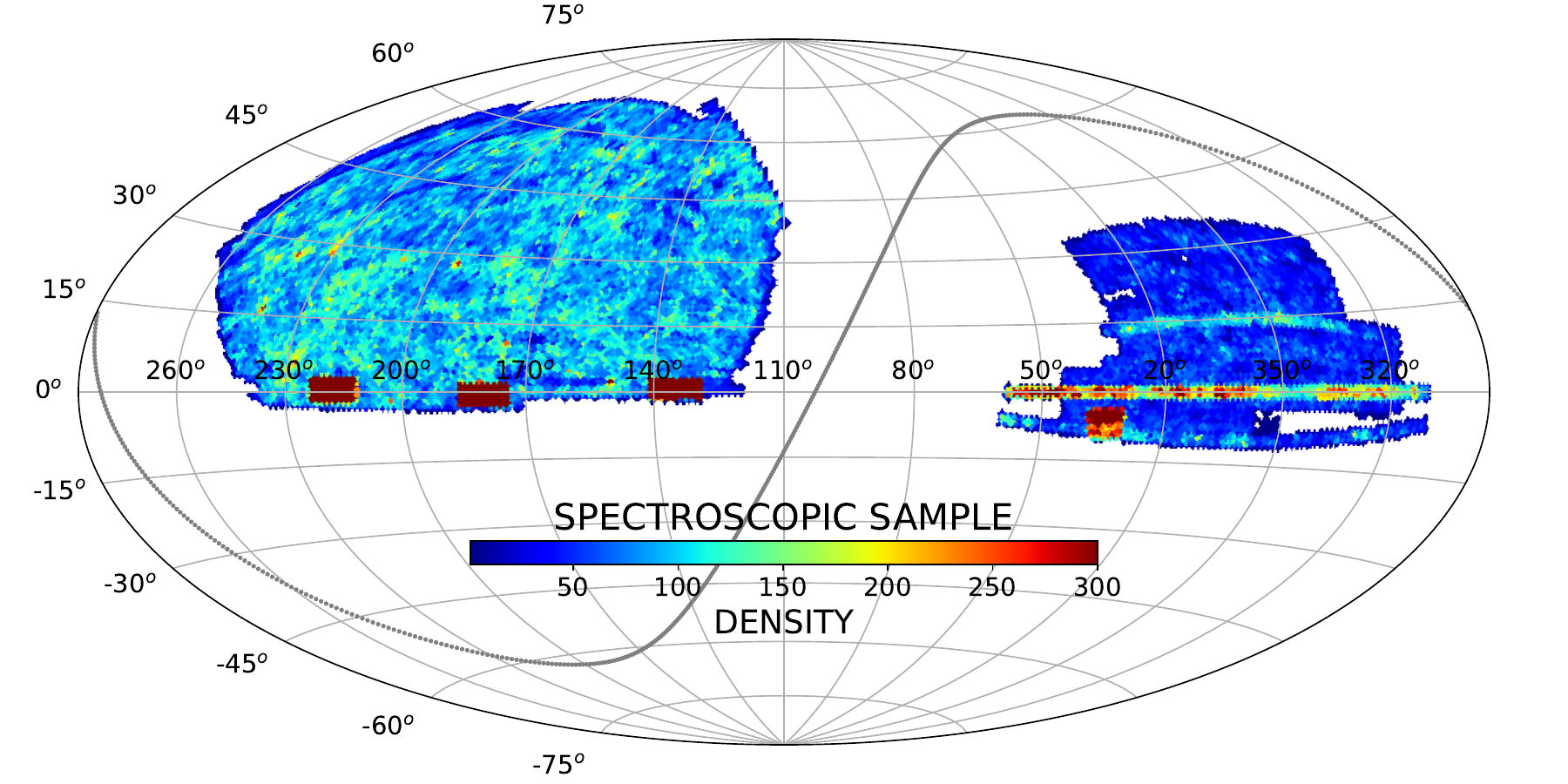}
\caption{
{\bf Top:} Mollweide projection of the SDSS photometric sample at $r\leq20$ without spectroscopy ($\sim$13.8M sources). The color code indicates the source density in HEALPix cells with nside=140 (0.18deg$^2$/pixel). The sky coverage is $\sim$11,529 deg$^2$.
{\bf Bottom:} Mollweide projection of the spectroscopic sample at $r\leq 20$ ($\sim$1.5M galaxies). The color code indicates the source density in HEALPix cells with nside=64 (0.84deg$^2$/pixel). The sky coverage is $\sim$11,029 deg$^2$. 
%The sky coverage is $\sim$9,950 deg$^2$. 
The 4 high density regions in red are the GAMA fields. 
%nside=128 (0.21deg$^2$/pixel). The sky coverage is $\sim$ 11,545deg$^2$.
}
\label{fig:maps}
\end{figure}

 Figure \ref{fig:maps} shows the sky distribution of the photometric and spectroscopic samples down to the dereddened petrosian magnitude of $r=20$ in the two main regions of the SDSS.

\section{Galaxy type classification}
\label{sec:galtypes}

\begin{figure}
\includegraphics[width=8.5cm]{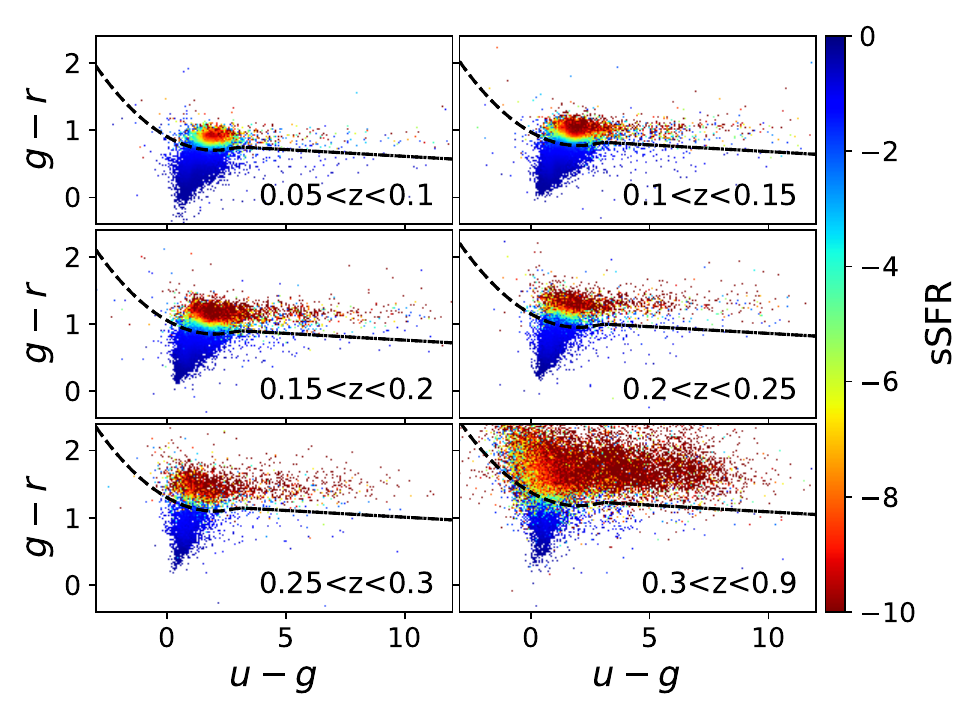}
\caption{The $(u-g)/(g-r)$) distribution of the SDSS spectroscopic sample at $r<20$, colored with the mean specific star-formation rate ($\rm M_\odot/yr$) per pixel. The narrow demarcations between the blue and the red zones highlighting the bimodal distribution of the sSSFR at a given redsfhit, are used to model boundaries between blue and red galaxies as a function of redshift.}
\label{fig:colcol_ssfr}
\end{figure}

To classify galaxies as either blue/star-forming or red/passive, we use the specific star-formation rate (sSFR, the star-formation rate per unit stellar mass) derived for the spectroscopic sample in the SDSS DR12 by \citet{Brinchmann04}. Figure \ref{fig:colcol_ssfr} shows this sSFR (mean value per pixel) in the observed $(u-g)$/$(g-r)$ plane in several bins of redshifts. The narrow greenish demarcations between the blue and the red zones highlight the bimodal distribution of the sSFR at a given redshift that prompts the distinction between blue and red galaxies. The black dashed lines running through these demarcations are modeled as followed: 

\begin{equation} \label{eq:ssfr_col}
\begin{split}
 y_{lim}[(u-g)<3] &=y_b+0.05((u-g)-2)^2 \\
 y_{lim}[(u-g)>3] &=y_b+0.05-0.02((u-g)-3)
\end{split}
\end{equation}
where: 
\begin{equation} \label{eq:ssfr_z}
\begin{split}
    y_b[z<0.32] &=0.65+1.59z+1.19z^2 \\
    y_b[z>0.32] &=1.28  
\end{split}
\end{equation}

We compare this classification to that derived for GAMA at $z<0.3$ by SED fitting the rich multi-band (UV to IR) photometry \citep{Treyer2018, Kraljic2018}. The limit between blue and passive galaxies was set at $\rm sSFR\approx -10.5 \rm M_\odot/yr$. The red completeness, defined as the number of galaxies classified as passive according to both criteria over the number of passive galaxies according to the GAMA sSFR criterion, is 87\%. The red purity, defined as the number of galaxies classified as passive according to both criteria over the number of passive galaxies according to the $(u-g)$/$(g-r)$/redshift criterion, is 75\%. Likewise, the blue completeness and purity are 87\% and 93\% respectively. The present recipe tends to overestimate red galaxies compared to the GAMA sSFR limit but the two classifications are in reasonably good agreement. 
It is however inadequate for the LRGs in the highest redshift bin, many of which would be classified as blue. BOSS galaxies are considered red regardless. Rough as it is, we will use this prescription to uncover statistical differences in CNN performance, if any, between the two types. Optimizing it is beyond the scope of this paper.

\section{The "dirty" photometric sample}
\label{sec:uncleanphoto}

\begin{figure}
\includegraphics[width=4.2cm]{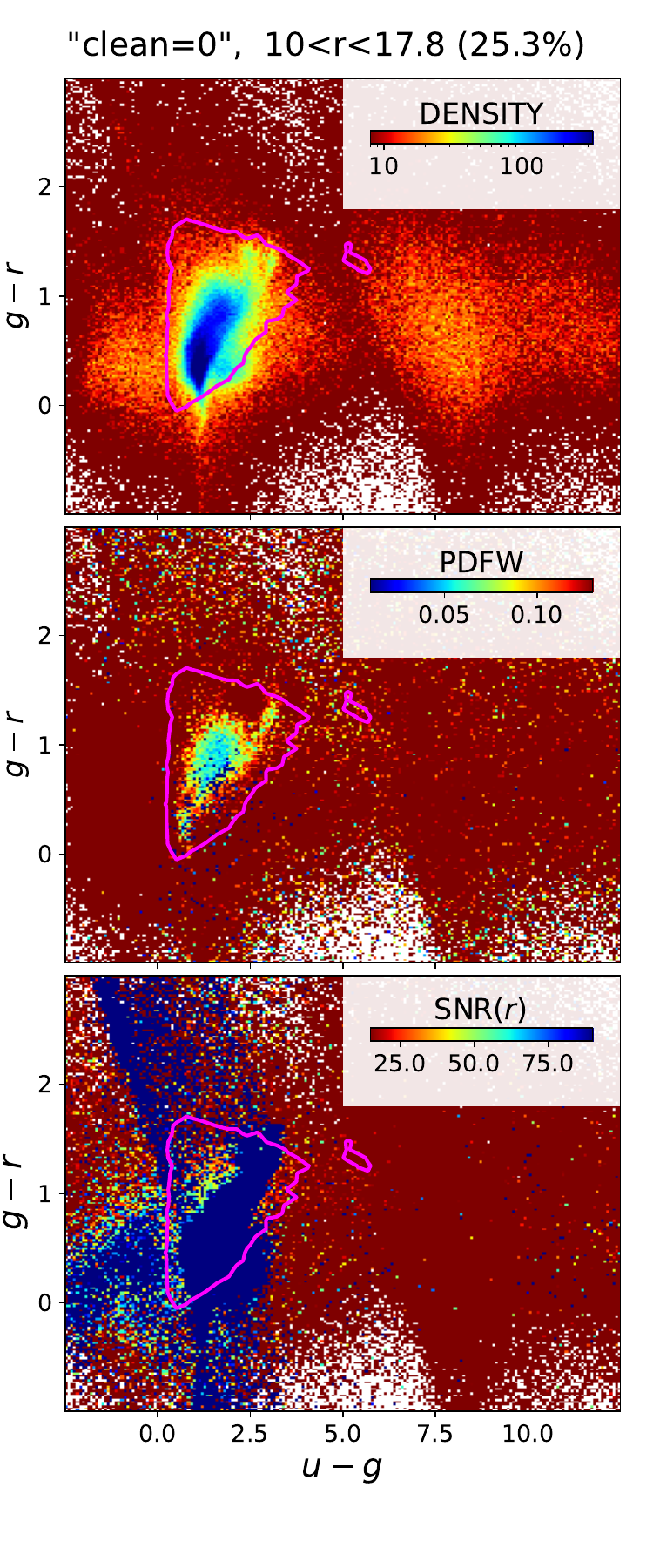}
\includegraphics[width=4.2cm]{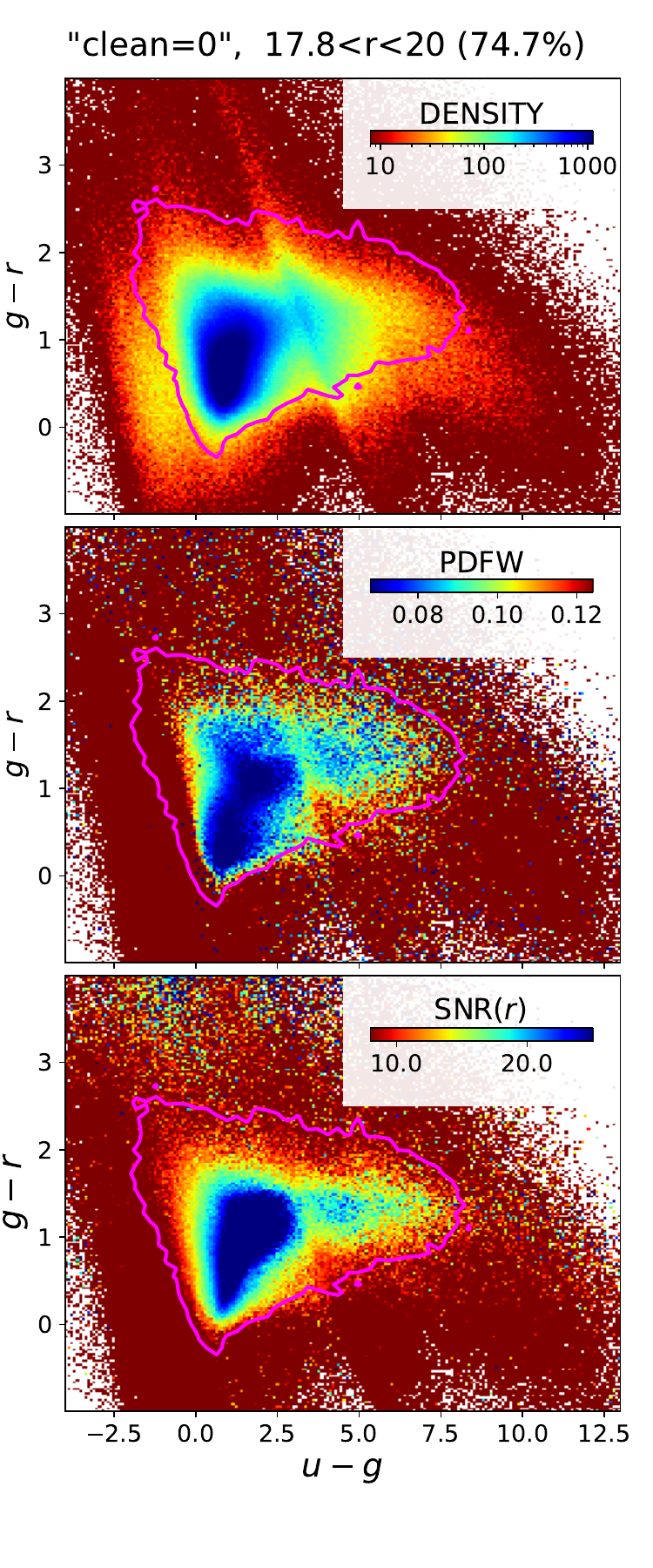}
\caption{The $(u-g)/(g-r)$ color distribution of the "{\bf clean}=0" photometric sample at $r<17.8$ (left panels) and $r>17.8$ (right panels), color-coded by the density (top), the mean PDF width (middle) and the $r-$band signal-to-noise ratio (bottom). The quality of the data is significantly inferior to the "{\bf clean}=1" sample shown in Fig. \ref{fig:colcol_inference_clean}.}
\label{fig:colcol_inference_notclean}
\end{figure}

Figures \ref{fig:colcol_inference_notclean} shows the $(u-g)/(g-r)$ color distribution of the "{\bf clean}=0" photometric sources at $r<17.8$ and $17.8<r<20$ (left and right panels respectively). The distributions are color-coded by the mean PDF width in the middle panels and by the mean $r-$band signal-to-noise ratio (SNR) in the bottom panels. We use the same color code as in Fig. \ref{fig:colcol_inference_clean} to emphasize the differences between the two samples. The SNR and PDF quality are very degraded nearly everywhere compared to the "clean" sample, included within the training contours. The striking star sequence at $r<17.8$ in Fig. \ref{fig:colcol_inference_clean} is drowned in sources with equally poor PDFs.

\begin{figure}
\includegraphics[width=8.8cm]{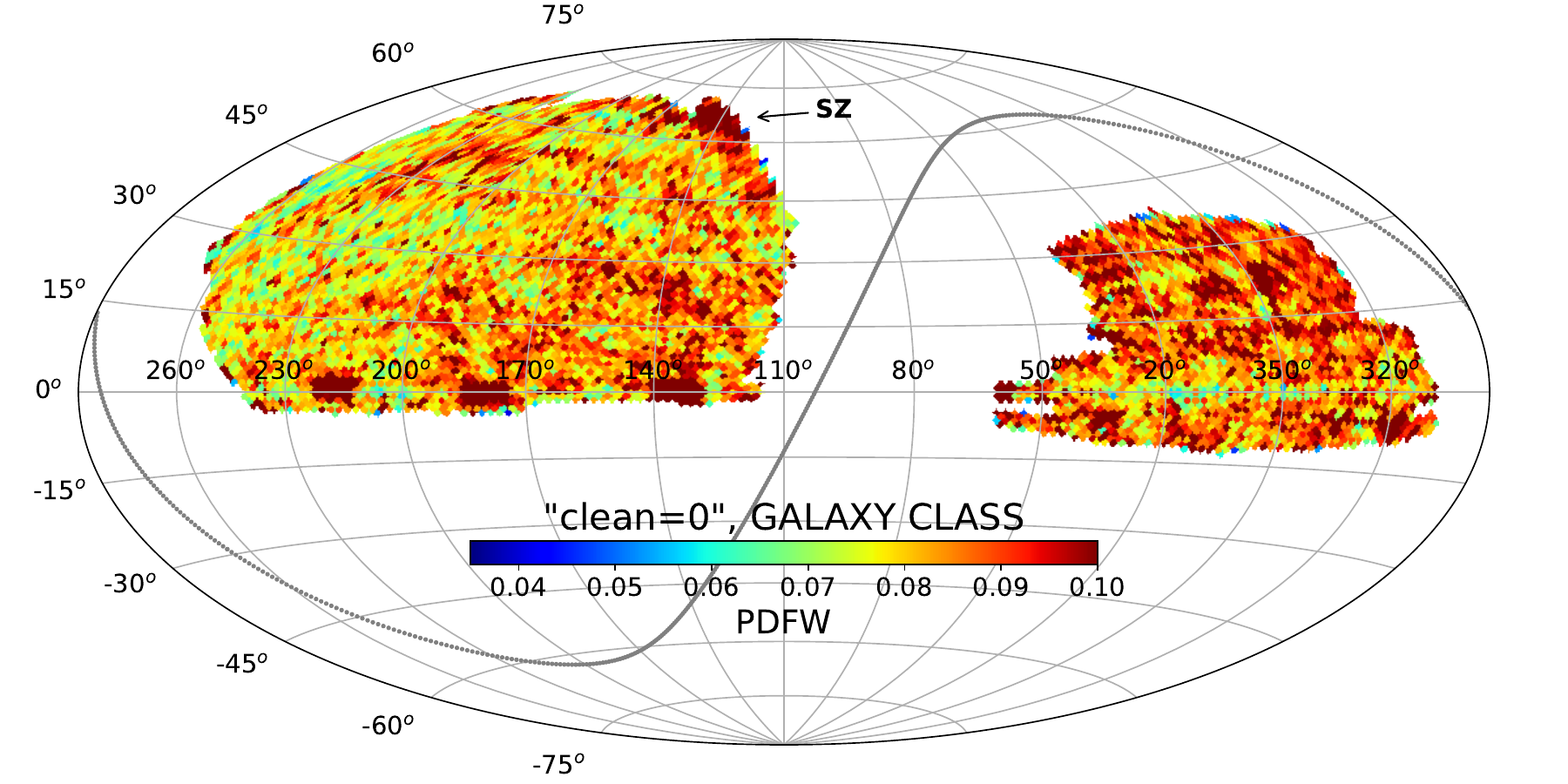}
\caption{PDF width colored map of sources classified as galaxies in the "{\bf clean}=0" photometric sample, to be compared with Fig. \ref{fig:map_clean_pdfw}. The difference between the two samples is all the more striking that the clean data are deeper. The larger than average PDF widths in the GAMA regions are due to their deeper average magnitude. The higher image quality in Stripe 82 (greenish equatorial stripe in the South) remains visible.}
\label{fig:map_notclean_pdfw}
\end{figure}

Figure \ref{fig:map_notclean_pdfw} shows the PDF width colored sky map of the sources classified as galaxies, to be compared with Fig. \ref{fig:map_clean_pdfw}. The difference between the two samples is all the more striking that the present data are brighter (<$r$>=18.81) than the "clean" data (<$r$>=19.24). The "suspect zone" identified by P19 is much more prominent in this data set, which also contains many other similarly degraded areas, especially in the southern region. The enhanced image quality in Stripe 82 (greenish equatorial stripe in the South) remains visible.

\begin{figure}
\includegraphics[width=4.2cm]{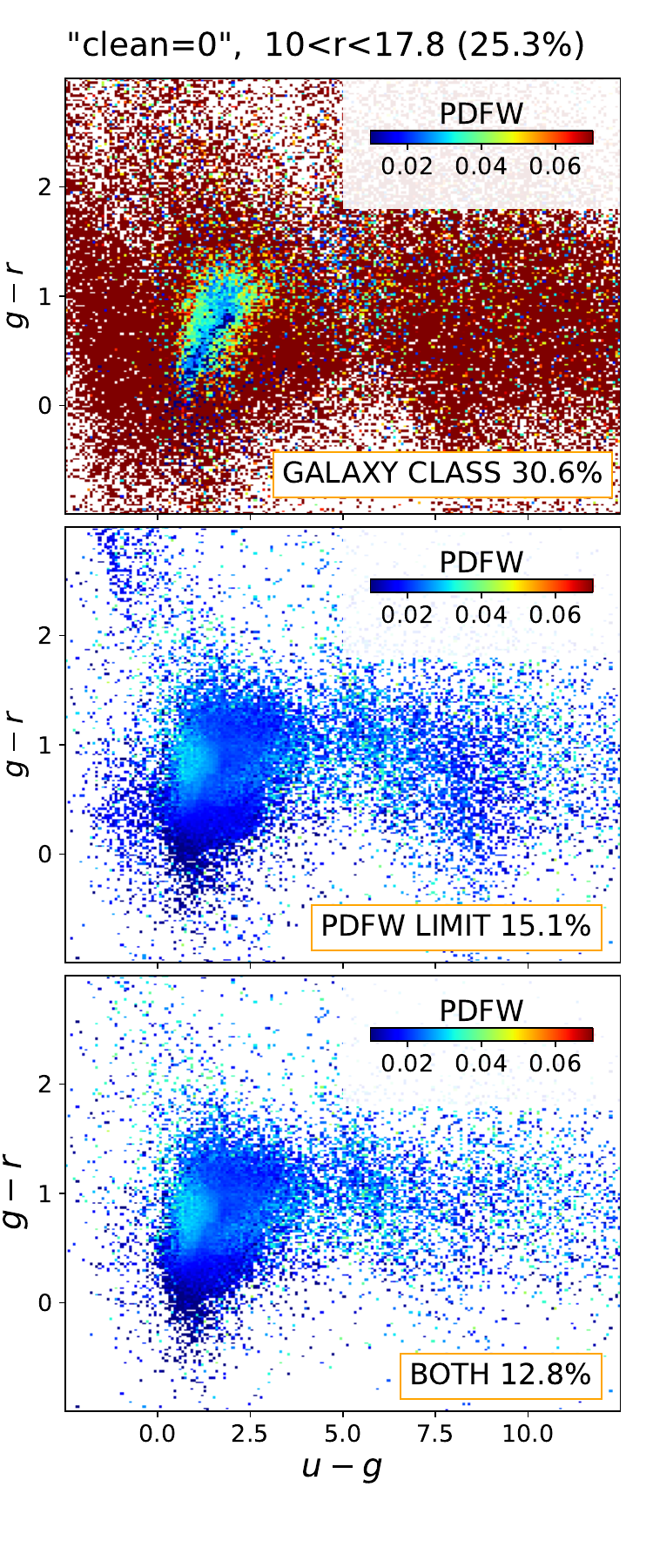}
\includegraphics[width=4.2cm]{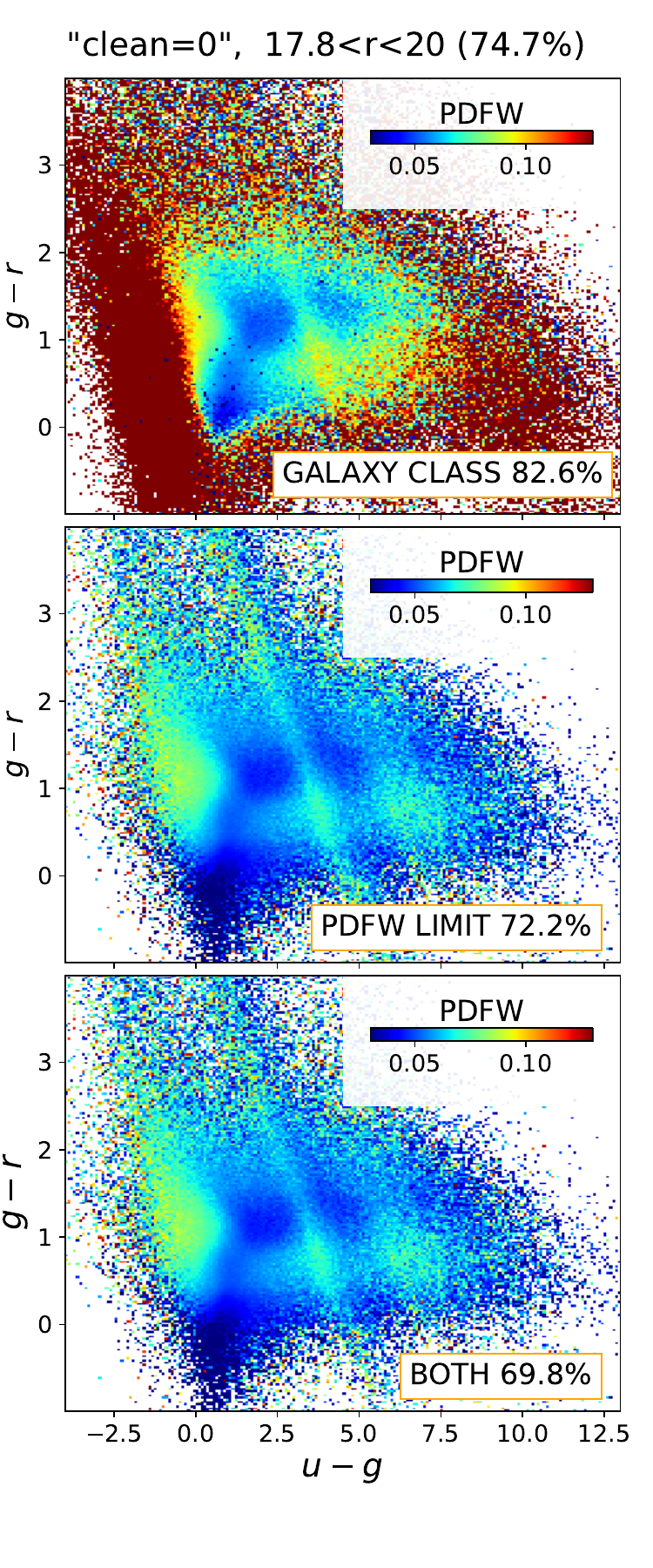}
\caption{The $(u-g)/(g-r)$ color distribution of the "{\bf clean}=0" sample color-coded by the mean PDF width at $r<17.8$ in the left panels and $r>17.8$ in the right panels. The top panels restrict the sample to sources classified as galaxies, the middle panels apply the PDF width threshold, and the bottom panels use both constraints.}
\label{fig:colcol_inference_notclean_cleaned}
\end{figure}

Figure \ref{fig:colcol_inference_notclean_cleaned} shows the $(u-g)/(g-r)$ color distributions of the sample color-coded by the mean PDF width and split into the bright and faint magnitude intervals, following the three cleansing procedures: the top panels restrict the samples to sources classified as galaxies (30.6\% at $r<17.8$, 82.6\% at $r>17.8$), the middle panels to sources with PDF widths below the threshold (15.1\% at $r<17.8$, 72.2\% at $r>17.8$), and finally the bottom panels to classified galaxies with PDF widths below the threshold (12.8\% at $r<17.8$, 69.8\% at $r>17.8$). This final procedure rejects 44.5\% of the initial data (30.5\% classified as stars or QSOs, 41.8\% with PDF widths above the threshold). 

\section{Tomographer}
\label{sec:tomographer_apx}

We run Tomographer on the spectroscopic sample in order to test the level of accuracy of the output distributions.
%Many of the galaxies in our spectroscopic sample are likely to be used by Tomographer, therefore we expect this test to be a best case scenario, 
The results are shown as red dots in Fig. \ref{fig:tomo_spectro} for 12 narrow intervals of magnitude ($\Delta \rm mag=0.2$), with the spectroscopic redshift distributions as shaded histograms. These need to be normalized to match the Tomographer outputs. Given the many negative and unrealistically high data points in the high redshift tail, we choose to ignore everything at $z>1$ and to normalize the spectroscopic redshift distributions by the Tomographer counts at $z<1$, which allows for a much better, though far from perfect agreement in the redshift ranges of interest. This comparison gauges the accuracy we may expect for unknown distributions.

\begin{figure}
\includegraphics[width=8.7cm]{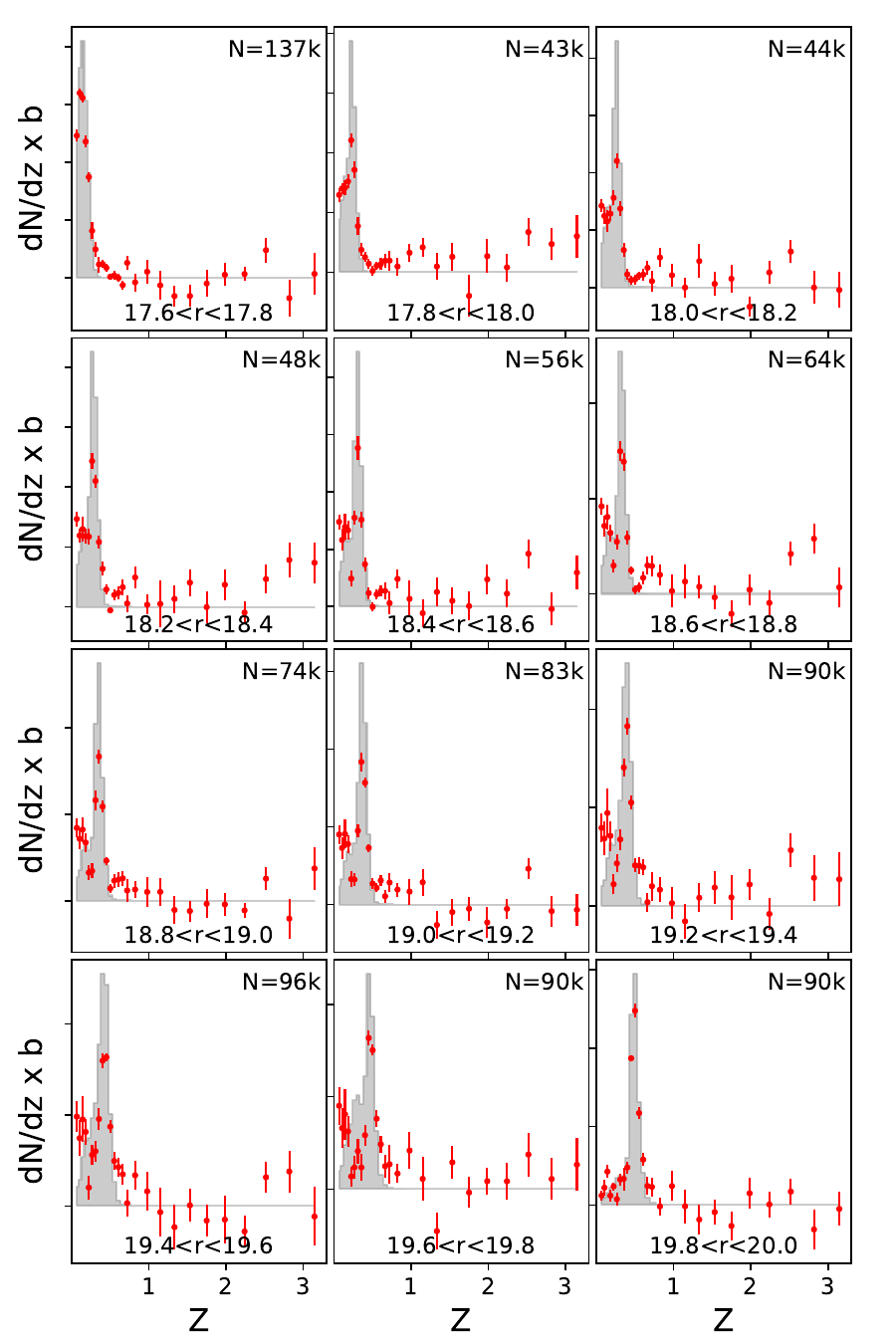}
\caption{The normalized redshift distributions of the spectroscopic sample in 12 bins of magnitude (gray shaded histograms) compared to the distributions derived from Tomographer (red dots). The normalization ignores the tail at $z>1$, which allows for a better agreement at $z<1$. 
}
\label{fig:tomo_spectro}
\end{figure}

\iffalse
\begin{figure*}
\includegraphics[width=\textwidth]{tomo_spectro.pdf}
\caption{The normalized redshift distributions of the spectroscopic sample in bins of magnitude (gray shaded histograms) compared to the Tomographer-derived distributions (red dots). The normalization ignores the tail at $z>1$, which allows for a better agreement at $z<1$. 
}
\label{fig:tomo_spectro}
\end{figure*}
\fi

\section{Alternative experiments}
\label{sec:alternatives}

The following experiments are conducted under the final conditions of this work: we average the outputs of 15 networks trained with a redshift bin width $\delta z=$0.009. 

\subsection{Classification versus regression}
\label{subsec:regression}

Table \ref{table:binsreg} shows the CNN performance on the test sample for different training strategies: classification + regression with a RMSE loss function, classification + regression with a MAE loss function, classification alone without regression, regressions alone with either a RMSE or a MAE loss function. The classification alone provides better statistics than both regressions, especially in terms of $\sigma_{\rm MAD}$. Adding a regression to the classification has a minor positive impact on the classification but a significant one on the $\sigma_{\rm MAD}$ of the regression. The preferred strategy, used in this work, is highlighted in bold.

\begin{table*}
\caption{The CNN performance on the test sample (Section \ref{subsec:s20go}) for different training strategies: classification + regression with RMSE loss (used in the present work), classification + regression with MAE loss, classification alone without regression, regressions alone with either an RMSE or MAE loss. The best statistics are highlighted in bold.
} 
\label{table:binsreg}
\centering
\begin{tabular}{ c|ccc|ccc|cc|c|c|}
\hline
 & \multicolumn{3}{|c|}{PDF+REG(RMSE)} & \multicolumn{3}{|c|}{PDF+REG(MAE)} & \multicolumn{2}{|c|}{PDF} & {REG(RMSE)} & {REG(MAE)} \\
 \hline
& $z_{mean}$ & $z_{med}$ & $z_{reg}$ &  $z_{mean}$ & $z_{med}$ & $z_{reg}$ & $z_{mean}$ & $z_{med}$ & $z_{reg}$ & $z_{reg}$\\
\hline
{$10^{5} \sigma_{\rm MAD}$} & 1466 & \bf{1421} & 1481 & 1464 & 1431 & 1461 & 1470 &  1444 & 1586 & 1517 \\  
{$10^{5}$ <$\Delta z$>} & 205 & \bf{153} & 228  &   209& 154 & 196  &   218 &    159 & 240  & 189 \\
{$\eta(>0.05)(\%)$} & 4.1 & \bf{3.99} & 4.22 & 4.05& 4.06& 4.26&  4.18 &  4.16&  4.2 & 4.15 \\
\hline \\
\end{tabular}
\end{table*}

\subsection{Randomness strategy}
\label{subsec:randomness}

Instead of training several samples, with the goal of feeding
the CNN the largest variety of galaxies from the spectroscopic sample, we test the alternative strategy of training a single sample several times. This leaves twice as many galaxies for testing.  We pick one of the 15 trained samples at random and retrain it 14 times.

\begin{figure}
\includegraphics[width=8.5cm]{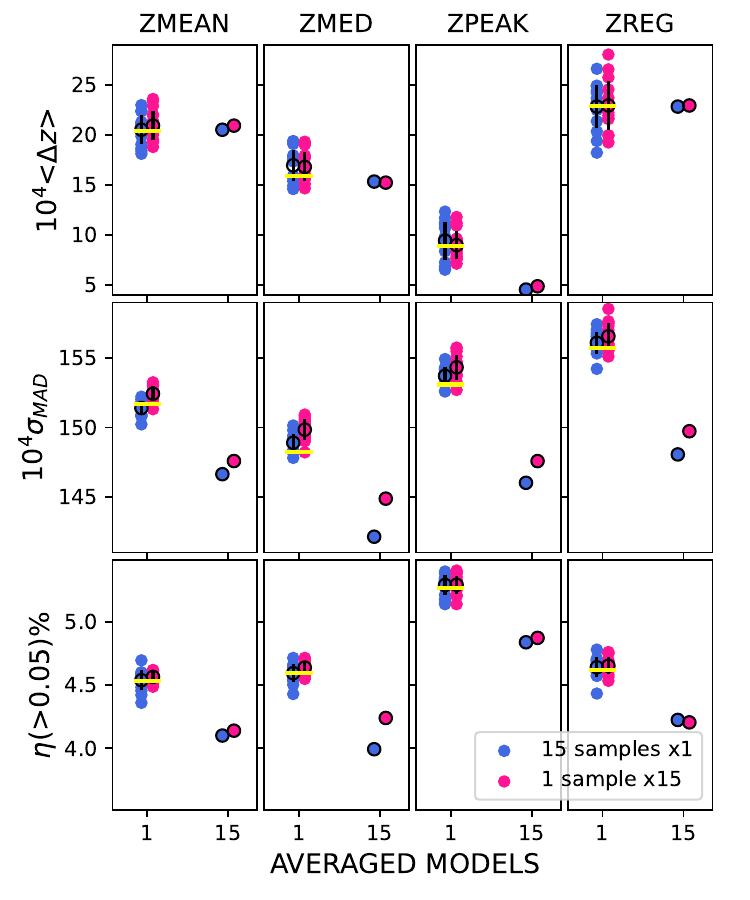}
\caption{The mean bias, $\sigma_{\rm MAD}$ and rate of catastrophic failures of the 4 point estimates in the test sample for the 15 networks trained under two strategies: 15 samples trained once (blue points), 1 sample trained 15 times (pink points). The metric of the joint network is marked by a yellow line. The black circles and vertical lines show the mean and standard deviations. The right most points in each panel result from averaging the 15 outputs.}
\label{fig:metrics_15samples_15runs}
\end{figure}

Figure \ref{fig:metrics_15samples_15runs} shows the different metrics for the different point estimates for the 15 networks trained under the two strategies: 15 samples trained once (blue points), 1 sample trained 15 times (pink points). The yellow lines mark the metrics of the joint network. The scattered points to the left of each panel (N=1) show that the second strategy generates as much variation as the original strategy for $\sigma_{\rm MAD}$ and the bias, but less for the catastrophic failures (except for $z_{peak}$). The general similarity may not be surprising given that the 15 training samples are designed to contain similar galaxies, if not the same. But the overlap between 2 samples is in many cases less than 50\% so the impact of randomly initializing the weights at the start of training is as large as replacing half of the training set with different but similar sources. The right most points in each panel (N=15) show that the metrics resulting from averaging the 15 outputs are slightly poorer in the second scenario. Even if the reverse could presumably have happened, it seems that randomizing 15 samples had a higher chance of reaching lower metrics than randomizing the initial training weights of a unique sample. 

\begin{figure}
\includegraphics[width=8.6cm]{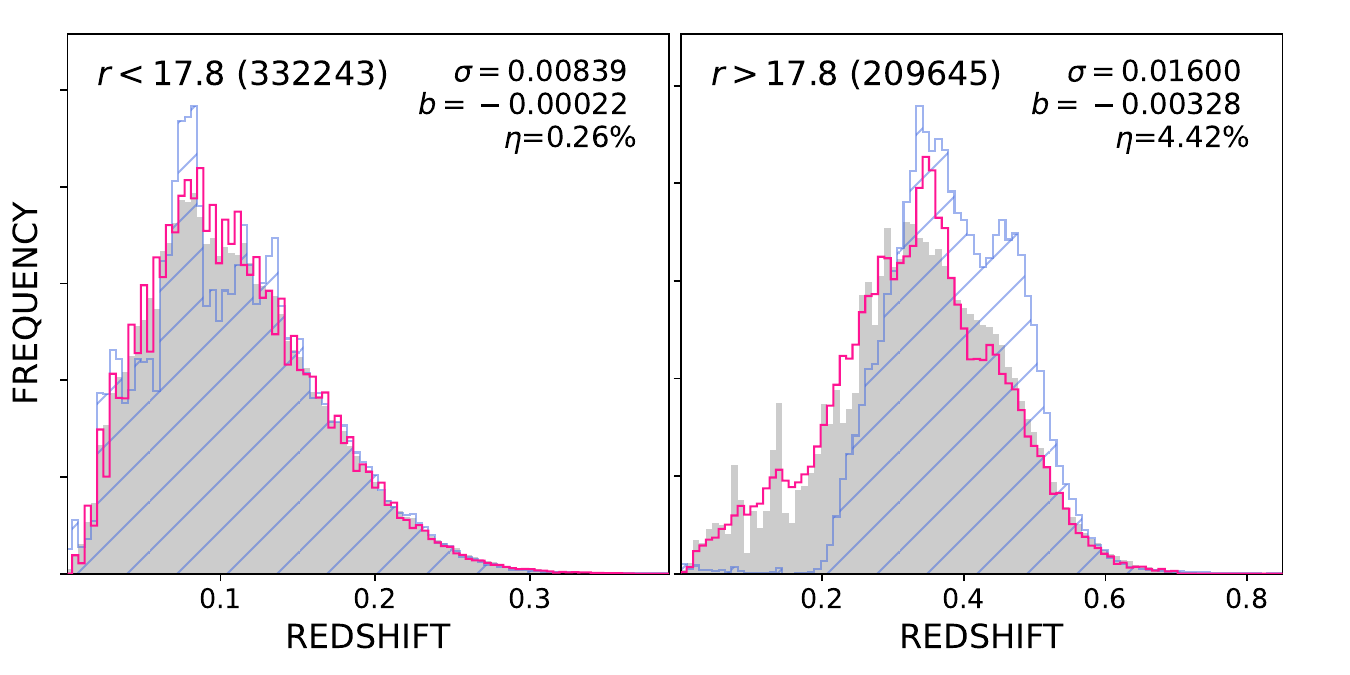}
\caption{The $z_{spec}$ and $z_{med}$ distributions of the additional leftover galaxies at $r<17.8$ (left) and $17.8<r<20$ (right), in gray and pink respectively, with the corresponding metrics. The hatched histograms are the $z_{spec}$ distributions of the leftover galaxies common to both training strategies (shown in Fig. \ref{fig:nz_leftover}).}
\label{fig:nz_leftover_train0_extra}
\end{figure}

As expected from the above results, the metrics found for the leftover galaxies common to both training scenarios (the sample shown in Fig. \ref{fig:nz_leftover}) are also slightly degraded.
%Figure \ref{fig:nz_leftover_train0_incommon} shows the spectroscopic redshift and $z_{med}$ distributions of the leftover galaxies common to both training scenarios (Fig. \ref{fig:nz_leftover}), in blue and pink respectively, with the corresponding metrics in the direction of the arrow. As expected from Fig. \ref{fig:metrics_15samples_15runs}, the metrics are slightly degraded in the second scenario. 
Figure \ref{fig:nz_leftover_train0_extra} shows the normalized spectroscopic redshift and $z_{med}$ distributions of the additional $\sim$542k galaxies left over from the second scenario. The hatched histograms are the spectroscopic redshift distributions of the leftover galaxies in common, for comparison. The bright fraction of this additional sample is smoother than the hatched one, very slightly improving the predictions. On the contrary the faint fraction contains galaxies in the strong peaks of the GAMA redshift distribution on the left flank and a larger fraction of red galaxies in the region of color degeneracy than of LRGs. The bias is reduced, compared to the common sample shown in Fig. \ref{fig:nz_leftover}, but the deviation and rate of catastrophic failure are degraded. While it may be interesting to see the effect of mixing heterogeneous populations, the twice as large number of leftover galaxies does not allow for more informative testing. It confirms that the performance of the network is poor for red galaxies at "normal" redshifts due to the color degeneracy, very poor for LRGs that are deliberately relegated to the leftover sample, and that the predictions are not able to capture strong redshift structures, also deliberately smoothed in the training samples. 

%\begin{figure}
%\includegraphics[width=8.6cm]{nz_leftovers_train0_incommon.pdf}
%\caption{The redshift distributions of the leftover galaxies common to the two training strategies, at $r<17.8$ (left) and $17.8<r<20$ (right). The gray shaded histograms are the spectroscopic redshift distributions. The blue histogram is the $z_{med}$ distribution for the 15 samples trained once, the pink histogram is the $z_{med}$ distribution for 1 sample trained 15 times, with the corresponding metrics in the direction of the arrow. As expected from Fig. \ref{fig:metrics_15samples_15runs}, the metrics are slightly degraded in the second scenario.}
%\label{fig:nz_leftover_train0_incommon}
%\end{figure}

\subsection{Luminous Red Galaxies}
\label{subsec:LRG}

\begin{figure}
\includegraphics[width=8.6cm]{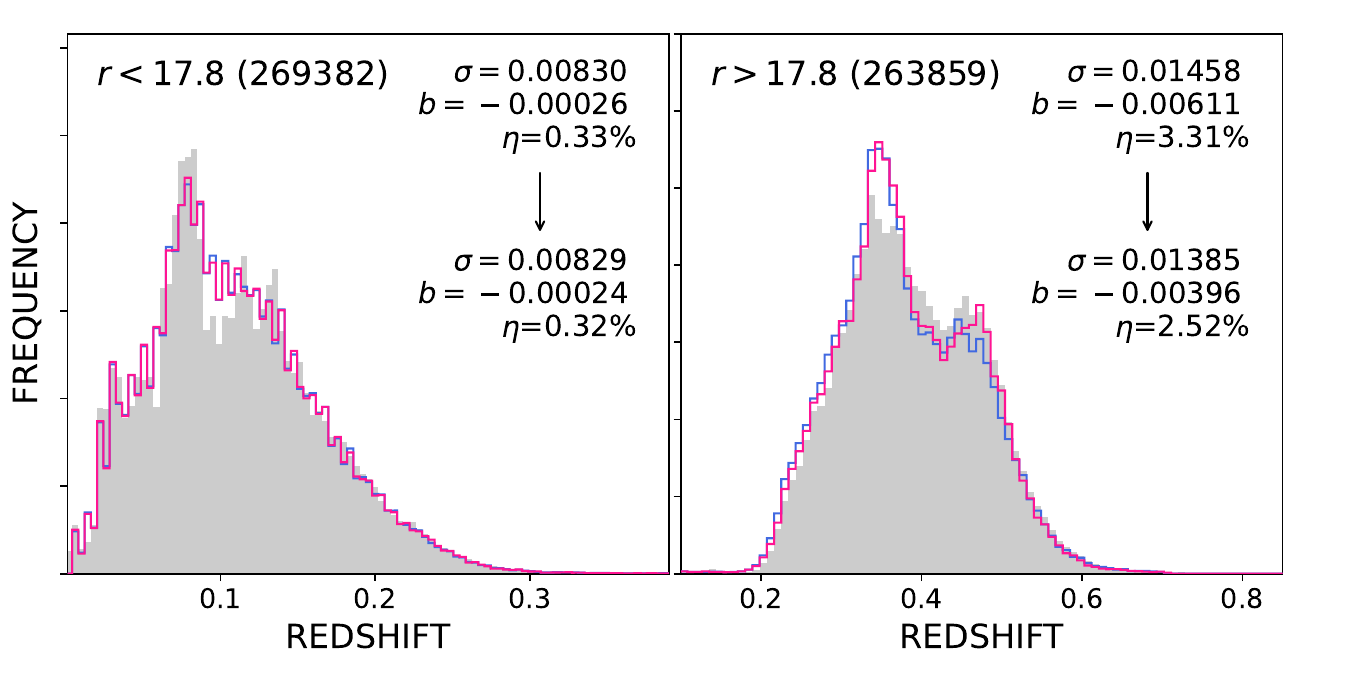}
\caption{The $z_{med}$ distributions of the leftover galaxies at $r<17.8$ and $17.8<r<20$ before and after adding 50k LRGs to the training samples, in blue and pink respectively, with the corresponding metrics in the direction of the arrow. The gray shaded histograms are the $z_{spec}$ distributions. The bright sample is unchanged, the bias in the faint interval is reduced by 35\%. The deviation and rate of catastrophic failure are also improved.}
\label{fig:nz_leftovers_withLRG}
\end{figure}

\begin{figure}
\includegraphics[width=8.6cm]{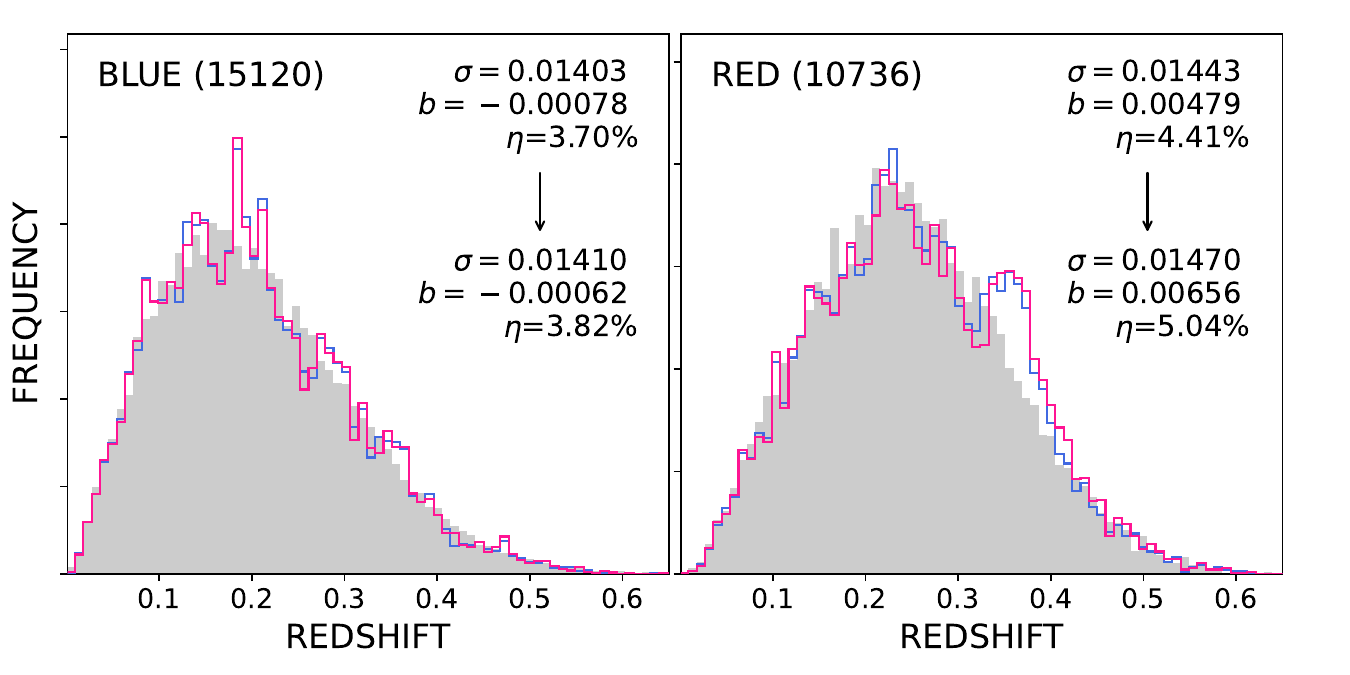}
\caption{The $z_{med}$ distributions of the blue and red galaxies in the test sample before and after adding 50k LRGs to the training samples, in blue and pink respectively, with the corresponding metrics in the direction of the arrow. The gray shaded histograms are the $z_{spec}$ distributions. The blue sample is unchanged, the red galaxy bias is increased by 37\%. The deviation and rate of catastrophic failure are also degraded.}
\label{fig:nz_test_withLRG}
\end{figure}

Figure \ref{fig:nz_leftovers_withLRG} shows the $z_{med}$ distributions of the leftover sample (shown in Fig. \ref{fig:nz_leftover}) at $r<17.8$ and $17.8<r<20$ before and after the addition of 50k LRGs subtracted from it in the training samples. The corresponding metrics are indicated in the direction of the arrow. The bright sample is unchanged, while the bias in the faint interval is reduced by 35\%. The deviation and rate of catastrophic failure are also improved.
However this improvement is at the cost of degrading "normal" galaxies. 
Figure \ref{fig:nz_test_withLRG} shows the $z_{med}$ distributions of the blue and red galaxies in the test sample before and after the LRG addition. The blue sample is unchanged, while the red galaxy bias is increased by 37\%. The deviation and rate of catastrophic failure are also degraded. 

This demonstrates the utmost importance of matching the training and test samples and that more does not necessarily mean better. Although the spectroscopic sample contains 1.5M galaxies, it cannot be used as is for lack of representativeness. Randomly splitting it for training/validation, e.g. 80\%/20\%, as is usually done to evaluate the performance of a network, would yield very misleading results. We choose 50\%/50\% (i.e. 750k randomly selected galaxies for training) to illustrate the point. This is the strategy employed by B16. It more than doubles computing resources (memory and time) compared to the smaller, more representative training samples we adopt. It also doubles the bias for the red population in the test sample ($b=0096$) and further degrades the MAD and rate of catastrophic failure. Meanwhile the metrics on the validation sample are significantly better than on the test sample ($b=-0.00015$, $\sigma_{MAD}=0.011$, $\eta=2\%$) as it matches the training sample by design, in particular the LRG population. The same applies to the performance reported by B16 on their validation sample, which is significantly lower than this however, save for their even smaller bias. We note that LRGs could be used to train an independent network that would improve their redshift estimates in the inference sample, provided we were able to identify them.

\section{CNN architecture}
\label{sec:architecture_apdx}

\begin{figure*}
\includegraphics[width=\textwidth]{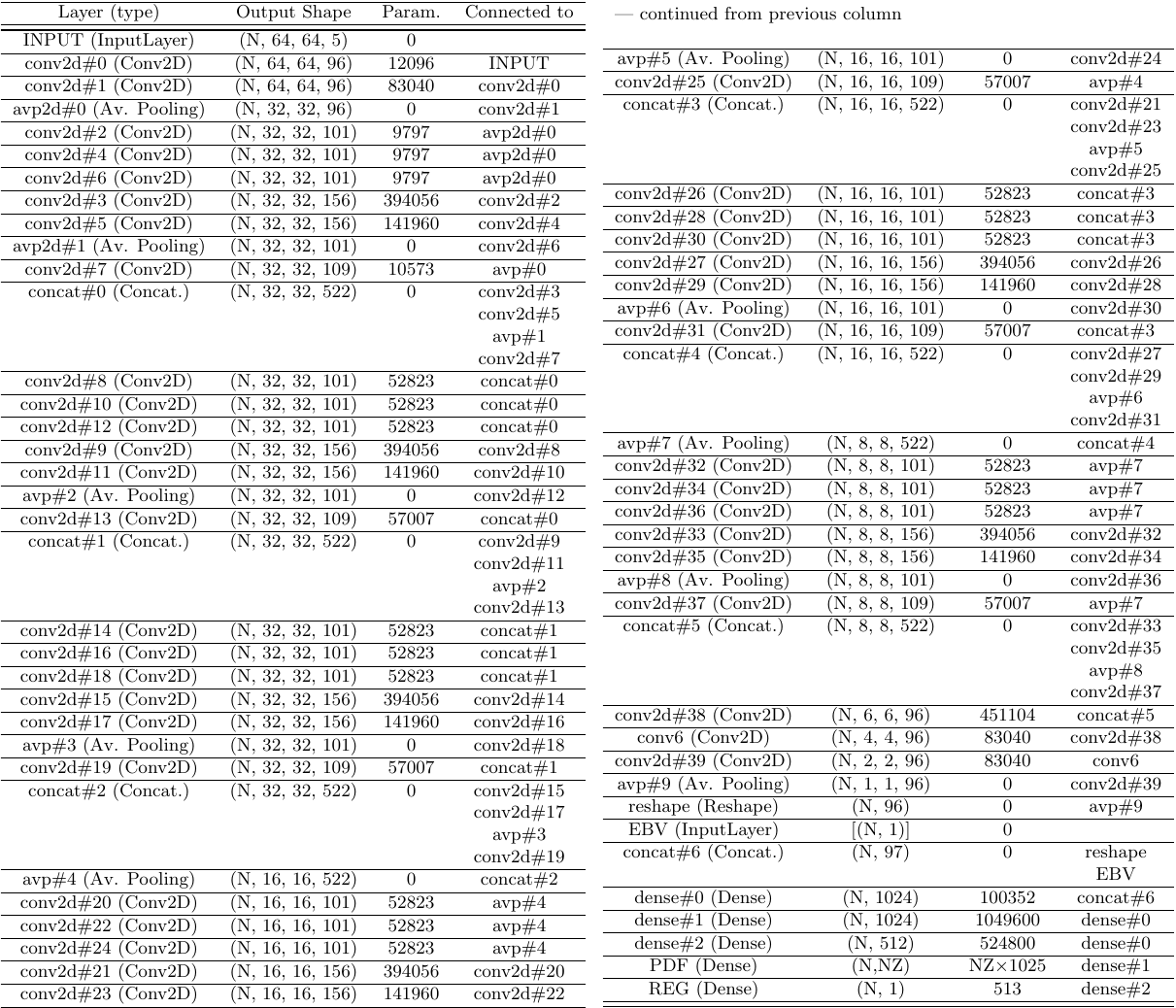}
\caption{Successive CNN layers with their output dimension, number of trainable parameters and the layer(s) that connect(s) to them. The CNN is diagrammed in Fig. \ref{fig:cnn} and \ref{fig:inception}.}
\label{fig:CNN_layers}
\end{figure*}

Figure \ref{fig:CNN_layers} lists all the CNN layers with their type, shape, number of parameters and the layer(s) they are connected to upstream. N is the number of galaxies in a batch (32 for training), NZ in the "PDF" output layer is the number of redshift classes. 

In the case of the SDSS at $r<17.8$ (Section \ref{subsec:sdss}), the CNN is trained for 45 epochs, with a learning rate of $10^{-4}$ from epoch 1 to 30, decreasing by a factor of 10 at epoch 30 and 40. At $r<20$ (Section \ref{subsec:s20go}), the network is trained for 50 epochs, with a learning rate of $10^{-4}$ from epoch 1 to 35, decreasing by a factor of 10 at epoch 35 and 45. 

\bsp	% typesetting comment
\label{lastpage}
\end{document}